\documentclass[12pt]{article}
\usepackage{refcheck}
\makeatletter \@addtoreset{equation}{section} \makeatother
\renewcommand{\theequation}{\thesection.\arabic{equation}}
\newcommand{\ba}{\begin{array}}
\newcommand{\ea}{\end{array}}
\newcommand{\beq}{\begin{equation}}
\newcommand{\eeq}{\end{equation}}
\newcommand{\bea}{\begin{eqnarray}}
\newcommand{\eea}{\end{eqnarray}}

\def\bce{\begin{center}}
\def\ece{\end{center}}

\def\nonu{\nonumber}

\def\pa{\partial}
\def\al{\alpha}
\def\be{\beta}
\def\ga{\gamma}

\def\si{\sigma}

\def\eps6{{\displaystyle \mathop{\epsilon}^{6}}{}}
\def\g6{{\displaystyle \mathop{g}^{6}}{}}

\def\nab6{{\displaystyle \mathop{\nabla}^{6}}{}}

\def\0{{\sst{(0)}}}
\def\1{{\sst{(1)}}}
\def\2{{\sst{(2)}}}
\def\3{{\sst{(3)}}}
\def\4{{\sst{(4)}}}
\def\5{{\sst{(5)}}}
\def\6{{\sst{(6)}}}
\def\7{{\sst{(7)}}}
\def\8{{\sst{(8)}}}

\def\ba{\begin{array}}
\def\ea{\end{array}}
\def\beq{\begin{equation}}
\def\eeq{\end{equation}}
\def\be{\begin{equation}}
\def\ee{\end{equation}}

\def\eps{\epsilon}

\def\ba{\begin{array}}
\def\ea{\end{array}}
\def\beq{\begin{equation}}
\def\eeq{\end{equation}}
\def\be{\begin{equation}}
\def\ee{\end{equation}}

\def\eps{\epsilon}

\def\eps6{{\displaystyle \mathop{\epsilon}^{6}}{}}

\def\nab6{{\displaystyle \mathop{\nabla}^{6}}{}}

\newcommand{\bean}{\begin{eqnarray*}}
\newcommand{\eean}{\end{eqnarray*}}

\begin{document}
\thispagestyle{empty} \addtocounter{page}{-1}
   \begin{flushright}
\end{flushright}

\vspace*{1.3cm}
  
\centerline{ \Large \bf   
Higher Spin Currents} 
\vspace*{0.3cm}
\centerline{ \Large \bf 
in the Enhanced ${\cal N}=3$ Kazama-Suzuki Model } 
\vspace*{1.5cm}
\centerline{{\bf Changhyun Ahn } and {\bf Hyunsu Kim}
} 
\vspace*{1.0cm} 
\centerline{\it 
Department of Physics, Kyungpook National University, Taegu
41566, Korea} 
\vspace*{0.8cm} 
\centerline{\tt ahn@knu.ac.kr, \qquad kimhyun@knu.ac.kr 
} 
\vskip2cm

\centerline{\bf Abstract}
\vspace*{0.5cm}

The ${\cal N}=3$ Kazama-Suzuki model at the `critical' level
has been found by Creutzig, Hikida and Ronne.
We construct the lowest higher spin currents 
of spins $(\frac{3}{2}, 2,2,2,\frac{5}{2}, \frac{5}{2}, \frac{5}{2}, 3)$
in terms of various fermions.  In order to obtain 
the operator product expansions (OPEs) between these higher spin currents,
we describe three ${\cal N}=2$ OPEs between the two 
${\cal N}=2$ higher spin currents denoted by $(\frac{3}{2}, 2, 2, 
\frac{5}{2})$ and $(2, \frac{5}{2}, \frac{5}{2}, 3)$ (corresponding  
$36$ OPEs in the component approach).    
Using the various Jacobi identities, the coefficient functions appearing on 
the right hand side of these ${\cal N}=2$ OPEs are determined 
in terms of central charge completely.
Then we describe them as one single ${\cal N}=3$ OPE in the ${\cal N}=3$
superspace. The right hand side of this ${\cal N}=3$ OPE
contains the $SO(3)$-singlet ${\cal N}=3$ higher spin multiplet 
of spins $(2, \frac{5}{2}, \frac{5}{2},
\frac{5}{2}, 3,3,3, \frac{7}{2})$, 
the $SO(3)$-singlet ${\cal N}=3$ higher spin multiplet 
of spins $(\frac{5}{2}, 3,3,3, \frac{7}{2}, 
\frac{7}{2}, \frac{7}{2}, 4)$,   
and the $SO(3)$-triplet ${\cal N}=3$ higher spin multiplets
where each multiplet 
has the spins 
$(3, \frac{7}{2}, \frac{7}{2}, \frac{7}{2}, 4,4,4, \frac{9}{2})$,
in addition to ${\cal N}=3$ superconformal family of the identity 
operator. 
Finally, by factoring out the spin-$\frac{1}{2}$ current 
of ${\cal N}=3$ linear superconformal algebra generated by 
eight currents of spins $(\frac{1}{2}, 1,1,1, \frac{3}{2}, \frac{3}{2},
\frac{3}{2}, 2)$, we obtain the 
extension of so-called $SO(3)$ nonlinear 
Knizhnik Bershadsky algebra.  

\baselineskip=18pt
\newpage
\renewcommand{\theequation}
{\arabic{section}\mbox{.}\arabic{equation}}

\tableofcontents

\section{Introduction}

One of the remarkable aspects of WZW models is that 
WZW primary fields are also Virasoro primary fields \cite{BS,DMS}. 
The Virasoro zeromode acting on the primary state corresponding to 
the WZW primary field is proportional to the quadratic Casimir operator 
of the finite Lie algebra. Associating a conformal weight (or spin) 
to the primary state,
we find the conformal weight (or spin) which is equal to the one half times 
the quadratic Casimir 
eigenvalues divided by the sum of the level and the dual Coxeter number of the
finite Lie algebra \cite{BS,DMS}.
In particular, the adjoint representation at the `critical' level 
(which is equal to the dual Coxeter number)
has conformal weight $\frac{1}{2}$. For example, 
for $SU(N)$, the quadratic Casimir
eigenvalue is given by $2N$ and the dual Coxeter number is $N$. Then we are 
left with the overall numerical factor $\frac{1}{2}$
which is the spin of adjoint fermion.
For the diagonal coset theory \cite{BS}, the conformal spin-$\frac{3}{2}$
current (${\cal N}=1$ supersymmetry generator) 
that commutes with the diagonal spin-$1$ current can be determined
\cite{Douglas,BBSScon,GS88} and is given by the linear combination of
two kinds of spin-$1$ currents and adjoint fermions. 
For $SU(2)$ case, this leads to the well known coset construction of
the ${\cal N}=1$ superconformal algebra \cite{GKO}.
For $SU(3)$ case, this leads to the coset construction of ${\cal N}=1$ 
$W_3$ algebra \cite{HR,ASS,SS,Ahn1211}.
Moreover, for $SU(N)$, the ${\cal N}=1$ higher spin multiplets
are found in \cite{Ahn1305}.
One can go one step further.
By taking the adjoint spin-$\frac{1}{2}$ 
fermions in the second factor in the numerator
of the diagonal coset model \cite{BS},  
 the coset construction of
the ${\cal N}=2$ superconformal algebra
is obtained \cite{BFK} and the higher spin currents are observed and 
determined in 
\cite{GHKSS,Ahn1604}.

One can consider the above feature in the different coset model.
There exists other type of coset model, so-called Kazama-Suzuki model
\cite{KS1,KS2}
which is described as 
\bea
\frac{\hat{SU}(N+M)_{k} \oplus \hat{SO}(2 N M)_1}{\hat{SU}(N)_{k+M} 
\oplus \hat{SU}(M)_{k+N} 
\oplus \hat{U}(1)_{NM(N+M)(k+N+M)}}, 
\qquad c =\frac{3 k N M}{(k+N+M)}.
\label{KScoset}
\eea
The ${\cal N}=2$ superconformal algebra,
which is generated by the ${\cal N}=2$ multiplet of spins
$(1, \frac{3}{2}, \frac{3}{2}, 2)$,
is realized for arbitrary level $k$ in the coset model (\ref{KScoset}).
When $M=2$, the above coset  is similar to 
the Wolf space coset \cite{Van} 
where the $\hat{SU}(M=2)_{k+N}$ factor in the denominator
of (\ref{KScoset}) is not present.  
At the ``critical'' level where the level $k$ of spin-$1$ current in 
the $SU(N+M)$ factor in the numerator (\ref{KScoset}) is equal to 
the dual Coxeter number $(N+M)$,
\bea
 k =N+M, \qquad c= \frac{3}{2} N M,
\label{critical}
\eea
the above coset model (\ref{KScoset})
has ${\cal N}=3$ supersymmetry \cite{CHR1406,HR1503,CH1506}.
According to the previous analysis, there exist $(N+M)^2-1$ adjoint 
spin-$\frac{1}{2}$ 
fermions
residing in the $SU(N+M)$ factor of the numerator.
Under the decomposition of $SU(N+M)$ into the $SU(N) \times SU(M)$,
the adjoint representation of $SU(N+M)$ breaks into as follows:
${\bf (N+M)^2-1} \rightarrow ({\bf N^2-1},{\bf 1}) \oplus 
({\bf 1}, {\bf M^2-1}) \oplus ({\bf 1},{\bf 1}) \oplus ({\bf N}, 
{\bf \overline{M}}) \oplus ({\bf \overline{N} },{\bf M})$.
The adjoint fermion in the trivial representation $({\bf 1},{\bf 1})$ 
of both $SU(N)$ and $SU(M)$ plays the important role in the construction of
the `extra' currents
consisting of the extra ${\cal N}=2$ multiplet of spins 
$(\frac{1}{2}, 1, 1, \frac{3}{2})$ 
besides the above ${\cal N}=2$ current  
 of spins
$(1, \frac{3}{2}, \frac{3}{2}, 2)$
\cite{Ademolloetal,Ademolloetal1,CK,GS,Schoutens88,BO}.

One way to obtain these `extra' currents is as follows.
Starting with the above adjoint fermion corresponding to 
$({\bf 1},{\bf 1})$, one can calculate the OPEs between this single fermion
(which is the lowest spin-$\frac{1}{2}$ current in the above 
${\cal N}=2$ multiplet) 
and the spin-$\frac{3}{2}$ currents of ${\cal N}=2$ supersymmetry assuming 
that the above ${\cal N}=2$ multiplet of spins $(1, \frac{3}{2}, 
\frac{3}{2}, 2)$ is known explicitly in the coset model (\ref{KScoset}) 
together with (\ref{critical}).
Via the operator product expansions (OPEs) of ${\cal N}=3$
superconformal algebra, one can determine the above two spin-$1$ currents 
residing in the ${\cal N}=2$ multiplet $(\frac{1}{2}, 1, 1, \frac{3}{2})$.  
In other words, the first-order poles of the above OPE 
between the adjoint single fermion and the two spin-$\frac{3}{2}$ currents
provide the two spin-$1$ currents respectively.
Furthermore, by taking one of the spin-$1$ currents and one of the 
spin-$\frac{3}{2}$ currents of ${\cal N}=2$ superconformal algebra
and calculating the OPE between them (their $SO(3)$ indices are different 
from each other), one obtains the first-order pole 
which provides the spin-$\frac{3}{2}$ current residing in the above 
${\cal N}=2$ multiplet $(\frac{1}{2}, 1, 1, \frac{3}{2})$.
Therefore, the ${\cal N}=3$ superconformal algebra generated by the currents
of spins $(\frac{1}{2}, 1, 1, 1, \frac{3}{2}, \frac{3}{2}, \frac{3}{2}, 2)$
by adding the two ${\cal N}=2$ multiplets $(1,\frac{3}{2},\frac{3}{2},2)$ 
and $(\frac{1}{2}, 1,1, \frac{3}{2})$
and computing the OPEs between them
is realized in the enhanced ${\cal N}=3$ 
Kazama-Suzuki model (\ref{KScoset}) 
together with (\ref{critical}) \cite{CHR1406}.  

In this paper,
we study the higher spin currents in the coset model (\ref{KScoset}).
For the ${\cal N}=3$ holography \cite{CHR1406}, 
the deformation breaks the higher spin symmetry 
and induces the mass to the higher spin fields \cite{HR1503,CH1506}.
The masses are not generated for the $SO(3)_R$ singlet higher spin fields 
at the leading order of $\frac{1}{c}$
while the mass formula for the $SO(3)_R$ triplet higher spin fields 
looks like the Regge trajectory on the flat spacetime. 
So far it is not known what is the higher spin 
symmetry algebra
for the higher spin currents together with ${\cal N}=3$ superconformal 
algebra. 
It would be interesting to see the higher spin symmetry algebra
between the low higher spin currents
explicitly. For finite $(N,M)$ (or finite $c$) in the coset model,
we would like to observe the marginal operator which breaks the 
higher spin symmetry keeping the ${\cal N}=3$ supersymmetry.
Furthermore, we should obtain the explicit higher spin symmetry algebra,
where the structure constants on the right hand side of OPEs depend on 
$(N,M)$ explicitly,
in order to calculate the mass formula as in the large $c$ limit 
\cite{HR1503,CH1506}.

We expect that 
 the ${\cal N}=3$ lowest higher spin multiplet
of spins $(\frac{3}{2}, 2, 2, 2, \frac{5}{2}, \frac{5}{2}, \frac{5}{2}, 3)$
can be obtained 
by adding the two ${\cal N}=2$ higher spin 
multiplets $(\frac{3}{2},2,2,\frac{5}{2})$ 
and $(2, \frac{5}{2}, \frac{5}{2}, 3)$ (or by adding the spin one 
to the ${\cal N}=3$ multiplet of ${\cal N}=3$ superconformal algebra
we have described above).
Once we know the higher spin currents explicitly, then
we can perform the OPEs between them and obtain the 
higher spin algebra for low higher spin currents.
Then how we can obtain the higher spin currents explicitly besides
the ${\cal N}=3$ currents generated by 
${\cal N}=3$ superconformal algebra? 
As in the first paragraph, we return to the construction of 
WZW currents in the coset model (\ref{KScoset}).
One of the usefulness of this construction is 
that we can obtain the higher spin currents directly and 
due to the ${\cal N}=3$ supersymmetry, we can determine 
the other higher spin currents after the lowest higher spin current
is fixed in the given ${\cal N}=3$ multiplet. 
For example, once the higher spin-$\frac{3}{2}$ current 
is determined completely, then the higher spin-$2, \frac{5}{2}$
and $3$ currents can be obtained with the help of spin-$\frac{3}{2}$
currents of ${\cal N}=3$ superconformal algebra.

Furthermore, the general feature in the OPE between any two 
quasiprimary currents is used \cite{BS}.
Because the left hand side of any OPE    
can be calculated from the WZW currents explicitly,
the pole structures of the OPEs
are known. From these, we should express them 
in terms of the known ${\cal N}=3$ currents and the known 
higher spin currents by assuming that the right hand sides of 
the OPEs contain any multiple products between them.
If we cannot describe the poles of the OPEs in terms of 
the known (higher spin) currents, then   
we should make sure that the extra terms
should transform as a new (quasi)primary current. 
This will consist of the component of next higher spin multiplet.
When the poles of the OPEs are described by low spin, then 
it is easy to figure out the right candidate for the composite currents 
at each pole of the OPEs. 
As the spins of the left hand side of the OPEs increase, then 
it is not easy to write down all possible terms correctly.
We use the $SO(3)$ index structure in the (higher spin) currents
and according to the $SO(3)$ index structure of the left hand side
of the OPEs, the right hand side of the OPEs should preserve 
the $SO(3)$ invariance. In other words, 
if the left hand side of the OPEs transforms as a singlet, then 
the right hand side of the OPEs should be a singlet under the 
$SO(3)$. Similarly, the $SO(3)$ vector (free) index 
can arise both sides of the 
OPEs and we will see the appearance of the new higher spin currents with
$SO(3)$ vector index.

We would like to construct the 
complete $36$ OPEs between the above 
eight higher spin currents for generic central charge.
Now we can proceed to the ${\cal N}=2$ superspace from 
these component results and all the expressions are given for $(N,M)=(2,2)$.
Of course we can stay at the component approach but we should 
introduce more undetermined quantities we should determine.
Let us replace the structure constants
with  arbitrary coefficients. Then we have the complete OPEs in the 
${\cal N}=2$ superspace with undetermined structure constants.
 We 
use the Jacobi identities to fix the structure constants.
In general, the new ${\cal N}=2$ primary higher spin 
current transforming as 
a primary current under the ${\cal N}=2$ stress energy tensor  
can appear on the right-hand side of the OPEs as the spins of the currents 
increase. The above $8$ higher spin currents can be represented by 
two ${\cal N}=2$ multiplets. Similarly, the $8$ currents of the 
${\cal N}=3$ (linear) superconformal algebra can be combined into 
two ${\cal N}=2$ multiplets (as described before). 
Then we can use the Jacobi identities 
by choosing 
one ${\cal N}=2$ current and two ${\cal N}=2$ higher spin currents.        
We cannot use the Jacobi identities by taking  
three ${\cal N}=2$ higher spin 
currents because, if we consider the OPE between 
any ${\cal N}=2$ higher spin currents and another new ${\cal N}=2$ higher spin 
current, we do not know this OPE at this level.
Therefore, the three quantities used for the Jacobi identities
are given by one ${\cal N}=2$ current and two ${\cal N}=2$ higher spin
currents. We can also consider the combination of one ${\cal N}=2$ higher
spin current and two ${\cal N}=2$ currents, 
but this will do not produce any nontrivial equations
for the unknown coefficients. They are satisfied trivially.

After we obtain the complete three ${\cal N}=2$ OPEs, then 
it is straightforward to write down them as a single 
${\cal N}=3$ OPE (or as the component results). 
We observe that on the right hand side of the 
${\cal N}=3$ OPE, there exist 
three types of ${\cal N}=3$ higher spin multiplets.
There are two $SO(3)$ singlets of spins 
$(2,\frac{5}{2}, \frac{5}{2}, \frac{5}{2},3,3,3,\frac{7}{2})$
and $(\frac{5}{2}, 3,3,3, \frac{7}{2}, \frac{7}{2}, \frac{7}{2},4)$
and one $SO(3)$ triplet 
where  each multiplet 
has the spins 
$(3, \frac{7}{2}, \frac{7}{2}, \frac{7}{2}, 4,4,4, \frac{9}{2})$. 
The presence of $SO(3)$ triplet higher spin multiplet 
is crucial to the ${\cal N}=3$ OPE.
We observe that the $SO(3)$ vector index for the 
last ${\cal N}=3$ higher spin multiplet  
is contracted with the one 
appearing in the fermionic coordinates of ${\cal N}=3$ superspace. 
In other words, one can use the $SO(3)$ invariant tensor 
of rank $3$ and make a contraction 
with both two fermionic coordinates and the above 
${\cal N}=3$ $SO(3)$ triplet higher spin current. 
Because the spin of the first-order pole with two fermionic 
coordinates is given by zero, the sum of the two spin $\frac{3}{2}$
of the left hand side should appear on the right hand side.   
Note that the above lowest higher spin multiplet 
of spins 
$(\frac{3}{2}, 2,2,2, \frac{5}{2}, \frac{5}{2}, \frac{5}{2}, 3)$
is a $SO(3)$ singlet.

In section $2$,
we review the ${\cal N}=3$ stress energy tensor, 
the primary higher spin multiplets and the realization
of the ${\cal N}=3$ superconformal algebra in the above coset model. 

In section $3$,
we construct the lowest eight higher spin currents 
for generic central charge $c$ explicitly.

In section $4$,
we obtain the fundamental OPEs between the higher spin currents 
found in previous section for $(N,M)=(2,2)$ case. 

In section $5$,
we present how we can determine the next higher spin currents.

In section $6$,
we calculate the Jacobi identities for the three ${\cal N}=2$ OPEs
and determine the structure constants completely.

In section $7$,
based on the previous section, the component result can be 
obtained. Furthermore, we describe its ${\cal N}=3$ OPE.

In section $8$,
by factoring out the spin-$\frac{1}{2}$ current 
of ${\cal N}=3$ superconformal algebra, we obtain 
the (minimally) 
extended Knizhnik Bershadsky algebra \cite{Knizhnik,Bershadsky}.  

In Appendices $A, B, \cdots, H$,
we describe some details which are necessary to the previous sections.

The packages \cite{Thielemans,KT} are used all the times.

Some of the relevant works in the context of \cite{KS1,KS2}
are given in \cite{GK1604}-\cite{CHR1111}. 

\section{The eight currents of ${\cal N}=3$ superconformal algebra
in the coset model: Review}

In this section, we describe the $8$ currents of the 
${\cal N}=3$ (linear) superconformal algebra 
in the ${\cal N}=3$ superspace, where $SO(3)$ symmetry 
is manifest. 
Then the corresponding 
${\cal N}=3$ superconformal algebra, which consists of 
$9$ nontrivial OPEs  in the component approach (in Appendix $A$), 
can be expressed in terms of
a single ${\cal N}=3$ (super) OPE.
We describe the ${\cal N}=3$ (super) primary higher spin current,
in an $SO(3)$ symmetric way,
under the ${\cal N}=3$ stress energy tensor.
The ${\cal N}=3$ higher spin current, in general,  
transforms as a nontrivial 
representation under  group $SO(3)$.
Furthermore, the superspin is, in general, 
given by the positive integer or half integer $\Delta$, 
but its lowest value $\Delta=\frac{3}{2}$
will be considered later 
when the OPEs between them are calculated for generic central charge. 
The OPEs between the $8$ currents   and the 
$8$ higher spin currents in the component approach are also given
(in Appendix $B$).

\subsection{ The ${\cal N}=3$ stress energy tensor}

The ${\cal N}=3$ stress energy tensor can be described as \cite{CK}
\bea
{\bf J}(Z) & = & \frac{1}{2} i \, \Psi(z) + \theta^i \, \frac{i}{2} J^i(z) +
\theta^{3-i} \, \frac{1}{2} G^i(z) + \theta^{3-0} \, T(z)
\nonu \\
& = &  \frac{1}{2} i \, \Psi(z) + \theta^1 \, \frac{i}{2} J^1(z) 
+   \theta^2 \, \frac{i}{2} J^2(z)
+  \theta^3 \, \frac{i}{2} J^3(z) \nonu \\
& + & \theta^1 \theta^2 \, \frac{1}{2} G^3(z) + \theta^2 \theta^3 \, 
\frac{1}{2} G^1(z) +
\theta^3 \theta^1 \, \frac{1}{2} G^2(z) + \theta^1 \theta^2 \theta^3 \, T(z)
\nonu \\
 & \equiv & \Bigg( \frac{i}{2}  \Psi, \quad \frac{i}{2} J^i, \quad
\frac{1}{2} G^i, \quad T \Bigg). 
\label{j3}
\eea
The  ${\cal N}=3$  superspace 
coordinates  can be described as 
$(Z, \overline{Z}),$ where 
$Z=(z, \theta^i)$, $\overline{Z} =(\bar{z},$ $\bar{\theta}^i)$,
and $i =1,2,3$ and the index $i$ is the $SO(3)$-vector index. 
The left covariant spinor derivative 
is given by 
$ D^i = \theta^i \frac{\pa}{\pa z } +  \frac{\pa}{\pa {\theta^i}}$
and satisfies 
the anticommutators: 
$
\{ D^i, D^j \} = 2 \delta^{ij} \frac{\pa}{\pa z}$
where the Kronecker delta $\delta^{ij}$ is the rank $2$ $SO(3)$
symmetric invariant tensor.
In the first line of (\ref{j3}), the summation over 
repeated indices (note  
that the ${\cal N}=3$ stress energy tensor 
${\bf J}(Z)$ is an $SO(3)$-singlet) 
is taken.\footnote{
We use  boldface notation for the ${\cal N}=3$ or ${\cal N}=2$ 
multiplet to emphasize 
the fact that the corresponding multiplet has many component currents.
For the ${\cal N}=3$ multiplet,  $8$ independent component currents arise
while, for the ${\cal N}=2$ multiplet, $4$ independent 
components arise.} 
The simplified notation  
$\theta^{3-0}$ is used for 
$\theta^1 \,\theta^2 \,\theta^3$. 
The complement $3-i$ is defined such that $\theta^1 \, \theta^2 \, \theta^3 
= \theta^{3-i}\,
\theta^{i}$ (no sum over $i$).
In the second line of (\ref{j3}), 
the complete $8$ currents for the ${\cal N}=3$ stress energy tensor 
are described in an expansion of Grassmann coordinates 
completely. The quartic- and higher-order terms in $\theta^i$
vanish owing to the property of $\theta^i$. 
The $8$ currents are given by
a single spin-$\frac{1}{2}$ current $\Psi(z)$, 
three spin-$1$ currents
$J^i(z)$ transforming as  a vector representation under  $SO(3)$,
three spin-$\frac{3}{2}$ currents $G^{i}(z)$
transforming as a vector representation under  
$SO(3)$, and 
the spin-$2$ current $T(z)$. 
In particular, the spin-$\frac{1}{2}$ and spin-$2$ 
currents are $SO(3)$ singlets.
The spin of $\theta^i$ is given by 
$-\frac{1}{2}$ (and the covariant spinor derivative
$D^i$ has spin $\frac{1}{2}$)
and therefore the ${\cal N}=3$ (super)spin of the stress energy tensor
${\bf J}(Z)$ 
is equal to $\frac{1}{2}$.
Each term in (\ref{j3}) has a spin-$\frac{1}{2}$ value. 

The ${\cal N}=3$ OPE between the ${\cal N}=3$
stress energy tensor and itself  can be summarized by 
\cite{Ademolloetal,Ademolloetal1,CK,Schoutens88}
{\small
\bea
{\bf J}(Z_1) \, {\bf J}(Z_2) &=& 
-
\frac{1}{z_{12}} \, \frac{c}{12}+
\frac{\theta_{12}^{3-0}}{z_{12}^2} \, \frac{1}{2} {\bf J}(Z_2) +
\frac{\theta_{12}^{3-i}}{z_{12}} \, \frac{1}{2} D^i {\bf J}(Z_2) +
\frac{\theta_{12}^{3-0}}{z_{12}} \,  \pa {\bf J}(Z_2)  +\cdots,
\label{j3j3}
\eea}
where  summation over the 
repeated indices is assumed (the OPE between the 
$SO(3)$-singlet current and itself),  
the fermionic coordinate difference for given index $i$ is defined as
 $ \theta_{12}^i = \theta_1^i-\theta_2^i$, and the bosonic
coordinate difference is given by $z_{12} = z_1 -z_2 -
\theta_1^i \theta_2^i$.
Note that there exists ${\bf J}(Z_2)$-term on the right-hand side 
of (\ref{j3j3}). 
Also the explicit component results (which will be described in
next section) will be given in Appendix $A$.
\footnote{
\label{projection}
By assuming that there are four types of OPEs
between the four component currents and the spin-$\frac{1}{2}$ current, 
$\Psi(z) \, \Psi(w)$,
$J^i(z) \, \Psi(w)$, $G^i(z) \, \Psi(w)$, and $T(z) \, \Psi(w)$, we 
can also write down the ${\cal N}=3$ OPE  in (\ref{j3j3}).
By simply taking $\theta_1^i = \theta_2^i =0$ in the equation (\ref{j3j3}),
we observe that the coefficient of the first term of (\ref{j3j3})
can be obtained from the OPE $\Psi(z) \, \Psi(w)$. 
The third term of the right hand side of (\ref{j3j3})
by acting the differential operator $D_1^{3-i}$
and setting $\theta_1^i = \theta_2^i =0$
can be obtained from the OPE $G^i(z) \, \Psi(w)$. 
Similarly, by acting $D_1^1 D_1^2 D_1^3$ on the equation (\ref{j3j3})
and putting $\theta_1^i = \theta_2^i =0$, the singular terms 
can be obtained from the OPE $T(z) \, \Psi(w)$. 
The regularity of the OPE $J^i(z) \, \Psi(w)$ implies that there is no 
linear in $\theta_{12}^i$ in the equation (\ref{j3j3}).  }

\subsection{ The ${\cal N}=3$ primary higher spin multiplet}

For general superspin $\Delta$
with nontrivial representation $\alpha$ for the 
$SO(3)$, the ${\cal N}=3$ (higher spin) multiplet 
can be described as 
\bea
{\bf \Phi}_{\Delta}^{\alpha}(Z)  & = &  
 \frac{i}{2}  \, \psi_{\Delta}^{\alpha}(z) + \theta^i \, \frac{i}{2} 
\phi^{i,\alpha}_{\Delta+\frac{1}{2}}(z) +
\theta^{3-i} \, \frac{1}{2} \psi_{\Delta+1}^{i,\alpha}(z) + 
\theta^{3-0} \, \phi^{\alpha}_{\Delta+\frac{3}{2}}(z)
\nonu \\
&=& 
 \frac{i}{2}  \, \psi_{\Delta}^{\alpha}(z) + \theta^1 \, \frac{i}{2} 
\phi^{1,\alpha}_{\Delta+\frac{1}{2}}(z) +
 \theta^2 \, \frac{i}{2} 
\phi^{2,\alpha}_{\Delta+\frac{1}{2}}(z) +
 \theta^3 \, \frac{i}{2} 
\phi^{3,\alpha}_{\Delta+\frac{1}{2}}(z)
\nonu \\
& + & \theta^1 \theta^2 \frac{1}{2} 
\, \psi^{3,\alpha}_{\Delta+1}(z)
+ \theta^2 \theta^3 \frac{1}{2} 
\, \psi^{1,\alpha}_{\Delta+1}(z)
+ \theta^3 \theta^1 \frac{1}{2} 
\, \psi^{2,\alpha}_{\Delta+1}(z)
+ \theta^1 \theta^2 \theta^3 
 \, \phi^{\alpha}_{\Delta+\frac{3}{2}}(z)
\nonu \\
& \equiv &
\Bigg(\frac{i}{2} \psi_{\Delta}^{\alpha}, 
\quad 
\frac{i}{2} \phi_{\Delta+\frac{1}{2}}^{i,\alpha}, \quad
\frac{1}{2} \psi_{\Delta+1}^{i,\alpha}, 
\quad \phi_{\Delta+\frac{3}{2}}^{\alpha} \Bigg)
\label{BigPhiexpression}
\eea
In components, 
there exist
a single higher spin-$\Delta$ current $\psi_{\Delta}^{\alpha}(z)$, three higher 
spin-$(\Delta+\frac{1}{2})$ 
currents
$\phi_{\Delta+\frac{1}{2}}^{i, \alpha}(z)$ 
transforming as  a vector representation under  $SO(3)$,
three higher spin-$(\Delta+1)$ currents $\psi_{\Delta +1}^{i,\alpha}(z)$
transforming as a vector representation under  
$SO(3)$, and the  higher spin-$(\Delta+\frac{3}{2})$ 
currents $\phi_{\Delta+\frac{3}{2}}^{\alpha}(z)$ for given $\alpha$ 
representation. \footnote{It will turn out that 
we obtain the $SO(3)$ vector representation for the index $\alpha$. } 
Depending on the superspin $\Delta$, 
the above ${\cal N}=3$ (higher spin) multiplet 
is a bosonic higher spin current for integer spin $\Delta$ or 
a fermionic higher spin current for half-integer spin $\Delta$.

Because the superspin of ${\bf J}(Z_1)$ 
is $\frac{1}{2}$, the right-hand side 
of the OPE 
${\bf J}(Z_{1}) \, {\bf \Phi}_{\Delta}^{\alpha}(Z_{2})$
has a superspin $(\Delta+ \frac{1}{2})$. The pole structure of the 
linear term in $ {\bf \Phi}_{\Delta}^{\alpha}(Z_{2})$ on the
right-hand side should have  spin $\frac{1}{2}$ without any $SO(3)$ indices.
This implies that the structure should be $\frac{\theta_{12}^{3-0}}{z_{12}^{2}}$,
where the spin of $\frac{1}{z_{12}}$ 
is equal to $1$ and the spin of $\theta_{12}$
is equal to $-\frac{1}{2}$. The ordinary derivative term can occur
at the singular term $\frac{\theta_{12}^{3-0}}{z_{12}}$. Furthermore, 
the spinor derivative terms (descendant terms) 
with a quadratic product of $\theta_{12}$ arise.
One should also consider the ${\cal N}=3$ higher spin multiplet
${\bf \Phi}^{\beta}_{\Delta}(Z_2)$ 
with different index $\beta$ with the contraction of
$SO(3)$ generator ${\bf T}^i$. \footnote{We have 
them explicitly as follows:
${\bf T}^1=
\left(
\begin{array}{ccc}
 0 & 0 & 0 \nonu \\
0 & 0 & -i \nonu \\
0 & i & 0 \nonu
\end{array}
\right)$, 
${\bf T}^2=
\left(
\begin{array}{ccc}
 0 & 0 & i \nonu \\
0 & 0 & 0 \nonu \\
-i & 0 & 0 \nonu
\end{array}
\right)
$,
${\bf T}^3=
\left(
\begin{array}{ccc}
 0 & -i & 0 \nonu \\
i & 0 & 0 \nonu \\
0 & 0 & 0 \nonu
\end{array}
\right)
$ with the relation $[{\bf T}^i, {\bf T}^j] = i \epsilon^{ijk} {\bf T}^k$.}
Due to the $SO(3)$ index $i$ of ${\bf T}^i$, we 
should consider the linear $\theta_{12}^i$ term 
which contracts with the above ${\bf T}^i$.
Finally,    we obtain the following ${\cal N}=3$ primary condition
for the $8$ higher spin currents in the ${\cal N}=3$ superspace 
\cite{CK,Schoutens88}:
{\small
\bea
{\bf J}(Z_1) \, {\bf \Phi}_{\Delta}^{\alpha}(Z_2) &=& 
\frac{\theta_{12}^{3-0}}{z_{12}^2} \,  \Delta \, {\bf \Phi}_{\Delta}^{\alpha}(Z_2) +
\frac{\theta_{12}^{3-i}}{z_{12}} \, \frac{1}{2} D^i 
{\bf \Phi}_{\Delta}^{\alpha}(Z_2) +
\frac{\theta_{12}^{3-0}}{z_{12}} \,  \pa {\bf \Phi}_{\Delta}^{\alpha}(Z_2) \nonu \\
& - &
\frac{\theta_{12}^{i}}{z_{12}} \, \frac{i}{2} \, ({\bf T}^i)^{\alpha \beta} \, 
{\bf \Phi}_{\Delta}^{\beta}(Z_2)  +\cdots.
\label{jphi}
\eea}
where $\theta_{12}^{3-0} = \theta_{12}^1 \, \theta_{12}^2 \,
\theta_{12}^3$.
By assuming that there exist four types of OPEs
between the four component currents and the spin-$\frac{1}{2}$ current, 
$\Psi(z) \, \psi_{\Delta}^{\alpha}(w)$,
$J^i(z) \, \psi_{\Delta}^{\alpha}(w)$, 
$G^i(z) \, \psi_{\Delta}^{\alpha}(w)$, 
and $T(z) \, \psi_{\Delta}^{\alpha}(w)$, we 
can write down the ${\cal N}=3$ OPE  in (\ref{jphi})
along the line of the footnote \ref{projection}.
By simply taking $\theta_1^i = \theta_2^i =0$ in the equation (\ref{jphi}),
we observe that the right hand side vanishes
which can be seen from the OPE $\Psi(z) \, \psi_{\Delta}^{\alpha}(w)$
which is regular in Appendix $B$. 
The second term of the right hand side of (\ref{jphi})
by acting the differential operator $D_1^{3-i}$
and setting $\theta_1^i = \theta_2^i =0$
can be obtained from the OPE $G^i(z) \, \psi_{\Delta}^{\alpha}(w)$
in Appendix $B$. 
Similarly, by acting $D_1^1 D_1^2 D_1^3$ on the equation (\ref{jphi})
and putting $\theta_1^i = \theta_2^i =0$, the singular terms 
can be obtained from the OPE $T(z) \, \psi_{\Delta}^{\alpha}(w)$
in Appendix $B$. Finally, 
  by acting $D_1^i $ on the equation (\ref{jphi})
and putting $\theta_1^i = \theta_2^i =0$, the singular term of 
the last term in (\ref{jphi}) 
can be seen from the OPE $ J^i(z) \, \psi_{\Delta}^{\alpha}(w)$
in Appendix $B$.

We use the following notations for the 
$SO(3)$-singlet and $SO(3)$-triplet ${\cal N}=3$ higher spin multiplet
respectively as follows: 
\bea
{\bf \Phi}_{\Delta}^{\alpha =0}(Z) \rightarrow {\bf \Phi}^{(\Delta)}(Z),
\qquad {\bf \Phi}_{\Delta}^{\alpha}(Z) \rightarrow {\bf \Phi}^{(\Delta), \alpha}(Z).
\label{othernotation}
\eea
Furthermore, we have the component currents, from (\ref{BigPhiexpression})
and (\ref{othernotation}), 
\bea
{\bf \Phi}^{(\frac{3}{2})}(Z)  & = &  
 \frac{i}{2}  \, \psi^{(\frac{3}{2})}(z) + \theta^i \, \frac{i}{2} 
\phi^{(2),i}(z) +
\theta^{3-i} \, \frac{1}{2} \psi^{(\frac{5}{2}),i}(z) + 
\theta^{3-0} \, \phi^{(3)}(z),
\nonu \\
{\bf \Phi}^{(2)}(Z)  & = &  
 \frac{i}{2}  \, \psi^{(2)}(z) + \theta^i \, \frac{i}{2} 
\phi^{(\frac{5}{2}),i}(z) +
\theta^{3-i} \, \frac{1}{2} \psi^{(3),i}(z) + 
\theta^{3-0} \, \phi^{(\frac{7}{2})}(z),
\nonu \\
{\bf \Phi}^{(\frac{5}{2})}(Z)  & = &  
 \frac{i}{2}  \, \psi^{(\frac{5}{2})}(z) + \theta^i \, \frac{i}{2} 
\phi^{(3),i}(z) +
\theta^{3-i} \, \frac{1}{2} \psi^{(\frac{7}{2}),i}(z) + 
\theta^{3-0} \, \phi^{(4)}(z),
\nonu \\
{\bf \Phi}^{(3), \alpha}(Z)  & = &  
 \frac{i}{2}  \, \psi^{(3),\alpha}(z) + \theta^i \, \frac{i}{2} 
\phi^{(\frac{7}{2}),i,\alpha}(z) +
\theta^{3-i} \, \frac{1}{2} \psi^{(4),i,\alpha}(z) + 
\theta^{3-0} \, \phi^{(\frac{9}{2}),\alpha}(z).
\label{BigPhis}
\eea
We will see that the OPE
between the first higher spin multiplet and itself 
leads to the right hand side containing the remaining 
higher spin multiplets in (\ref{BigPhis}) only.
We expect that the other higher spin multiplets beyond the 
above ones will appear in the OPEs between the next lowest 
higher spin multiplets.

\subsection{ The realization of ${\cal N}=3$ superconformal algebra }

As described in the introduction,
we follow the notations used in \cite{CHR1406}.
The $\al =1, 2, \cdots, N^2-1$ stands for the adjoint representation of
$SU(N)$. The $a =1,2, \cdots, N$ stands for the fundamental representation
of $SU(N)$. Similarly,
$\bar{b} =1,2, \cdots, N$ stands for the anti-fundamental representation
of $SU(N)$.
The $\rho=1, 2, \cdots, M^2-1$ denotes 
the adjoint representation of $SU(M)$, 
   the $i =1,2, \cdots, M$ denotes the fundamental representation
of $SU(M)$ and  
the $\bar{j} =1,2, \cdots, N$ denotes  
the anti-fundamental representation
of $SU(M)$.
The OPEs between the spin-$\frac{1}{2}$ currents 
and the spin-$1$ currents are presented in Appendix $C$.
The adjoint fermions of $SU(N+M)$ are 
denoted by $\Psi^{\al}(z)$, $\Psi^{a\bar{i}}(z)$, $\Psi^{\bar{a} i}(z)$,
$\Psi^{\rho}(z)$ and $\Psi^{u(1)}(z)$
while the vector representation fermions of $SO(2 N M)$  
are denoted by $\psi^{a\bar{i}}(z)$ and $\psi^{\bar{a} i}(z)$.

The realization of the ${\cal N}=3$ (linear) superconformal algebra 
has been obtained in \cite{CHR1406} 
and is summarized by 
{\small
\bea
\Psi(z) & = & \sqrt{\frac{N M}{2}} \, \Psi^{u(1)}(z), 
\nonu \\
J^1(z) & = & \frac{1}{2} 
\Bigg( \delta_{a\bar{b}} \, \delta_{\bar{i} j} \, 
\Psi^{a\bar{i}} \, \psi^{\bar{b} j} +  \delta_{a\bar{b}} \, \delta_{\bar{i} j} \, 
\psi^{a\bar{i}} \, \Psi^{\bar{b} j} \Bigg)(z), 
\nonu  \\
J^2(z) & = & -\frac{i}{2} 
\Bigg( \delta_{a\bar{b}} \, \delta_{\bar{i} j} \, 
\Psi^{a\bar{i}} \, \psi^{\bar{b} j} -  \delta_{a\bar{b}} \, \delta_{\bar{i} j} \, 
\psi^{a\bar{i}} \, \Psi^{\bar{b} j} \Bigg)(z), 
\nonu \\
J^3(z) & = & 
\frac{1}{2}  \delta_{a\bar{b}} \, \delta_{\bar{i} j}
\Bigg( 
\Psi^{a\bar{i}} \, \Psi^{\bar{b} j}  - \psi^{a\bar{i}} \, \psi^{\bar{b} j}
\Bigg)(z),
\nonu \\
G^1(z) & = & \frac{1}{\sqrt{2(N+M)}} \Bigg( 
\delta_{a\bar{b}} \, \delta_{\bar{i} j} \, 
J^{a\bar{i}} \, \psi^{\bar{b} j} +  \delta_{ \bar{a} b} \, \delta_{i \bar{j}} \, 
J^{ \bar{a} i} \, \psi^{b \bar{j}}
\Bigg)(z),
\nonu \\
G^2(z) & = & -\frac{i}{\sqrt{2(N+M)}} \Bigg( 
\delta_{a\bar{b}} \, \delta_{\bar{i} j} \, 
J^{a\bar{i}} \, \psi^{\bar{b} j} -  \delta_{ \bar{a} b} \, \delta_{i \bar{j}} \, 
J^{ \bar{a} i} \, \psi^{b \bar{j}}
\Bigg)(z),
\nonu \\
G^3(z) & = & \frac{1}{\sqrt{2(N+M)}} \Bigg( 
\Psi^{\alpha} (J_2^{\alpha} - j^{\alpha}) + \Psi^{\rho}(J_2^{\rho}-j^{\rho})\Bigg)
+  \frac{1}{\sqrt{2N M}} \Psi^{u(1)} (\hat{J}^{u(1)} -j^{u(1)})(z),
\nonu \\
T(z) & = & \frac{1}{4(N+M)} \Bigg(
J^{\alpha} J^{\alpha}  + J^{\rho} J^{\rho} + 
J^{a\bar{i}} \,  J^{\bar{a} i} 
+ J^{\bar{a} i} \,  J^{a \bar{i}}
+ J^{u(1)} J^{u(1)} \Bigg)
\label{n3sca}
\\
& 
- & \frac{1}{2} \delta_{a\bar{b}} \, \delta_{\bar{i} j}  \Bigg( 
\psi^{a\bar{i}} \,  \pa \psi^{\bar{b} j} - \pa 
\psi^{a\bar{i}} \,   \psi^{\bar{b} j}
\Bigg)(z)
-\frac{1}{4(N+M)} (J^{\alpha}+j^{\alpha})(J^{\alpha}+j^{\alpha})(z)
\nonu \\
& - & 
\frac{1}{4(N+M)} (J^{\rho}+j^{\rho})(J^{\rho}+j^{\rho})(z)
-\frac{1}{4 N M} (\hat{J}^{u(1)}+j^{u(1)})(\hat{J}^{u(1)}+j^{u(1)})(z).
\nonu
\eea}
Among these currents in (\ref{n3sca}), 
the ${\cal N}=3$ supersymmetry spin-$\frac{3}{2}$ currents 
$G^i(z)$ are used frequently in order to determine 
the $8$ higher spin currents in next section. 
The various spin-$1$ currents are defined in Appendix $C$.

\section{The lowest eight higher spin currents}

We would like to construct the lowest eight higher spin currents 
in terms of coset fermions in the spirit of 
\cite{Ahn1111,Ahn1202,AK1308,AK1411,AK1506,AKP1510}.
We have checked that there is no nontrivial  higher spin-$1$ current.

\subsection{The higher spin-$\frac{3}{2}$ current}

Let us consider the following 
ansatz for the lowest higher spin-$\frac{3}{2}$ current,  
based on the several $(N,M)$ cases,  
\bea
\bar{\psi}^{(\frac{3}{2})}(z)= \Bigg[  
a_1 J_1^{\alpha} \Psi^{\alpha}+a_2 J_2^{\alpha} \Psi^{\alpha}+a_3 j^{\alpha} \Psi^{\alpha}
+  b_1 J_1^{\rho} \Psi^{\rho} +b_2 J_2^{\rho} \Psi^{\rho}+ b_3 j^{\rho} \Psi^{\rho} 
\Bigg](z).
\label{psiansatz}
\eea
The spin-$1$ currents in terms of fermions 
are presented in Appendix $C$.
The relative coefficients 
should be determined.  
We will determine the normalized lowest higher spin-$\frac{3}{2}$ current 
later.
This ansatz should satisfy the ${\cal N}=3$ primary conditions 
and the regular (with denominator currents in the coset model)  conditions. 
One of the ${\cal N}=3$ primary conditions (that is, the fifth equation of 
Appendix (\ref{jphicomp})) is given by  
\bea
J^i(z) \, \bar{\psi}^{(\frac{3}{2})}(w)= + \cdots.
\label{jpsicondition}
\eea
This condition (\ref{jpsicondition}) requires 
the relations between the coefficients $a_3=a_2$ and $b_3=b_2$. 
Then the above expression (\ref{psiansatz})
can be written as 
\bea
\bar{\psi}^{(\frac{3}{2})}(z)
&=& b_2 \Bigg( \frac{a_2}{b_2} \Bigg[  
\frac{a_1}{a_2} J_1^{\alpha} \Psi^{\alpha}+ J_2^{\alpha} \Psi^{\alpha}+ 
j^{\alpha} \Psi^{\alpha} \Bigg]
+  \Bigg[ \frac{b_1}{b_2} J_1^{\rho} \Psi^{\rho} + J_2^{\rho} \Psi^{\rho}+  j^{\rho} \Psi^{\rho} \Bigg] \Bigg)(z)
\nonu \\
&\sim&   d \Bigg(  a J_1^{\alpha} \Psi^{\alpha}+ J_2^{\alpha} \Psi^{\alpha}+ 
j^{\alpha} \Psi^{\alpha} \Bigg)(z)
+  \Bigg( 
b J_1^{\rho} \Psi^{\rho} + J_2^{\rho} \Psi^{\rho}+  j^{\rho} \Psi^{\rho} \Bigg)(z).
\label{interexpression}
\eea
When we ignore the overall factor $b_2$, then 
we have three unknown coefficients $a(N,M), b(N,M)$, and $d(N,M)$ 
we should fix.

On the other hands, the regular conditions
are given by
\bea
(J_1^{\alpha} + J_2^{\alpha} +j^{\alpha})(z) \, 
\bar{\psi}^{(\frac{3}{2})}(w) & = & + \cdots,
\nonu \\
(J_1^{\rho} + J_2^{\rho} +j^{\rho})(z) \, 
\bar{\psi}^{(\frac{3}{2})}(w) & = & + \cdots.
\label{suNregularsuMregular}
\eea
Note that the OPE between the expression (\ref{psiansatz}) 
and the $U(1)$ current of the denominator in the coset (\ref{KScoset}) 
is regular from Appendix $C$.  
Then we can determine the coefficients 
$a(N,M)$ and $b(N,M)$, using the conditions (\ref{suNregularsuMregular}),
as follows:
\bea
a(N,M)= -\frac{2M}{3N}, \qquad b(N,M)= -\frac{2N}{3M}.
\label{abcoeff}
\eea
There is a $N \leftrightarrow M$ symmetry between the 
two coefficients in (\ref{abcoeff}).

The OPE between the lowest higher spin-$\frac{3}{2}$
current and itself, via the OPEs in Appendix $C$, 
can be obtained as follows:
{\small
\bea
\bar{\psi}^{(\frac{3}{2})}(z) \,
\bar{\psi}^{(\frac{3}{2})}(w)&=&\frac{1}{(z-w)^3} \Bigg[ 
d^2 \frac{2M}{3N}(2M+3N)(N^2-1)+\frac{2N}{3M} (2N+3M)(M^2-1) \Bigg]
\nonu \\
&+& \frac{1}{(z-w)} \Bigg[ d^2 \frac{2M}{3N}(2M+3N) \pa \Psi^{\alpha} \Psi^{\alpha}
+d^2 \frac{4 M^2}{3 N^2} J_1^{\alpha} J_1^{\alpha} 
\nonu \\
&+& d^2 \Bigg( -\frac{2(2M+N)}{N} J_1^{\alpha}+J_2^{\alpha}+j^{\alpha}\Bigg) 
(J_2^{\alpha}+j^{\alpha} ) 
+ \frac{2N}{3M}(2N+3M) \pa \Psi^{\rho} \Psi^{\rho}
\nonu \\
&+&  \frac{4 N^2}{3 M^2} J_1^{\rho} J_1^{\rho}
+ \Bigg( -\frac{2(2N+M)}{M} J_1^{\rho}+J_2^{\rho}+j^{\rho}\Bigg) 
(J_2^{\rho}+j^{\rho})  \Bigg](w) + \cdots.
\label{psibarpsibar}
\eea}
The  normalized higher spin-$\frac{3}{2}$ current
can be determined by requiring that the central term should 
behave as $\frac{2c}{3}$ where the central charge $c$ 
is given by (\ref{critical})
 \bea
\psi^{(\frac{3}{2})}(z)
=\sqrt{ \frac{N M}{d^2 \frac{2M}{3N}(2M+3N)(N^2-1)+\frac{2N}{3M} (2N+3M)(M^2-1)}} \, \bar{\psi}^{(\frac{3}{2})}(z).
\label{newdef}
\eea
By substituting (\ref{newdef}) into  
(\ref{psibarpsibar}), we obtain 
\bea
 \psi^{(\frac{3}{2})}(z) \,
\psi^{(\frac{3}{2})}(w) = \frac{1}{(z-w)^3} \, \frac{2}{3} c + \mbox{
other singular terms} +
\cdots.
\label{newope} 
\eea
Furthermore,
the lowest higher spin-$2$ current of the next higher spin multiplet
can be obtained from the following expression
which will be described in next section
\bea
\psi^{(2)}(w) =\frac{1}{C^{(2)}_{(\frac{3}{2}) (\frac{3}{2}) } }  \left[ \psi^{(\frac{3}{2})}(z) \, \psi^{(\frac{3}{2})}(w) \Bigg|_{\frac{1}{(z-w)}} +
\frac{6c}{(c+1)(2c-3)} J^i J^i(w)
\right.
\nonu \\
+ \left. \frac{6(c+3)}{(c+1)(2c-3)} \pa \Psi \Psi(w) - \frac{4c(c+3)}{(c+1)(2c-3)} T(w) \right].
\label{somerelation}
\eea
The first term in (\ref{somerelation})
can be read off from (\ref{psibarpsibar}) or (\ref{newope}).
We can determine the unknown coefficient 
$d(N,M)$ by using the various ${\cal N}=3$ primary conditions of 
the next higher spin multiplet including $\psi^{(2)}(z)$.  
In particular, we have used 
the fact that  $G^{+}(z) \, \psi^{(2)}(w) |_{\frac{1}{(z-w)^2}}=0$
from Appendix $B$. See also (\ref{Gphis}).
The second-order pole of this OPE contains 
the cubic fermions with the specific index structure.
Then by focusing on the coefficient appearing in the 
particular independent term,
we finally obtain the coefficient appearing in (\ref{interexpression})
\bea
d^2(N,M) = \frac{N(3M+2N)}{M(3N+2M)}.
\label{dvalue}
\eea
By choosing  the positive solution in (\ref{dvalue}), the normalized 
lowest higher spin-$\frac{3}{2}$ current
is given by
\bea
\psi^{(\frac{3}{2})}(z)&=&\sqrt{\frac{3N^2 M}{2(M+N)(2M+3N)(MN-1)}} 
\Bigg(   
-\frac{2M}{3N} J_1^{\alpha} +J_2^{\alpha} +j^{\alpha} \Bigg)\Psi^{\alpha}(z)
\nonu \\
&+& \sqrt{\frac{3M^2 N}{2(M+N)(2N+3M)(MN-1)}} \Bigg(   
-\frac{2N}{3M} J_1^{\rho} +J_2^{\rho} +j^{\rho} \Bigg) \Psi^{\rho}(z).
\label{spin3halfexpression}
\eea
We observe that under the exchange of 
$N \leftrightarrow M$ and $\al \leftrightarrow \rho$
this higher spin-$\frac{3}{2}$ current is invariant.
This symmetry also appears in the coset (\ref{KScoset}). 
Some of the terms in (\ref{spin3halfexpression})
appear in the spin-$\frac{3}{2}$ current $G^3(z)$ in (\ref{n3sca}).
In next subsection, we can obtain the remaining seven higher spin currents 
with the information of (\ref{spin3halfexpression}). 

\subsection{The higher spin-$2$ currents}
Due to the ${\cal N}=3$ supersymmetry, we
can determine other higher spin currents 
from the lowest one and the spin-$\frac{3}{2}$ currents 
in (\ref{n3sca}).
Using the following relation (coming from Appendix $B$),
\bea
G^i(z) \, \psi^{(\frac{3}{2})}(w) & = & \frac{1}{(z-w)} \, \phi^{(2),i}(w)
+ \cdots,
\label{g3halfope}
\eea
we can obtain three higher spin-$2$ currents
by calculating the left hand side of (\ref{g3halfope}) together with
(\ref{n3sca}) and (\ref{spin3halfexpression}). 
Note that
\bea
G^{\pm}(z) \equiv \frac{1}{\sqrt{2}} (G^1  \pm i G^2)(z), \qquad
\phi^{(2), \pm}(z) \equiv \frac{1}{\sqrt{2}} 
(\phi^{(2), 1} \pm i \phi^{(2), 2})(z). 
\label{Gphis}
\eea
It is better to use the base (\ref{Gphis})
because $G^{\pm}(z)$ have simple term rather than $G^1(z)$ or $G^2(z)$
from (\ref{n3sca}).
The former will take less time to calculate the OPE manually.
By selecting the first-order pole of this OPE,
we obtain the higher spin-$2$ currents $\phi^{(2), \pm}(w)$ directly.
The remaining higher spin-$2$ current $\phi^{(2),3}(w)$
can be obtained from the OPE (\ref{g3halfope}).
We present the final expression in Appendix (\ref{2expression}).

\subsection{The higher spin-$\frac{5}{2}$ currents}

Because we have obtained the higher spin-$2$ currents in previous subsection,
we can continue to calculate the next higher spin-currents.
The other defining relation in Appendix 
$B$ leads to the following expression
\bea
G^i(z) \, \phi^{(2),j}(w) \Bigg|_{\frac{1}{(z-w)}} & = & 
 \Bigg( 
\delta^{ij} \pa \psi^{(\frac{3}{2})} + 
i \epsilon^{ijk} \, \psi^{(\frac{5}{2}),k} \Bigg)(w).
\label{Gphiother}
\eea
Let us introduce the following quantities as before
\bea
\psi^{(\frac{5}{2}), \pm}(z) \equiv \frac{1}{\sqrt{2}} 
(\psi^{(\frac{5}{2}), 1} \pm i 
\psi^{(\frac{5}{2}), 2})(z). 
\label{plusminusother}
\eea
We will see that these preserve the $U(1)$ charge 
of ${\cal N}=2$ superconformal algebra later.
By using (\ref{Gphiother}) and (\ref{plusminusother}),
we can write down the higher spin-$\frac{5}{2}$ currents, 
together with (\ref{n3sca}), (\ref{Gphis}) , 
Appendix (\ref{2expression}) 
and (\ref{spin3halfexpression}), as follows:
\bea
\psi^{(\frac{5}{2}),+}(w)&=& -G^{+}(z) \, \phi^{(2),3}(w) \Bigg|_{\frac{1}{(z-w)}},
\nonu \\
\psi^{(\frac{5}{2}),-}(w)&=& G^{-}(z) \, \phi^{(2),3}(w) \Bigg|_{\frac{1}{(z-w)}},
\nonu \\
\psi^{(\frac{5}{2}),3}(w)&=& G^{+}(z) \, \phi^{(2),-}(w) \Bigg|_{\frac{1}{(z-w)}} 
- \pa \psi^{(\frac{3}{2})}(w).
\label{5halfdefinition}
\eea 
The last term of the last equation in 
(\ref{5halfdefinition}) can be obtained from (\ref{spin3halfexpression}).
It turns out that 
the final results for the higher spin-$\frac{5}{2}$ currents 
are given in Appendix (\ref{5halfexpression}).

\subsection{The higher spin-$3$ current}

From the relation which can be obtained from Appendix $B$,
\bea
G^i(z) \, \psi^{(\frac{5}{2}),j}(w) \Bigg|_{\frac{1}{(z-w)}} & = &  
\Bigg( 2 \delta^{ij} \, \phi^{(3)} + i \epsilon^{ijk}
\pa \phi^{(2),k} \Bigg)(w), 
\label{spin3inter}
\eea
we can read off the higher spin-$3$ current with (\ref{Gphis}) and 
(\ref{plusminusother}) as follows:
\bea
\phi^{(3)}(w)=\frac{1}{2} G^{+}(z) \, \psi^{(\frac{5}{2}),-}(w) 
\Bigg|_{\frac{1}{(z-w)}}  - \frac{1}{2} \pa \phi^{(2),3}(w).
\label{spin3defintion}
\eea
The last term for the explicit form 
can be seen from the previous subsection.
We summarize 
this higher spin-$3$ current in Appendix (\ref{3expression}).

Therefore, the lowest eight higher spin currents are obtained
in terms of various fermions in (\ref{KScoset}). 
For the next higher spin multiplets, one can 
obtain the explicit forms in terms of fermions by using the 
methods in this section.

\section{The OPEs between the lowest eight higher spin currents}

In this section, we consider the four types of OPEs 
between the higher spin currents for fixed $(N,M)=(2,2)$
where the central charge is given by $c=6$.
Based on the results of this section which are valid for $c=6$ only 
(although we put the central charge as $c$), 
we can go to the ${\cal N}=2$ superspace
approach in next section where all the undetermined coefficient functions
will be fixed and can be written in terms of the arbitrary central charge. 

\subsection{The OPE between the higher spin-$\frac{3}{2}$ current
and itself}

Let us consider the simplest OPE and calculate the following OPE
with the description of previous section.
{\small 
\bea
\psi^{(\frac{3}{2})}(z) \, \psi^{(\frac{3}{2})}(w) & = & 
\frac{1}{(z-w)^3} \, \frac{2c}{3} \nonu \\
& + & \frac{1}{(z-w)} \, \Bigg[ \frac{1 }{(c+1) (2 c-3)} \Bigg(
-6 c  J^i J^i - 
6 (c+3) \pa \Psi \Psi +
4 c (c+3) T  \Bigg) \nonu \\
& + &  C_{(\frac{3}{2}) (\frac{3}{2})}^{(2)} \psi^{(2)} 
\Bigg](w) + 
\cdots,
\label{psipsi} 
\eea}
where the normalization for the lowest higher spin-$\frac{3}{2}$ current
is fixed as $\frac{2c}{3}$ as in (\ref{psipsi}).
Each three term in the first-order pole of (\ref{psipsi})
is a quasiprimary current in which the third-order pole with the stress 
energy tensor $T(z)$ does not have any singular term.
It turns out that we are left with a new primary higher spin-$2$ current 
$\psi^{(2)}(w)$ (which is the lowest component of ${\cal N}=3$ higher spin 
multiplet ${\bf \Phi}^{(2)}(Z_2)$ in (\ref{BigPhis})) 
with a structure constant $ C_{(\frac{3}{2}) (\frac{3}{2})}^{(2)}$.   

\subsection{The OPEs between the higher spin-$\frac{3}{2}$ current
and the higher spin-$2$ currents}

Let us consider the second type of OPE with the preliminary results in 
previous section where the explicit forms for the 
higher spin-$2$ currents are known.
It turns out that we have
{\small 
\bea
\psi^{(\frac{3}{2})}(z) \, \phi^{(2),i}(w) & = &
\frac{1}{(z-w)^2} \, \frac{1}{(2c-3)} \Bigg[ 6 c G^i 
-18  \Psi J^i
\Bigg](w) 
\nonu \\
&+& \frac{1}{(z-w)} \Bigg[ \frac{1}{3} \pa (\mbox{pole two})
+ \frac{1}{(c+1)(2c-3)} \Bigg( 6 i c  ( \epsilon^{ijk} J^j G^k
-\frac{2}{3} i \pa G^i)
\nonu \\
& - & 18 ( \pa \Psi  J^i -\frac{1}{3} \pa (\Psi J^i)) \Bigg)
- 
\frac{1}{2}  C_{(\frac{3}{2}) (\frac{3}{2})}^{(2)} \phi^{(\frac{5}{2}),i}
\Bigg](w) + \cdots,
\label{psiphi2}
\eea}
where the correct coefficient $\frac{1}{3}$ 
for the descendant term in the first-order pole
of (\ref{psiphi2}) is taken.
We can further examine the first-order pole in order to 
write down in terms of the sum of quasiprimary currents. 
In this case, there are two kinds of quasiprimary currents.
In addition to them, there exists a new primary current 
$\phi^{(\frac{5}{2}),i}(w)$ which belongs to
the previous ${\cal N}=3$ higher spin 
multiplet ${\bf \Phi}^{(2)}(Z_2)$ in (\ref{BigPhis}).
At the moment, it is not obvious how the structure constant 
appearing in the $\phi^{(\frac{5}{2}),i}(w)$
behaves as the one in (\ref{psiphi2}) but this 
will arise automatically 
after the analysis of Jacobi identity in next section.
In other words, that structure constant can be written in terms of
the one introduced in (\ref{psipsi}).   
Note that in the OPE (\ref{psiphi2}), the free index $i$ of $SO(3)$
group appears on the right hand side also.

\subsection{The OPE between the higher spin-$\frac{3}{2}$ current
and the higher spin-$\frac{5}{2}$ currents}

Let us consider the third type of OPE which will be the 
one of the main important results of this paper.
Again, based on the previous section, 
there are known higher spin-$\frac{5}{2}$ currents
in terms of coset fermions for fixed $c=6$.
We obtain the following OPE
{\small 
\bea
\psi^{(\frac{3}{2})}(z) \, \psi^{(\frac{5}{2}),i}(w) & = &
\frac{1}{(z-w)^3} \,  6 J^i(w)
-\frac{1}{(z-w)^2} \,  \frac{18 }{(2 c-3)} \Psi G^i(w)
\nonu \\
& + & \frac{1}{(z-w)} \Bigg[ 
 \frac{1}{4} \pa (\mbox{pole two}) \nonu \\
& + & 
 \frac{1}{(c+1) (c+6) (2 c-3)} \Bigg(
18 i (5 c+6) (  \epsilon^{ijk} \Psi
J^j G^k -\frac{2}{3} i \Psi \pa G^i) 
\nonu \\
& - & 18 i c (c+2)
(\epsilon^{ijk} G^j G^k -\frac{2}{3} i \pa^2 J^i)
+  36 (c^2+3 c+6)
(T J^i -\frac{1}{2} \pa^2 J^i)
\nonu \\  
& - & 54 (c+2) J^i J^j J^j
+  72 i c
(\epsilon^{ijk} \pa J^j J^k -\frac{1}{3} i \pa^2 J^i)
\nonu \\
& + &  
6 (c^2-17 c-42)
(\pa \Psi G^i -\frac{1}{4} \pa (\Psi G^i)) 
-72 c \pa \Psi \Psi J^i \Bigg)
\nonu \\
&+ &  C_{(\frac{3}{2}) (\frac{3}{2})}^{(2)} \Bigg(
\frac{3 (c+3)  }{5 (c-3) c} 
\Psi \phi^{(\frac{5}{2}), i} 
+ \frac{9 (3 c-1)  }{5 (c-3) c} 
J^i \psi^{(2)}
+ \frac{3 (c-7) }{10 (c-3)}
  \psi^{(3), i} \Bigg)
 \nonu \\
&+& \frac{2}{5}   C_{(\frac{3}{2}) (3)}^{(\frac{5}{2})}  \phi^{(3),i} +  
 C_{(\frac{3}{2}) (\frac{5}{2})}^{(3)} \psi^{(3),\alpha=i}
\Bigg](w) +\cdots.
\label{psipsi5half}
\eea}
After subtracting the descendant term with the coefficient
$\frac{1}{4}$, there exist various quasiprimary currents and 
primary currents as in (\ref{psipsi5half}).
We can easily check the above seven quasiprimary currents 
do not have any singular term in the third-order pole 
in the OPE with $T(z)$. 
Furthermore, there are three primary currents with structure constant
$ C_{(\frac{3}{2}) (\frac{3}{2})}^{(2)}$ whose presence 
is not clear at the moment.
Among them, the higher spin-$3$ current $\psi^{(3),i}$
belongs to  previous ${\cal N}=3$ higher spin 
multiplet ${\bf \Phi}^{(2)}(Z_2)$ in (\ref{BigPhis}).
Furthermore, there exists a primary current 
with structure constant     
$C_{(\frac{3}{2}) (3)}^{(\frac{5}{2})}$
which will be defined in next OPE soon.
This primary current belongs to 
 the ${\cal N}=3$ higher spin 
multiplet ${\bf \Phi}^{(\frac{5}{2})}(Z_2)$ in (\ref{BigPhis}).
Now the crucial point is the presence of the last term
with the structure constant 
$ C_{(\frac{3}{2}) (\frac{5}{2})}^{(3)}$ we introduce (or define). 
We can check that the first-order pole subtracted by 
the descendant term, seven quasiprimary current terms, and four
primary current terms (which is given in terms of the coset fermions 
explicitly), denoted by $\psi^{(3),\al=i}$ with above structure 
constant, transforms as a primary current under the stress energy tensor
$T(z)$. Furthermore, the other defining equations, the first, the fifth
and the ninth equations, 
presented in 
Appendix $B$ are satisfied.    
Note that the higher spin-$3$ current $\psi^{(3),\al=i}$
belongs to the ${\cal N}=3$ higher spin multiplet 
 ${\bf \Phi}^{(3),\al}(Z_2)$ in (\ref{BigPhis}). 
Because the left hand side has a free index $i$ for $SO(3)$, 
we identify that the representation $\al$ corresponds to 
the $SO(3)$  vector index.  

\subsection{The OPE between the higher spin-$\frac{3}{2}$ current
and the higher spin-$3$ current}

Let us describe the fourth type of OPE.
Again, based on the previous section, 
we can calculate the following OPE
and it turns out that
{\small
\bea
\psi^{(\frac{3}{2})}(z) \, \phi^{(3)}(w) & = &
\frac{1}{(z-w)^4} \, 3 \Psi(w) -
\frac{1}{(z-w)^3} \, 3 \pa \Psi(w) 
\nonu \\
& + & \frac{1}{(z-w)^2} \Bigg[ \frac{1}{(c+1) (2 c-3)}
\Bigg(
6 (c+3) (T \Psi - \frac{3}{4} \pa^2 \Psi)
+ 
36 \frac{c(c+1)}{(c+6)}
J^i G^i
\nonu \\
& - & \frac{9 (13 c+18)}{(c+6)} \Psi J^i J^i \Bigg)
+\frac{3  }{2 c} C_{(\frac{3}{2}) (\frac{3}{2})}^{(2)} \Psi \psi^{(2)}
+  C_{(\frac{3}{2}) (3)}^{(\frac{5}{2})} \psi^{(\frac{5}{2})}
\Bigg](w) \nonu \\
& + & \frac{1}{(z-w)}
\Bigg[ 
\frac{1}{5} \pa (\mbox{pole two})
+  \frac{1}{(c+1) (2 c-3)} \Bigg(
-24 (c+3) (\pa \Psi T -\frac{1}{5} \pa (\Psi T))
\nonu \\
& - & 12 c (\pa J^i G^i -\frac{2}{5} \pa (J^i G^i))
+ 
18
(\pa \Psi J^i J^i -\frac{1}{5}\pa (\Psi J^i J^i)) \Bigg)
\nonu \\
& + & 
 C_{(\frac{3}{2}) (\frac{3}{2})}^{(2)} \Bigg(-
\frac{9 }{2 (c-3)}
(\pa \Psi \psi^{(2)} -\frac{1}{5} \pa (\Psi \psi^{(2)}))
-\frac{3 }{2 (c-3)}  J^i \phi^{(\frac{5}{2}),i}
\nonu \\
&- & 
\frac{(c-12) }{5 (c-3)}   \phi^{(\frac{7}{2})} \Bigg)
-\frac{1 }{4}  C_{(\frac{3}{2}) (\frac{5}{2})}^{(3)} \phi^{(\frac{7}{2}),i,\alpha=i} 
\Bigg](w) 
+ \cdots.
\label{psiphi3} 
\eea}
At the second-order pole of (\ref{psiphi3}), 
there are three quasiprimary currents. Furthermore, there exists
a primary current $\psi^{(\frac{5}{2})}(w)$ 
which belongs to 
the ${\cal N}=3$ higher spin 
multiplet ${\bf \Phi}^{(\frac{5}{2})}(Z_2)$ in (\ref{BigPhis}) in addition to 
the composite current which is also primary current.
We introduce the structure constant $ C_{(\frac{3}{2}) (3)}^{(\frac{5}{2})}$.
At the first-order pole, 
we have the descendant term with coefficient $\frac{1}{5}$.
There are also three quasiprimary currents.
The quasiprimary current appears in the first term with the structure 
constant   $C_{(\frac{3}{2}) (\frac{3}{2})}^{(2)}$.
The next two terms are primary currents. 
Finally, the primary higher spin-$\frac{7}{2}$ currents,
which are the second components of 
 the ${\cal N}=3$ higher spin multiplet 
 ${\bf \Phi}^{(3),\al}(Z_2)$ in (\ref{BigPhis}), 
appear.  
There is a summation over the index $i$.

\subsection{The remaining other OPEs}

For the remaining OPEs between the higher spin currents, 
we have checked that they can be written in terms of known currents
and known higher spin currents for fixed central charge using the package
of \cite{Thielemans}. 

\section{The next higher spin currents}

The next lowest higher spin-$2$ current $\psi^{(2)}(w)$ was obtained from 
(\ref{psipsi}) by looking at the first-order pole.
In other words, by subtracting the 
three quasiprimary currents in the first-order pole from the 
left hand side of (\ref{psipsi}), 
we obtain the nontrivial expression which should 
satisfy the properties in Appendix $B$.
Because we have explicit form for the higher spin-$2$ current in terms of
fermions, we can also calculate 
the OPE between this higher spin-$2$ current and itself
and this will eventually determine the structure constant 
$C_{(\frac{3}{2}) (\frac{3}{2})}^{(2)}$
together with the normalization for the higher spin-$2$ current 
we fix.
Of course, for the $(N,M)=(2,2)$, we have the 
explicit expression for the higher spin-$2$ current
in terms of previous fermions.   
As done in section $3$, the corresponding remaining $7$ higher spin currents
residing on the ${\cal N}=3$ higher spin multiplet 
${\bf \Phi}^{(2)}(Z)$ in (\ref{BigPhis}) 
can be obtained using the ${\cal N}=3$ supersymmetry
spin-$\frac{3}{2}$ currents.

The next lowest higher spin-$\frac{5}{2}$ current $\psi^{(\frac{5}{2})}(w)$ 
has been observed from 
(\ref{psiphi3}) by looking at the second-order pole.
Again, there are three quasiprimary currents and the composite 
current with the previous structure constant.
Because the left hand side of (\ref{psiphi3})
can be calculated from the explicit form from the section $3$,
we can extract the higher spin-$\frac{5}{2}$ current with 
the structure constant   
$C_{(\frac{3}{2}) (3)}^{(\frac{5}{2})}$. Furthermore, 
in order to fix the normalization for 
the higher spin-$\frac{5}{2}$ current, we should calculate the OPE
between this higher spin-$\frac{5}{2}$ current and itself.
Then as we did before,
the corresponding remaining $7$ higher spin currents
residing on the ${\cal N}=3$ higher spin multiplet 
${\bf \Phi}^{(\frac{5}{2})}(Z)$ in (\ref{BigPhis}) 
can be determined using the ${\cal N}=3$ supersymmetry
spin-$\frac{3}{2}$ currents.

The next lowest higher spin-$3$ currents $\psi^{(3),\al=i}(w)$ were found from 
(\ref{psipsi5half}) by looking at the first-order pole.
Once again, 
for each index $i$, the higher spin-$3$ current 
with the structure constant $C_{(\frac{3}{2}) (\frac{5}{2})}^{(3)}$
can be determined 
from the first-order pole of (\ref{psipsi5half}) in terms of
fermions and the algebraic expressions on the right hand side of
(\ref{psipsi5half}).
Based on the higher spin-$3$ currents, 
the corresponding remaining each $7$ higher spin currents
residing on the ${\cal N}=3$ higher spin multiplet 
${\bf \Phi}^{(3),\al=i}(Z)$ in (\ref{BigPhis}) 
can be determined, in principle, using the 
previous ${\cal N}=3$ supersymmetry
spin-$\frac{3}{2}$ currents.

\section{The OPEs  between the lowest eight higher spin currents in 
${\cal N}=2$ superspace}

To obtain the complete OPEs between 
 the $8$  higher spin  currents  in the 
${\cal N}=2$ superspace, the complete composite fields 
appearing in the OPEs should be determined.
It is known that some of the OPEs between the 
$8$ higher spin currents  in the component approach
are found explicitly for $(N,M)=(2,2)$. 
Then we can move to the ${\cal N}=2$ superspace by collecting those OPEs 
in the component approach and rearranging them in an ${\cal N}=2$ 
supersymmetric 
way. So far, all the coefficients in the OPEs are given with fixed $N$
and $M$. Now we set these coefficients as  functions of 
$N$ and $M$ and  use Jacobi identities between the ${\cal N}=2$
currents or higher spin currents. Eventually, we obtain 
the complete structure constants with arbitrary central charge 
appearing 
in the complete OPEs in the ${\cal N}=2$ superspace.

Let us introduce the two ${\cal N}=2$ higher spin currents
\footnote{Although we use the same notation for the 
${\cal N}=2$ superspace coordinates as $Z(\overline{Z})$ for the 
${\cal N}=3$ superspace coordinates, it is understood that 
in ${\cal N}=2$ superspace, we have $\theta^3=\bar{\theta}^3=0$.
One can write down 
the ${\cal N}=3$ stress energy tensor (\ref{j3}) in terms of 
two ${\cal N}=2$ ones in Appendix (\ref{stressn2}) and 
Appendix (\ref{multipletn2}) as 
${\bf J}(Z) = \frac{i}{4} {\bf T}^{(\frac{1}{2})}(Z) +
\theta^3 \frac{i}{2} {\bf T}(Z)$ with $\theta \equiv \theta^1 - i \theta^2$
and $\bar{\theta} \equiv -\theta^1- i \theta^2$. 
Similarly, one sees 
${\bf \Phi}^{(\frac{3}{2})}(Z) = \frac{i}{4} {\bf W}^{(\frac{3}{2})}(Z) + 
\theta^3  \frac{i}{4}
{\bf W}^{(2)}(Z)$.}
\bea
{\bf W}^{(\frac{3}{2})}(Z) & = &  2 \psi^{(\frac{3}{2})}(z)
+ \theta \, 
 (\phi^{(2),1} + i \phi^{(2),2})(z)
+ \bar{\theta} \, 
( -\phi^{(2),1} + i \phi^{(2),2})(z)
+ \theta \bar{\theta} \, 
\psi^{(\frac{5}{2}),3}(z)
\nonu \\
& \equiv & 
\Bigg(  2 \psi^{(\frac{3}{2})}, \quad  \phi^{(2),1} + i \phi^{(2),2}, \quad 
 -\phi^{(2),1} + i \phi^{(2),2}, \quad \psi^{(\frac{5}{2}),3} \Bigg),
\label{comp3half}
\eea
and 
\bea
{\bf W}^{(2)}(Z) & = &  2 \phi^{(2),3}(z)
+ \theta \, 
 (-\psi^{(\frac{5}{2}),1} - i \psi^{(\frac{5}{2}),2})(z)
+ \bar{\theta} \, 
( -\psi^{(\frac{5}{2}),1} + i \psi^{(\frac{5}{2}),2})(z)
+ \theta \bar{\theta} \, 
2 \phi^{(3)}(z)
\nonu \\
& \equiv & 
\Bigg(  2 \phi^{(2),3}, \quad  -\psi^{(\frac{5}{2}),1} - 
i \psi^{(\frac{5}{2}),2}, \quad 
 -\psi^{(\frac{5}{2}),1} + i \psi^{(\frac{5}{2}),2}, \quad 2 \phi^{(3)} \Bigg).
\label{comp2}
\eea
The exact coefficients appearing in the component currents 
in (\ref{comp3half}) and (\ref{comp2})
can be fixed from Appendix $B$ (or its ${\cal N}=2$ version). 
For example, 
the ${\cal N}=2$ stress energy tensor is given by Appendix (\ref{stressn2})
and the ${\cal N}=2$ primary conditions are given by 
the first two equations of Appendix (\ref{primaryopes}) 
which determine 
the above coefficients exactly. 
As usual, each second component current of (\ref{comp3half}) and 
(\ref{comp2}) 
has $U(1)$ charge $+1$
while each third component of them  
has $U(1)$ charge $-1$.
This can be checked from the OPEs between the $J^3(z)$
current of ${\cal N}=2$ superconformal algebra and the corresponding 
currents above.
We would like to construct the three ${\cal N}=2$ OPEs
between these ${\cal N}=2$ higher spin currents.

\subsection{The OPE between the ${\cal N}=2$ 
higher spin-$\frac{3}{2}$ current
and itself}

Let us consider the OPE between 
$ {\bf W}^{(\frac{3}{2})}(Z_1)$ and $ {\bf W}^{(\frac{3}{2})}(Z_2)$.
That is, the OPE between the first ${\cal N}=2$ higher spin-$\frac{3}{2}$ 
multiplet and itself.
The corresponding component results for $(N,M)=(2,2)$ are 
obtained from section $4$.
Now one can introduce the arbitrary coefficients in the right hand side of the
OPE.  
Inside of the package \cite{KT}, one introduces the OPE 
in Appendix (\ref{threen2}) for the ${\cal N}=3$ superconformal algebra
in ${\cal N}=2$ superspace, the OPEs in Appendix (\ref{primaryopes})
where the OPEs are given by 
the two ${\cal N}=2$ currents, ${\bf T}(Z_1)$ and ${\bf T}^{(\frac{1}{2})}(Z_1)$, 
and the ${\cal N}=2$ 
higher spin multiplets: 
  ${\bf W}^{(\frac{3}{2})}(Z_2)$ and  
${\bf W}^{(2)}(Z_2)$ which have explicit component currents in 
(\ref{comp3half})and (\ref{comp2}),   
${\bf W}^{(2')}(Z_2)$ and  
${\bf W}^{(\frac{5}{2})}(Z_2)$,  ${\bf W}^{(\frac{5}{2}')}(Z_2)$ and  
${\bf W}^{(3)}(Z_2)$,  
 ${\bf W}^{(3),\al}(Z_2)$ and   
${\bf W}^{(\frac{7}{2}),\al}(Z_2)$.
We write down the component currents for the remaining 
${\cal N}=2$ higher spin multiplets as follows:
{\small
\bea
{\bf W}^{(2')}(Z) & = &  2 \psi^{(2)}(z)
+ \theta \, 
 (\phi^{(\frac{5}{2}),1} + i \phi^{(\frac{5}{2}),2})(z)
+ \bar{\theta} \, 
( -\phi^{(\frac{5}{2}),1} + i \phi^{(\frac{5}{2}),2})(z)
+ \theta \bar{\theta} \, 
\psi^{(3),3}(z)
\nonu \\
& \equiv & 
\Bigg(  2 \psi^{(2)}, \quad  \phi^{(\frac{5}{2}),1} + i \phi^{(\frac{5}{2}),2}, 
\quad 
 -\phi^{(\frac{5}{2}),1} + i \phi^{(\frac{5}{2}),2}, \quad \psi^{(3),3} \Bigg),
\nonu \\
{\bf W}^{(\frac{5}{2})}(Z) & = &  2 \phi^{(\frac{5}{2}),3}(z)
+ \theta \, 
 (-\psi^{(3),1} - i \psi^{(3),2})(z)
+ \bar{\theta} \, 
( -\psi^{(3),1} + i \psi^{(3),2})(z)
+ \theta \bar{\theta} \, 
2 \phi^{(\frac{7}{2})}(z)
\nonu \\
& \equiv & 
\Bigg(  2 \phi^{(\frac{5}{2}),3}, \quad  -\psi^{(3),1} - 
i \psi^{(3),2}, \quad 
 -\psi^{(3),1} + i \psi^{(3),2}, \quad 2 \phi^{(\frac{7}{2})} \Bigg),
\nonu \\
{\bf W}^{(\frac{5}{2}')}(Z) & = &  2 \psi^{(\frac{5}{2})}(z)
+ \theta \, 
 (\phi^{(3),1} + i \phi^{(3),2})(z)
+ \bar{\theta} \, 
( -\phi^{(3),1} + i \phi^{(3),2})(z)
+ \theta \bar{\theta} \, 
\psi^{(\frac{7}{2}),3}(z)
\nonu \\
& \equiv & 
\Bigg(  2 \psi^{(\frac{5}{2})}, \quad  \phi^{(3),1} + i \phi^{(3),2}, \quad 
 -\phi^{(3),1} + i \phi^{(3),2}, \quad \psi^{(\frac{7}{2}),3} \Bigg),
\nonu \\
{\bf W}^{(3)}(Z) & = &  2 \phi^{(3),3}(z)
+ \theta \, 
 (-\psi^{(\frac{7}{2}),1} - i \psi^{(\frac{7}{2}),2})(z)
+ \bar{\theta} \, 
( -\psi^{(\frac{7}{2}),1} + i \psi^{(\frac{7}{2}),2})(z)
+ \theta \bar{\theta} \, 
2 \phi^{(4)}(z)
\nonu \\
& \equiv & 
\Bigg(  2 \phi^{(3),3}, \quad  -\psi^{(\frac{7}{2}),1} - 
i \psi^{(\frac{7}{2}),2}, \quad 
 -\psi^{(\frac{7}{2}),1} + i \psi^{(\frac{7}{2}),2}, \quad 2 \phi^{(4)} \Bigg),
\nonu \\
{\bf W}^{(3),\al}(Z) & = &  2 \psi^{(3),\al}
+ \theta \, 
 (\phi^{(\frac{7}{2}),1,\al} + i \phi^{(\frac{7}{2}),2,\al})
+ \bar{\theta} \, 
( -\phi^{(\frac{7}{2}),1,\al} + i \phi^{(\frac{7}{2}),2,\al})
+ \theta \bar{\theta} \, 
\psi^{(4),3,\al}
\nonu \\
& \equiv & 
\Bigg(  2 \psi^{(3),\al}, 
\quad  \phi^{(\frac{7}{2}),1,\al} + i \phi^{(\frac{7}{2}),2,\al}, \quad 
 -\phi^{(\frac{7}{2}),1,\al} + i \phi^{(\frac{7}{2}),2,\al}, \quad \psi^{(4),3,\al} 
\Bigg),
\nonu \\
{\bf W}^{(\frac{7}{2}),\al}(Z) & = &  2 \phi^{(\frac{7}{2}),3,\al}
+ \theta \, 
 (-\psi^{(4),1,\al} - i \psi^{(4),2,\al})
+ \bar{\theta} \, 
( -\psi^{(4),1,\al} + i \psi^{(4),2,\al})
+ \theta \bar{\theta} \, 
2 \phi^{(\frac{9}{2}),\al}
\nonu \\
& \equiv & 
\Bigg(  2 \phi^{(\frac{7}{2}),3,\al}, \quad  -\psi^{(4),1,\al} - 
i \psi^{(4),2,\al}, \quad 
 -\psi^{(4),1,\al} + i \psi^{(4),2,\al}, \quad 2 \phi^{(\frac{9}{2}),\al} \Bigg).
\label{compothers}
\eea}
As described before, each 
second component current for the first four higher spin multiplets
in (\ref{compothers}) has $U(1)$ charge $+1$
and 
each third component current for them 
has $U(1)$ charge $-1$.
For the other two higher spin multiplets,
the $U(1)$ charge is little different from the above 
assignments. \footnote{We list for the higher spin currents having 
$U(1)$ charge $+1$ as follows:
$ (\psi^{(3), \al=1} + i \psi^{(3), \al=2})(z)$, $(\phi^{(\frac{7}{2}),3,\al=1} +i
\phi^{(\frac{7}{2}),3,\al=2})(z)$, $(\phi^{(\frac{7}{2}),1,\al=3} +i
\phi^{(\frac{7}{2}),2,\al=3})(z)$, $(\psi^{(4),1,\al=3} +i
\psi^{(4),2,\al=3})(z)$, $(\psi^{(4),3,\al=1} +i
\psi^{(4),3,\al=2})(z)$, 
and $ (\phi^{(\frac{9}{2}), \al=1} + i \phi^{(\frac{9}{2}), \al=2})(z)$.
The corresponding higher spin currents 
with $U(1)$ charge $-1$ can be obtained by changing 
each second term of above ones with minus sign.
Furthermore, we have the higher spin currents 
$(\phi^{(\frac{7}{2}),1,\al=1} \pm i
\phi^{(\frac{7}{2}),1,\al=2} \pm  i \phi^{(\frac{7}{2}),2,\al=1} -
\phi^{(\frac{7}{2}),2,\al=2} )(z)$
with $U(1)$ charge $\pm 2$.
Similarly, 
there are the higher spin currents $(\psi^{(4),1,\al=1} \pm i
\psi^{(4),1,\al=2} \pm  i \psi^{(4),2,\al=1} -
\phi^{(4),2,\al=2} )(z)$
with $U(1)$ charge $\pm 2$.}
Then one can write down  $
  {\bf W}^{(\frac{3}{2})}(Z_1) \, {\bf W}^{(\frac{3}{2})}(Z_2)$,
 ${\bf W}^{(\frac{3}{2})}(Z_1) \, {\bf W}^{(2)}(Z_2)$,
and 
 ${\bf W}^{(2)}(Z_1) \, {\bf W}^{(2)}(Z_2)$,
with arbitrary coefficients.
From the component results, we can reduce the independent terms 
appearing on the right hands of these OPEs. 
If we do not know them, we should include  all possible terms
with correct spins at each singular terms.
We can consider the cubic terms in ${\cal N}=2$ currents,
${\bf T}(Z_2)$ and ${\bf T}^{(\frac{1}{2})}(Z_2)$ and the linear terms
in the ${\cal N}=2$ higher spin multiplets with the additions of 
the covariant derivatives, $D$, $\overline{D}$ and the partial derivative
$\pa$ (there are also mixed terms between them). 
The number of these derivatives ($D, \overline{D}$ and $\pa$) 
are constrained to satisfy 
the correct spin for the composite currents.

By using the Jacobi identity 
\footnote{The outcome of ${\tt OPEJacobi}$ is a double list of operators 
\cite{ThielemansPhd}.  It is better to analyze the elements at the 
end of the list first
  because 
the higher spin currents with large spin 
appear in the beginning of
this list while the higher spin currents with small spin 
appear at the end of this list.}
between the three (higher spin) currents
\bea
({\bf T},  {\bf W}^{(\frac{3}{2})}, {\bf W}^{(\frac{3}{2})}), \qquad
({\bf T}^{(\frac{1}{2})},  {\bf W}^{(\frac{3}{2})}, {\bf W}^{(\frac{3}{2})}),
\qquad
({\bf T},  {\bf W}^{(\frac{3}{2})}, {\bf W}^{(2)}), \qquad
({\bf T}^{(\frac{1}{2})},  {\bf W}^{(\frac{3}{2})}, {\bf W}^{(2)}),
\label{jacobi}
\eea
all the structure constants which depend on the central charge 
$c$ 
are determined except three unknown ones.
There are also 
the Jacobi identities 
between the higher spin currents,
$({\bf T},  {\bf W}^{(2)}, {\bf W}^{(2)})$ and $
({\bf T}^{(\frac{1}{2})},  {\bf W}^{(2)}, {\bf W}^{(2)})$,
but these are 
satisfied automatically after imposing the above 
Jacobi identities (\ref{jacobi}). 
The first OPE can be summarized by
{\small
\bea
{\bf W}^{(\frac{3}{2})}(Z_1) \, {\bf W}^{(\frac{3}{2})}(Z_2) & = & 
\frac{1}{z_{12}^3} \, \frac{8c}{3} +\frac{\theta_{12} \bar{\theta}_{12}}{z_{12}^3}
\, 12 {\bf T}(Z_2)
\nonu \\
& + & \frac{\theta_{12}}{z_{12}^2}  \frac{1}{(2 c-3)}
\Bigg[ -24 c D {\bf T}
-18 {\bf T}^{(\frac{1}{2})} D  {\bf T}^{(\frac{1}{2})}
 \Bigg](Z_2)
\nonu \\
& + & \frac{\bar{\theta}_{12}}{z_{12}^2} \frac{1}{(2 c-3)} 
\Bigg[ 24 c
\overline{D} {\bf T}
-18 {\bf T}^{(\frac{1}{2})} \overline{D}  {\bf T}^{(\frac{1}{2})}
 \Bigg](Z_2)
\nonu \\
& + & \frac{\theta_{12} \bar{\theta}_{12}}{z_{12}^2} \Bigg[ -
 \frac{9}{(2 c-3)}  {\bf T}^{(\frac{1}{2})} [D, \overline{D}]  
{\bf T}^{(\frac{1}{2})}
+ 12 \pa {\bf T}
\Bigg](Z_2)
\nonu \\
&+& \frac{1}{z_{12}} \Bigg[ \frac{1}{(c+1) (2 c-3)}
\Bigg(
-6 (c+3) \pa {\bf T}^{(\frac{1}{2})} {\bf T}^{(\frac{1}{2})} 
+  24 c 
\overline{D}  {\bf T}^{(\frac{1}{2})} D   {\bf T}^{(\frac{1}{2})}
\nonu \\
& - & 24 c  {\bf T} {\bf T}
- 
8 c (c+3)  [D, \overline{D}] {\bf T}
-24 c \pa  {\bf T}
\Bigg)
+ 
2 C_{(\frac{3}{2})(\frac{3}{2})}^{(2)} {\bf W}^{(2')}
\Bigg](Z_2)
\nonu \\
&+& \frac{\theta_{12}}{z_{12}} \Bigg[ 
\frac{1}{(c+1) (2 c-3)}
\Bigg(
-6 (2 c+3) 
\pa D  {\bf T}^{(\frac{1}{2})}  {\bf T}^{(\frac{1}{2})}
- 12 c 
\pa  {\bf T}^{(\frac{1}{2})} D  {\bf T}^{(\frac{1}{2})}
\nonu \\
& - & 8 c (2 c+3)
\pa D {\bf T}
+
6 c  [D, \overline{D}]  {\bf T}^{(\frac{1}{2})}
D {\bf T}^{(\frac{1}{2})}
-  24 c {\bf T} D {\bf T} 
\Bigg)
\nonu \\
& + &   C_{(\frac{3}{2})(\frac{3}{2})}^{(2)}
D {\bf W}^{(2')}
\Bigg](Z_2)
\nonu \\
&+& \frac{\bar{\theta}_{12}}{z_{12}} \Bigg[ 
\frac{1 }{(c+1) (2 c-3)}
\Bigg(
-6 (2 c+3) 
\pa \overline{D} {\bf T}^{(\frac{1}{2})} {\bf T}^{(\frac{1}{2})}
-
12 c 
\pa {\bf T}^{(\frac{1}{2})} \overline{D} {\bf T}^{(\frac{1}{2})}
\nonu \\
& + & 
16 c^2 
\pa \overline{D} {\bf T}
-6 c 
\overline{D} {\bf T}^{(\frac{1}{2})} [D, \overline{D}] {\bf T}^{(\frac{1}{2})}
- 
24 c 
{\bf T} \overline{D} {\bf T}
\Bigg)
\nonu \\
& + &
C_{(\frac{3}{2})(\frac{3}{2})}^{(2)}
\overline{D}
{\bf W}^{(2')}
\Bigg](Z_2)
\nonu \\
&+& 
\frac{\theta_{12} \bar{\theta}_{12}}{z_{12}} \Bigg[ 
\frac{1}{(c+1) (c+6) (2 c-3)}
\Bigg(
6 c^2
\pa [D, \overline{D}] {\bf T}^{(\frac{1}{2})} {\bf T}^{(\frac{1}{2})}
\nonu \\
& + &
18(c+6) 
\pa \overline{D} {\bf T}^{(\frac{1}{2})} D {\bf T}^{(\frac{1}{2})}
-  18 (c+6)
\pa D  {\bf T}^{(\frac{1}{2})}  \overline{D}  {\bf T}^{(\frac{1}{2})}
\nonu \\
& - & 9 (c-2) (c+3)
\pa  {\bf T}^{(\frac{1}{2})}  [D, \overline{D}]  {\bf T}^{(\frac{1}{2})} 
-  36 c 
{\bf T } \pa   {\bf T}^{(\frac{1}{2})}  {\bf T}^{(\frac{1}{2})}
\nonu \\
& - & 108 (c+2)
\pa {\bf T}  {\bf T}
\nonu \\
& + & 6 (2 c^3+9 c^2-9 c-18)
\pa^2 {\bf T}
-36 c (c+2) 
\pa [D, \overline{D} ] {\bf T}  
\nonu \\
& - & 36 (c^2+3 c+6) 
{\bf T} [D, \overline{D} ] {\bf T}  
+108 (c+2) 
{\bf T}  \overline{D}  {\bf T}^{(\frac{1}{2})}  D  {\bf T}^{(\frac{1}{2})} 
\nonu \\
& - & 18 (5 c+6) 
\overline{D} {\bf T}  {\bf T}^{(\frac{1}{2})}  D  {\bf T}^{(\frac{1}{2})} 
-  18 (5 c+6)
D {\bf T}  {\bf T}^{(\frac{1}{2})}  \overline{D}  {\bf T}^{(\frac{1}{2})} 
\nonu \\
& - & 144 (c+2) c 
\overline{D} {\bf T} D {\bf T}
-  108 (c+2)  {\bf T} {\bf T} {\bf T}
\Bigg)
\nonu \\
& + & 
C_{(\frac{3}{2})(\frac{3}{2})}^{(2)} \Bigg(
\frac{9 (3 c-1)  }{5 (c-3) c} 
{\bf T} {\bf W}^{(2')}
+\frac{3 (c+3)  }{10 (c-3) c} 
{\bf T}^{(\frac{1}{2})} {\bf W}^{(\frac{5}{2})}
\label{w3halfw3half}
\\
& - & \frac{3 (c-7) }{10 (c-3)}  
[D, \overline{D}]  {\bf W}^{(2')} \Bigg) +
\frac{2}{5}  C_{(\frac{3}{2})(3)}^{(\frac{5}{2})} {\bf W}^{(3)} +
C_{(\frac{3}{2})(\frac{5}{2})}^{(3)}  {\bf W}^{(3),3}
\Bigg](Z_2)
+\cdots.
\nonu
\eea}
As described before, 
the maximum power of 
${\cal N}=2$ currents is given by $3$
while the ${\cal N}=2$ higher spin multiplets
appear linearly (there are also mixed terms).
There are three unknown structure constants in the OPE.
Moreover, we describe the above OPE in simple form  
as
\bea
[{\bf W}^{(\frac{3}{2})} ] \cdot [{\bf W}^{(\frac{3}{2})} ]
= [ {\bf I} ] + [ {\bf W}^{(2')} ]
+ [ {\bf W}^{(\frac{5}{2})} ] + [ {\bf W}^{(3)} ]
+ [ {\bf W}^{(3), 3} ],
\label{simplexp}
\eea
where $[{\bf I}]$ in (\ref{simplexp}) 
stands for the ${\cal N}=3$ superconformal family
of identity operator. 
We have seen the presence of $\psi^{(3),\al=i}(w)$ in the 
OPE of (\ref{psipsi5half}). In particular, 
for $\al=i=3$, this higher spin-$3$ current 
is the first component of the ${\cal N}=2$ higher spin multiplet
${\bf W}^{(3),\al=3}(Z)$. 

\subsection{The OPE between the ${\cal N}=2$ 
higher spin-$\frac{3}{2}$ current
and the ${\cal N}=2$ higher spin-$2$ current}

The explicit OPE for this case is given by
Appendix (\ref{w3halfw2})
and can be summarized by 
\bea
[{\bf W}^{(\frac{3}{2})} ] \cdot [{\bf W}^{(2)} ]
& = & [ {\bf I} ] + [ {\bf W}^{(2')} ]
+ [ {\bf W}^{(\frac{5}{2})} ] +
[ {\bf W}^{(\frac{5}{2}')} ] 
+  [ {\bf W}^{(3), 1}] + [ {\bf W}^{(3), 2} ]
+ [ {\bf W}^{(\frac{7}{2}),3} ].
\label{sub1}
\eea
The ${\cal N}=3$ higher spin multiplets,
$ {\bf W}^{(3), 1}(Z_2)$ and ${\bf W}^{(3), 2}(Z_2)$,
appear in particular combinations in (\ref{sub1}). 
Because the spin of the left hand side 
is given by $\frac{7}{2}$, we can have the higher spin-$\frac{7}{2}$
current on the right hand side of (\ref{sub1}).
According to (\ref{compothers}), the three higher spin-$\frac{7}{2}$
currents $\phi^{(\frac{7}{2}), i, \al}(z)$ for each $\al$
are residing on the second and third components of ${\bf W}^{(3),\al}(Z)$
and the first component of ${\bf W}^{(\frac{7}{2}),\al}(Z)$.
When we look at the last line of the OPE in Appendix (\ref{w3halfw2}),
 we realize that the $\theta_i= \bar{\theta}_i=0$ 
projection (after multiplying
the derivatives $D_2^1 D_2^2$ (or $[D_2, \overline{D}_2]$) on both sides)
will give rise to the sum of the previous higher spin-$\frac{7}{2}$
currents, denoted by  $\phi^{(\frac{7}{2}), i, \al=i}(w)$
which appeared in the last term of the first-order pole 
of (\ref{psiphi3}).
In other words, the $\al$ index in (\ref{sub1})
associated with the last three terms 
is summed over the different ${\cal N}=2$ higher spin multiplets
and the other dummy index $i$ is hidden in them.
That is, $SO(3)$ index $i$ is contained in 
the second and third components of ${\bf W}^{(3),\al}(Z)$
and the first component of ${\bf W}^{(\frac{7}{2}),\al}(Z)$
for fixed $\al$.

\subsection{The OPE between the ${\cal N}=2$ 
higher spin-$2$ current
and itself}

The explicit OPE for this case is given by
Appendix (\ref{w2w2})
and can be summarized by 
\bea
[{\bf W}^{(2)} ] \cdot [{\bf W}^{(2)} ]
& = &  [ {\bf I} ] + [ {\bf W}^{(2')} ]
+
[ {\bf W}^{(\frac{5}{2})} ] + [ {\bf W}^{(3)} ]
+  [ {\bf W}^{(3), 3} ] 
+ [ {\bf W}^{(\frac{7}{2}), 1} ]
+ [ {\bf W}^{(\frac{7}{2}),2} ].
\label{sub2}
\eea
In this case,
the ${\cal N}=3$ higher spin multiplets,
$ {\bf W}^{(\frac{7}{2}), 1}(Z_2)$ and ${\bf W}^{(\frac{7}{2}), 2}(Z_2)$,
appear in particular combinations in (\ref{sub2}). 
Note that the $\theta, \bar{\theta}$ independent term (or the 
first component) of ${\bf W}^{(2)}(Z)$
in (\ref{comp2}) contains the $SO(3)$ index $3$.
Furthermore, the other two $SO(3)$ indices $1$ and $2$
appear in the second or third component of 
 ${\bf W}^{(2)}(Z)$
in (\ref{comp2}).
Then we can understand that the $\al$ indices 
$1$ and $2$ of the  last two terms in (\ref{sub2}) correspond to those
in the second or third component of  ${\bf W}^{(2)}(Z_2)$
in (\ref{comp2}) when we select the first component for the 
 ${\bf W}^{(2)}(Z_1)$ on the left hand side of (\ref{sub2}).
In the component approach, one can think of 
the OPE $\phi^{(2),3}(z) \, \psi^{(\frac{5}{2}),1}(w)$ or 
the OPE 
$\phi^{(2),3}(z) \, \psi^{(\frac{5}{2}),2}(w)$.
This will produce 
$\phi^{(\frac{7}{2}), 3, \al=1}(w)$ or $\phi^{(\frac{7}{2}), 3, \al=2}(w)$
respectively.
Similarly,
if we take the fourth component of    ${\bf W}^{(2)}(Z_2)$
in (\ref{comp2}) (there is no $SO(3)$ index) 
with the same the first component for the 
 ${\bf W}^{(2)}(Z_1)$, then  
the $\al$ index $3$ in the ${\bf W}^{(3),3}(Z_2)$ in (\ref{sub2})
originates from the above index $3$ in the first component of 
 the 
 ${\bf W}^{(2)}(Z_1)$.
In other words, in the component approach, 
the OPE 
$\phi^{(2),3}(z) \, \phi^{(3)}(w)$
will lead to 
$\psi^{(3), \al=3}(w)$ on the right hand side.

\section{The OPEs  between the lowest eight higher spin currents in 
${\cal N}=3$ superspace}

We summarize the previous results in the ${\cal N}=3$ superspace
along the line of \cite{AK1509}.

\subsection{The OPEs  between the lowest eight higher spin currents in 
the component approach}

In the previous section, the complete ${\cal N}=2$ OPEs with complete
structure constants are fixed.
Again, using the package in \cite{KT}, we can proceed to the component approach
where the $36$ OPEs are determined completely.
They are presented in Appendix $F$ with simplified notation.
We might ask whether or not there exists a possibility of 
having  new primary currents in these $36$ OPEs.
Because we do not check for them from the $8$ higher spin currents 
in  ${\cal N}=3$ Kazama-Suzuki coset model for generic $(N,M)$ 
manually, we should 
be careful about the occurrence of  new primary currents in the OPEs.
However, in the present case, such a feature
does not arise. 
We have confirmed that there are no extra primary currents in the 
basic $8$ OPEs between the higher spin-$\frac{3}{2}$ 
current and $8$ higher spin 
currents using the WZW currents for several $(N,M)$ values.
We believe that these basic $8$ OPEs are satisfied even if we try to 
calculate them manually.    

Let us present how we can read off the component result from its
${\cal N}=2$ version. Let us consider the simplest 
example given in (\ref{psipsi}).
Because 
the first component of ${\bf W}^{\frac{3}{2}}(Z)$ is given 
by $2 \psi^{(\frac{3}{2})}(z)$. We can obtain the following OPE.
At the final stage, we put 
$\theta_{i} =
\bar{\theta}_{i}=0$ in order to 
extract the corresponding component OPE.
It turns out that the following expression holds, from (\ref{w3halfw3half}),
{\small
\bea
\frac{1}{2} {\bf W}^{(\frac{3}{2})}(Z_1) \, \frac{1}{2} 
{\bf W}^{(\frac{3}{2})}(Z_2) \Bigg|_{\theta_{i} =
\bar{\theta}_{i}=0} & =&
\frac{1}{(z-w)^3} \, \frac{2c}{3}
\nonu \\
& + & \frac{1}{(z-w)} \Bigg[ \Bigg(
\frac{1}{2}  C_{(\frac{3}{2})(\frac{3}{2})}^{(2)} {\bf W}^{(2')}
-\frac{2 c (c+3) }{(c+1) (2 c-3)} [D, \overline{D}] {\bf T}
\nonu \\
& - & \frac{6 c }{(c+1) (2 c-3)} {\bf T} {\bf T}
+\frac{6 c }{(c+1) (2 c-3)} \overline{D}  {\bf T}^{(\frac{1}{2})}
D  {\bf T}^{(\frac{1}{2})} 
\nonu \\
& - & \frac{3 (c+3) }{2 (c+1) (2 c-3)} \pa {\bf T}^{(\frac{1}{2})}
{\bf T}^{(\frac{1}{2})}
\nonu \\
& - & \frac{6 c }{(c+1) (2 c-3)} \pa {\bf T} \Bigg)_{\theta_{i} =
\bar{\theta}_{i}=0}
\Bigg](w) + \cdots.
\label{compexpression}
\eea}
In the first-order pole of
(\ref{compexpression}), 
before we take the condition 
$\theta_{i} =
\bar{\theta}_{i}=0$, 
they are written in terms of ${\cal N}=2$ (higher spin) currents.
As we take this condition, each factor current 
in the composite currents reduces to its component current.   
Using the relations in Appendix (\ref{stressn2}), (\ref{comp3half})
and (\ref{compothers}) with the relation 
$[J^1,J^2](w)=i \pa J^3(w)$, we obtain 
the previous OPE in (\ref{psipsi}).

In this way we can obtain the 
remaining $35$ OPEs from its three ${\cal N}=2$ OPEs.
We present the $36$ component OPEs in Appendix $F$ for convenience:
Appendix (\ref{Fresult1}), Appendix (\ref{Fresult2}) and 
Appendix (\ref{compresult}).

\subsection{The ${\cal N}=3$ description}

The final  single ${\cal N}=3$ OPE 
between the ${\cal N}=3$ (higher spin current) multiplet 
of superspin $\frac{3}{2}$, with the help of Appendix (\ref{reverseorder})
and (\ref{psipsi}), can be described as
{\small 
\bea
 {\bf \Phi}^{(\frac{3}{2})}(Z_1) \, {\bf \Phi}^{(\frac{3}{2})}(Z_2)  & = & \frac{\theta_{12}^{3-0}}{z_{12}^4} \,
3  {\bf J}(Z_2) -
 \frac{1}{z_{12}^3} \, \frac{c}{6} 
+ \frac{\theta_{12}^{3-i}}{z_{12}^3} \, 3  D^i {\bf J}(Z_2) 
+ \frac{\theta_{12}^{3-0}}{z_{12}^3} \, 6  \pa {\bf J}(Z_2)
\nonu \\
& + &   
 \frac{\theta_{12}^{3-i}}{z_{12}^2} \, \Bigg[ 
-\frac{18}{(2 c-3)}  {\bf J } D^{3-i} {\bf J}+3 \pa D^i {\bf J} 
\Bigg](Z_2)  
\nonu \\
& + &  \frac{\theta_{12}^i}{z_{12}^2} 
\frac{1}{(2 c-3)}
\Bigg[ 
3c D^{3-i} {\bf J}  -18 {\bf J} D^i {\bf J}
\Bigg](Z_2)
\nonu \\
&+ &   
\frac{\theta_{12}^{3-0}}{z_{12}^2} \, \Bigg[ 
\frac{1}{(c+1) (2 c-3)} \Bigg(
-6 (c+3) 
(  D^{3-0} {\bf J} {\bf J} +\frac{3}{4} \pa^2 {\bf J})
-\frac{72c(c+1)}{(c+6)}
D^i {\bf J} D^{3-i} {\bf J}
\nonu \\
&+&   
\frac{36 (13 c+18)}{ (c+6)} 
{\bf J} D^i {\bf J} D^i {\bf J} \Bigg) 
 + \frac{9}{2}  \, \pa^2 {\bf J}
- 
\frac{3i}{c} C_{(\frac{3}{2})(\frac{3}{2})}^{(2)} 
{\bf J} {\bf \Phi}^{(2)} +
C_{(\frac{3}{2})(3)}^{(\frac{5}{2})} \,  
{\bf \Phi}^{(\frac{5}{2})}  
 \Bigg](Z_2) \nonu \\
&  + &   
\frac{1}{z_{12}} \, \Bigg[ 
\frac{1}{(c+1) (2 c-3)} \Bigg(
 c (c+3)
D^{3-0} {\bf J}
-6 c
 D^i {\bf J}
D^i {\bf J} 
-  6 (c+3)
\pa {\bf J}  {\bf J} \Bigg)   
\nonu \\
& + &  
\frac{1}{2} i C_{(\frac{3}{2})(\frac{3}{2})}^{(2)} \, {\bf \Phi}^{(2)}  \Bigg](Z_2)
\nonu \\
&  + & 
\frac{\theta_{12}^i}{z_{12}} \Bigg[ 
\frac{2}{3} \pa (\frac{\theta_{12}^{i}}{z_{12}^2}-\mbox{term})
+\frac{1}{(c+1) (2 c-3)}
\Bigg(
-6  c
(  \epsilon^{ijk} D^j {\bf J} D^{3-k} {\bf J} +\frac{1}{3} \pa 
D^{3-i} {\bf J} )
\nonu \\
&  + &
18
( \pa {\bf J} D^i {\bf J} -\frac{1}{3} \pa ({\bf J} D^i {\bf J}) 
) \Bigg) 
+\frac{i}{4}    C_{(\frac{3}{2})(\frac{3}{2})}^{(2)} D^i {\bf \Phi}^{(2)}
\Bigg](Z_2)
\nonu \\
&  + &  
\frac{\theta_{12}^{3-i}}{z_{12}} \, \Bigg[ 
\frac{3}{4} \pa (\frac{\theta_{12}^{3-i}}{z_{12}^2}-\mbox{term})
 -  \frac{3}{4} \pa^2 D^i {\bf J}
\nonu \\
& + &  
\frac{1}{(c+1) (c+6) (2 c-3)}
\Bigg(
-36  (5 c+6)
(   \epsilon^{ijk} {\bf J } D^j {\bf J} D^{3-k} {\bf J}
+\frac{1}{3}  {\bf J} \pa D^{3-i} {\bf J} )
    \nonu \\\
& + &   
 18  c (c+2)
(  \epsilon^{ijk} D^{3-j} {\bf J} D^{3-k} {\bf J}
-\frac{1}{3} \pa^2 D^i {\bf J} )
-  18 (c^2+3 c+6) 
( D^{3-0} {\bf J} D^i {\bf J} \nonu \\
& + & \frac{1}{2} \pa^2 D^i {\bf J})
\nonu \\
&  + & 
108 (c+2) 
D^i {\bf J} D^j {\bf J} D^j {\bf J}
+ 
36 c
( 2  \epsilon^{ijk} \pa D^j {\bf J} D^k {\bf J} +\frac{1}{3} \pa^2 D^i 
{\bf J} )
\nonu \\
&  - &
6 (c^2-17 c-42) 
( \pa {\bf J} D^{3-i} {\bf J} -\frac{1}{4} \pa ( {\bf J} D^{3-i} 
{\bf J}))
+
  144 c 
\pa {\bf J} {\bf J} D^i {\bf J} 
\Bigg)
\nonu \\
&  + & C_{(\frac{3}{2})(\frac{3}{2})}^{(2)}
\Bigg(-  
\frac{3i (c+3) }{5c (c-3) }  
{\bf J } D^i {\bf \Phi}^{(2)} - 
\frac{9i (3 c-1) }{5 c(c-3) }
D^i {\bf J } {\bf \Phi}^{(2)} 
-
\frac{3i (c-7) }{20 (c-3)}   
D^{3-i} {\bf \Phi}^{(2)} \Bigg)
\nonu \\
&  + &     
 \frac{1}{5}   C_{(\frac{3}{2})(3)}^{(\frac{5}{2})} D^i {\bf \Phi}^{(\frac{5}{2})} 
+\frac{1}{2} 
  C_{(\frac{3}{2})(\frac{5}{2})}^{(3)} \, {\bf \Phi}^{(3), \al=i}  \Bigg](Z_2)
\nonu \\
&  + &  
\frac{\theta_{12}^{3-0}}{z_{12}} \, \Bigg[ 
\frac{4}{5} \pa (\frac{\theta_{12}^{3-0}}{z_{12}^2}-\mbox{term})
 - \frac{8}{5} \pa^3 {\bf J}
\nonu \\
&   + &  
\frac{1}{(c+1) (2 c-3)} \Bigg(
-24 (c+3) 
( \pa {\bf J} D^{3-0}  {\bf J} -\frac{1}{5} \pa ({\bf J}
D^{3-0} {\bf J}) )
\nonu \\
& - &  
24 c 
( \pa D^i {\bf J} D^{3-i} {\bf J} -\frac{2}{5} \pa (D^i {\bf J} 
D^{3-i} {\bf J}) )   +  
 72
( \pa {\bf J} D^i {\bf J} D^i {\bf J} -\frac{1}{5} \pa 
({\bf J} D^i {\bf J} D^i {\bf J} ) )
\Bigg)
\nonu \\
&   + & C_{(\frac{3}{2})(\frac{3}{2})}^{(2)}
\Bigg(- 
\frac{9i }{ (c-3)} 
( \pa {\bf J} {\bf \Phi}^{(2)} -\frac{1}{5} \pa ({\bf J} {\bf \Phi}^{(2)}))
- \frac{3i }{ (c-3)}   
D^i {\bf J} D^i  {\bf \Phi}^{(2)}
 \nonu \\
& - &   
 \frac{i (c-12) }{10 (c-3)}  
D^{3-0}  {\bf \Phi}^{(2)} \Bigg)
+   \frac{1}{4}   C_{(\frac{3}{2})(\frac{5}{2})}^{(3)}  D^i {\bf \Phi}^{(3), \al=i}
\Bigg](Z_2)+ \cdots.
\label{finalPhiPhi}
\eea}
We can easily see that there are consistent $SO(3)$ index contractions 
with $SO(3)$-invariant tensors $\epsilon^{ijk}$ and $\delta^{ij}$ on the 
right-hand side of the OPE (\ref{finalPhiPhi}).\footnote{
Let us examine  how the 
${\cal N}=3$ OPE can be reduced to the corresponding 
OPE in the component approach.
Let us multiply both sides of (\ref{finalPhiPhi})
by the operator $D_2^1 \, D_2^2$ 
with the condition of $\theta_1^i=0=\theta^i_2$.
Then the left-hand side is given by $ \frac{i}{2} 
\psi^{(\frac{3}{2})}(z) \, (-1) \, 
\frac{1}{2} \psi^{(\frac{5}{2}),3}(w)$. 
In contrast, the right-hand side contains
$D_2^1 \, D_2^2 \, \frac{\theta_{12}^1 \, \theta_{12}^2}{z_{12}^2}$ 
with current-dependent
terms and this leads to the singular term $-\frac{1}{(z-w)^2}$.
Therefore, the nonlinear second-order term of the OPE 
$\psi^{(\frac{3}{2})}(z)  \, 
 \psi^{(\frac{5}{2}),3}(w)$
is given by $  \frac{18}{(2c-3)}  \,  {\bf J} D^{1} D^2 {\bf J}(Z_2)$ 
at vanishing $\theta_2^i$.
Furthermore, the additional contribution from 
the $\frac{\theta_{12}^i}{z_{12}^2}$ leads to 
   $ -\frac{36}{(2c-3)}  \,  {\bf J} D^{1} D^2 {\bf J}(Z_2)$ 
at vanishing $\theta_2^i$.
Then we obtain
 $ -\frac{18}{(2c-3)} \, \Psi G^3(w)$ which
is equal to the particular singular term in 
the corresponding OPE in Appendix $F$.} 

Therefore, the OPE for the 
lowest $8$ higher spin currents in the ${\cal N}=3$ superspace
can be characterized by
\bea
\left[ {\bf \Phi}^{(\frac{3}{2})} \right] \cdot \left[  {\bf \Phi}^{(\frac{3}{2})}
\right] =  \left[ {\bf I}\right] +
\left[ {\bf \Phi}^{(2)} \right] +  \left[ {\bf \Phi}^{(\frac{5}{2})} 
\right]
+  \theta^{3-i} \left[ {\bf \Phi}^{(3), i} \right],
\label{fusionfinalfinal}
\eea
where $[{\bf I}]$ denotes the large ${\cal N}=3$ linear superconformal family
of the identity operator.
In the last term of (\ref{fusionfinalfinal}), 
we put the quadratic fermionic coordinates in order to 
emphasize that the $SO(3)$ index $i$
is summed in (\ref{fusionfinalfinal}).
In ${\cal N}=3$ superspace, it is clear that 
the additional ${\cal N}=3$ higher spin-$3$ multiplet
should transform as the triplet of $SO(3)$. 
The above OPE (\ref{fusionfinalfinal}) 
is equivalent to the previous result given by 
(\ref{w3halfw3half}), Appendix (\ref{w3halfw2}), and Appendix 
(\ref{w2w2}) 
in the ${\cal N} =2$ superspace.
Note that 
the ${\cal N}=3$ multiplet 
${\bf \Phi^{(\frac{3}{2})}}(Z)$
stands for the 
two ${\cal N}=2$ higher spin multiplets 
${\bf W^{(\frac{3}{2})}}(Z)$,  and 
${\bf W^{(2)}}(Z)$.
\footnote{Let us describe how we can obtain the ${\cal N}=2$
superspace description starting from its ${\cal N}=3$ version in 
(\ref{finalPhiPhi}). Let us focus on the simplest OPE given by 
(\ref{w3halfw3half}). 
Setting $\theta_1^3 =\theta_2^3=0$ in (\ref{finalPhiPhi})
 gives rise to $ \frac{i}{4} {\bf W^{(\frac{3}{2})}}(Z_1) \, 
\frac{i}{4} {\bf W^{(\frac{3}{2})}}(Z_2)$ on the left-hand side. 
For the terms $\theta_{12}^{3-i}$ with $i=3$, we do have any 
nonzero contributions.
Let us focus on the third-order pole of (\ref{finalPhiPhi}).
We have the relation $\frac{\theta_{12}^1 \, \theta_{12}^2}{z_{12}^3}=
\frac{i}{2} \,
\frac{\theta_{12} \, \overline{\theta}_{12}}{z_{12}^3}$ and we obtain that
the coefficient of $\frac{\theta_{12} \, \overline{\theta}_{12}}{z_{12}^3}$
is given by $ -\frac{3}{4} {\bf T} (Z_2)$ 
where we used the fact that $D^3 {\bf J}(Z_2)$ at vanishing $\theta_2^3$
is equal to $\frac{i}{2} {\bf T}(Z_2)$.
This provides the corresponding
term    
in (\ref{w3halfw3half}), as we expect.
It is straightforward to check the other remaining  terms explicitly.  } 

\section{ The extension of $SO({\cal N}=3 )$ nonlinear
 Knizhnik Bershadsky algebra }

So far we have considered the ${\cal N}=3$ linear superconformal algebra
and its extension. Among the eight currents of the 
${\cal N}=3$ linear superconformal algebra, we can decouple the 
spin-$\frac{1}{2}$ current $\Psi(z)$ from the other remaining 
seven currents, along the lines of \cite{Ahn1311,Ahn1408,Ahn1504}.
In this section, we would like to obtain the 
lowest eight higher spin currents and their OPEs after
factoring out the above spin-$\frac{1}{2}$ current.     

\subsection{Knizhnik Bershadsky algebra}

According to the OPEs in Appendix $A$, 
the spin-$1$ currents of the 
${\cal N}=3$ superconformal algebra do not have any singular terms 
with the spin-$\frac{1}{2}$ current. 
Then it is natural to take the  spin-$1$
currents as the previous one $J^i(z)$.
For the spin-$\frac{3}{2}$ currents and the spin-$2$ current,
we should obtain the new spin-$\frac{3}{2}$ currents and spin-$2$ current.
It turns out that the new seven currents are described as
\cite{GS,Schoutens}
\bea
\hat{T}(z) & = & T(z) + \frac{3}{2c} \Psi \pa \Psi(z), \nonu \\
\hat{G}^i(z)  & = & G^i(z) -\frac{3}{c} J^i \Psi(z), \nonu \\
\hat{J}^i(z) & = & J^i(z).
\label{tgj}
\eea
The relative coefficients appearing on the right hand side 
of (\ref{tgj})
can be fixed by requiring that the following conditions   
should satisfy
\bea
\Psi(z) \; \hat{T}(w)   & = &  + \cdots, \nonu \\
\Psi(z) \; \hat{G}^i(w) & = & + \cdots,  \nonu \\
\Psi(z) \; \hat{J}^i(w) & = & +\cdots.
\label{conditions}
\eea

Then we can calculate the OPEs between the new seven currents 
using (\ref{tgj}) and they can be summarized by \cite{Knizhnik,Bershadsky}
{\small
\bea
\hat{J}^i(z) \, \hat{J}^j(w) & = & \frac{1}{(z-w)^2} \, 
\frac{c}{3} \delta^{ij} +
\frac{1}{(z-w)} \, i \epsilon^{ijk} \hat{J}^k(w) +\cdots,
\nonu \\
 \hat{J}^i(z) \, \hat{G}^j(w) & = & 
\frac{1}{(z-w)} \, i \epsilon^{ijk} \hat{G}^k(w) +\cdots,
\nonu \\
\hat{G}^i(z) \, \hat{G}^j(w) & = & 
\frac{1}{(z-w)^3} \, \frac{1}{3} (2 c-3) \delta^{ij}
+ \frac{1}{(z-w)^2} \,  \frac{i (2 c-3) }{c}
\epsilon^{ijk} \hat{J}^k(w) \nonu \\
& + &  \frac{1}{(z-w)}
\left[ 2 \delta^{ij} \hat{T} + i \epsilon^{ijk} \pa \hat{J}^k 
-\frac{3}{c} \hat{J}^i \hat{J}^j
\right](w) +\cdots,
\nonu \\
\hat{T}(z) \, \hat{J}^i(w) & = &  
\frac{1}{(z-w)^2} \,  \hat{J}^i(w) + \frac{1}{(z-w)} \pa \hat{J}^i(w) + \cdots,
\nonu \\
\hat{T}(z) \, \hat{G}^i(w) & = &  
\frac{1}{(z-w)^2} \, \frac{3}{2} \hat{G}^i(w) + 
\frac{1}{(z-w)} \pa \hat{G}^i(w) + \cdots,
\nonu \\
\hat{T}(z) \, \hat{T}(w) & = & \frac{1}{(z-w)^4} \, \frac{1}{4} (2 c-1) + 
\frac{1}{(z-w)^2} \, 2 \hat{T}(w) + \frac{1}{(z-w)} \pa \hat{T}(w) + \cdots.
\label{nonKB}
\eea}
Note that there exists a nonlinear term in the OPE 
between the spin-$\frac{3}{2}$ currents.  
One can easily see that the above OPEs (\ref{nonKB})
become the one in \cite{Knizhnik} (where there is a typo in the OPE) 
if one changes $i \hat{J}^i(z) =J_K^i(z)$
and other currents remain unchanged.

\subsection{The extension of Knizhnik Bershadsky current}

According to the first OPE in Appendix $B$, 
the lowest higher spin current of any 
${\cal N}=3$ multiplet does not have any singular terms 
with the spin-$\frac{1}{2}$ current. 
Then it is natural to take the lowest higher spin-$\frac{3}{2}$
current as the previous one $\psi^{(\frac{3}{2})}(z)$.
Similarly, the second higher spin-$2$ currents residing the 
$SO(3)$-singlet ${\cal N}=3$ multiplet do not 
have any singular terms due to the second OPE 
in Appendix $B$.   
Then we do not have to modify the higher spin-$2$ currents.

On the other hand, for the $SO(3)$-triplet 
${\cal N}=3$ higher spin multiplet, the second 
higher spin-$\frac{7}{2}$ currents do have the singular terms 
with the spin-$\frac{1}{2}$ current 
from the second OPE in Appendix $B$.
Therefore we should find the new higher spin-$\frac{7}{2}, 4, \frac{9}{2}$ 
currents. 
The final results for the new higher spin currents 
are given by
\bea
\hat{\psi}^{(\frac{3}{2})}(z) & = & \psi^{(\frac{3}{2})}(z),
\nonu \\
\hat{\phi}^{(2),i}(z) & = & \phi^{(2),i}(z), \nonu \\
\hat{\psi}^{(\frac{5}{2}),i}(z) & = & \psi^{(\frac{5}{2}),i}(z) -\frac{3}{c} 
\phi^{(2),i} \Psi(z), \nonu \\
\hat{\phi}^{(3)}(z) & = & \phi^{(3)}(z) -\frac{3}{2c} \, \pa \psi^{(\frac{3}{2})} 
\Psi(z) + \frac{9}{2c} \, \psi^{(\frac{3}{2})} \pa \Psi(z),
\nonu \\
\hat{\psi}^{(2)}(z) & = & \psi^{(2)}(z),
\nonu \\
\hat{\phi}^{(\frac{5}{2}),i}(z) & = & \phi^{(\frac{5}{2}),i}(z), 
\nonu \\
\hat{\psi}^{(3),i}(z) & = & \psi^{(3),i}(z) -\frac{3}{c} 
\, \Psi \phi^{(\frac{5}{2}),i} (z), \nonu \\
\hat{\phi}^{(\frac{7}{2})}(z) & = & 
\phi^{(\frac{7}{2})}(z) + \frac{3}{2c} \, \Psi \pa \psi^{(2)}(z) 
- \frac{6}{c} \, \pa \Psi \psi^{(2)}(z),
\nonu \\
\hat{\psi}^{(\frac{5}{2})}(z) & = & \psi^{(\frac{5}{2})}(z),
\nonu \\
\hat{\phi}^{(3),i}(z) & = & \phi^{(3),i}(z), 
\nonu \\
\hat{\psi}^{(\frac{7}{2}),i}(z) & = & \psi^{(\frac{7}{2}),i}(z) -\frac{3}{c} 
\, \Psi \phi^{(3),i} (z), \nonu \\
\hat{\phi}^{(4)}(z) & = & \phi^{(4)}(z) + \frac{3}{2c} \, \Psi \pa 
\psi^{(\frac{5}{2})}(z) - \frac{15}{2c} \, \pa \Psi \psi^{(\frac{5}{2})}(z),
\nonu \\
\hat{\psi}^{(3),\alpha}(z) & = & \psi^{(3),\alpha}(z),
\nonu \\
\hat{\phi}^{(\frac{7}{2}),i,\alpha=j}(z) & = &  
\phi^{(\frac{7}{2}),i,\alpha=j}(z) 
-\frac{3}{c} \, i  \epsilon^{ijk}
\Psi \psi^{(3),\alpha=k} (z), 
\nonu \\
\hat{\psi}^{(4), i, \alpha=j}(z) & = &  \psi^{(4),i,\alpha=j}(z) 
- \frac{3}{c}  \, \Psi 
\left(  \phi^{(\frac{7}{2}),i,\alpha=j} 
+ 
\phi^{(\frac{7}{2}),j,\alpha=i}
-  \delta^{ij}  \phi^{(\frac{7}{2}),k,\alpha=k} \right)(z),
\label{relation}
\\
\hat{\phi}^{(\frac{9}{2}),\alpha=i}(z) & = & 
\phi^{(\frac{9}{2}),\alpha=i}(z) 
+\frac{3}{2c} i \, \epsilon^{ijk}
\Psi \psi^{(4),j, \alpha=k} (z) +\frac{3}{2c} \, 
\Psi \pa \psi^{(3),\alpha=i}(z)
 -\frac{9}{c} \, \pa \Psi  \psi^{(3),\alpha=i}(z). 
\nonu
\eea
Of course, these are regular with the spin-$\frac{1}{2}$
current $\Psi(z)$ as in (\ref{conditions}).
The $\Psi$ dependent terms in the third and fourth equations 
in (\ref{relation}) leads to the nonlinear higher spin currents in their
OPEs. 
Note that for the higher spin currents with $\al$ representation 
the lowest higher spin currents remain unchanged only. See the last 
four equations of (\ref{relation}). 

\subsection{The OPEs between the eight lowest higher spin currents}

Using the explicit expressions in (\ref{relation}), 
we can calculate the OPEs between the higher spin currents.
We should reexpress the right hand sides of these OPEs 
in terms of new ${\cal N}=3$ 
higher spin currents in addition to the new ${\cal N}=3$ 
currents.

We summarize the OPEs as follows:
\bea
\left[ \hat{\psi}^{(\frac{3}{2})} \right] 
\cdot \left[ \hat{\psi}^{(\frac{3}{2})} \right] & = & \left[ I \right] +
\left[ \hat{\psi}^{(2)} \right], 
\nonu \\
\left[ \hat{\psi}^{(\frac{3}{2})} \right] 
\cdot \left[ \hat{\phi}^{(2),i} \right] & = & \left[ I \right] +
\left[ \hat{\phi}^{(\frac{5}{2}),i} \right], 
\nonu \\
\left[ \hat{\psi}^{(\frac{3}{2})} \right] 
\cdot \left[ \hat{\psi}^{(\frac{5}{2}),i} \right] & = & \left[ I \right] +
\left[ \hat{\psi}^{(2)} \right] + \left[ \hat{\psi}^{(3),i} \right] +
\left[ \hat{\phi}^{(3),i} \right] + \left[ \hat{\psi}^{(3),\al=i} \right], 
\nonu \\
\left[ \hat{\psi}^{(\frac{3}{2})} \right] 
\cdot \left[ \hat{\phi}^{(3)} \right] & = & \left[ I \right] +
\left[ \hat{\psi}^{(\frac{5}{2})} \right]
+ \left[ \hat{\phi}^{(\frac{5}{2}),i} \right]
+ \left[ \hat{\phi}^{(\frac{7}{2})} \right]
+ \left[ \hat{\phi}^{\frac{7}{2}, i, \al=i} \right],
\nonu \\
\left[ \hat{\phi}^{(2),i} \right] \cdot \left[ \hat{\phi}^{(2),j}\right] & = & 
\left[ I \right] + \delta^{ij} \left[ \hat{\psi}^{(2)}\right] + 
\epsilon^{ijk} \left[ \hat{\psi}^{(3),k}\right]
+ \epsilon^{ijk} \left[ \hat{\phi}^{(3),k}\right] + \epsilon^{ijk} 
\left[ \hat{\psi}^{(3),\al=k}\right],
\nonu \\
\left[ \hat{\phi}^{(2),i} \right] \cdot \left[ \hat{\psi}^{(\frac{5}{2}),j}\right] 
& = & \left[ I \right] +  \delta^{ij} \left[ \hat{\psi}^{(\frac{5}{2})} \right]
+  (\epsilon^{ijk} +\delta^{ij}) 
\left[ \hat{\phi}^{(\frac{5}{2}),k} \right]
+ \left[ \hat{\phi}^{(\frac{5}{2}),i} \right]
+ \left[ \hat{\phi}^{(\frac{5}{2}),j} \right]
\nonu \\
& + &  \delta^{ij} \left[ \hat{\phi}^{(\frac{7}{2})} \right]
+  \epsilon^{ijk} \left[ \hat{\psi}^{(2)} \right]
+   \delta^{ij}  \left[ \hat{\phi}^{(\frac{7}{2}),k, \al=k} \right]
+  \left[ \hat{\phi}^{(\frac{7}{2}),i, \al=j} \right]
+  \epsilon^{ijk}
  \left[ \hat{\psi}^{(\frac{7}{2}),k} \right],
\nonu \\
\left[ \hat{\phi}^{(2),i} \right] \cdot \left[ \hat{\phi}^{(3)}\right] & = & 
\left[ I \right]
+ \left[ \hat{\psi}^{(3),i}\right] 
+ \left[ \hat{\psi}^{(2)}\right]
+ \left[ \hat{\phi}^{(3),i}\right]
+ \left[ \hat{\psi}^{(3),\al=i}\right]
+ \epsilon^{ijk} \left[ \hat{\phi}^{(\frac{5}{2}),k}\right],
\nonu \\
&+& \epsilon^{ijk} \left[ \hat{\psi}^{(3),k}\right]
+ \epsilon^{ijk} \left[ \hat{\psi}^{(4),j, \al=k}\right],  
\nonu \\
\left[ \hat{\phi}^{(\frac{5}{2}),i} \right] \cdot \left[ 
\hat{\phi}^{(\frac{5}{2}),j}\right] & = & 
\left[ I \right] + ( \epsilon^{ijk}+ \delta^{ij} )
\left[ \hat{\psi}^{(2)}\right]+
(\epsilon^{ijk} +\delta^{ij}) \left[ \hat{\psi}^{(3),k}\right]
+ \epsilon^{ijk} \left[ \hat{\phi}^{(3),k}\right]
\nonu \\
& + &  \epsilon^{ijk} \left[ \hat{\psi}^{(3),\al=k}\right]
+ \delta^{ij}
\left[ \hat{\phi}^{(\frac{5}{2}),k}\right]
+ \left[ \hat{\phi}^{(\frac{5}{2}),i}\right]
+ \left[ \hat{\phi}^{(\frac{5}{2}),j}\right]
+\left[ \hat{\psi}^{(3),i}\right]
\nonu \\
& + &  \left[ \hat{\psi}^{(3),j}\right]
+  
\delta^{ij} \left[ \hat{\phi}^{(4)} \right]
+ \left[ \hat{\psi}^{(4),i,\al=j} \right]
+ \left[ \hat{\psi}^{(4), j, \al=i} \right]
+ \left[ \hat{\phi}^{(2),i}  \hat{\phi}^{(2),j} \right],
\nonu \\
\left[ \hat{\phi}^{(\frac{5}{2}),i} \right] \cdot \left[ 
\hat{\phi}^{(3)}\right] & = & 
\left[ I \right] +
 \left[ \hat{\phi}^{(\frac{5}{2}),i}\right]+ 
\epsilon^{ijk} \left[ \hat{\phi}^{(\frac{5}{2}),k}\right]+
\left[ \hat{\psi}^{(2)}\right]
+ \left[ \hat{\psi}^{(\frac{7}{2}),i} \right]
+  \epsilon^{ijk} \left[ \hat{\phi}^{(\frac{7}{2}),j,\al=k} \right]
\nonu \\
& + & \epsilon^{ijk}
 \left[ \hat{\psi}^{(3),k}\right]
+ \left[ \hat{\phi}^{(\frac{7}{2})} \right]
+  \left[ \hat{\phi}^{(\frac{9}{2}),\al=i} \right]
+
\left[ \hat{\psi}^{(\frac{3}{2})}  \hat{\phi}^{(2),i} \right],
\label{lastopeope}
\\
\left[ \hat{\phi}^{(3)} \right] \cdot \left[ 
\hat{\phi}^{(3)}\right] & = & 
\left[ I \right] +
\left[ \hat{\psi}^{(2)}\right]
+ \left[ \hat{\phi}^{(\frac{5}{2}),i}\right]+
\left[ \hat{\psi}^{(3),i}\right]
+ \left[ \hat{\phi}^{(4)} \right]
+ \left[ \hat{\psi}^{(4),i,\al=i} \right] 
+\left[ \pa \hat{\psi}^{(\frac{3}{2})} \hat{\psi}^{(\frac{3}{2})}  \right].
\nonu
\eea
Here we use the simplified notations where
the indices appearing in the ${\cal N}=3$ (nonlinear) 
superconformal currents are ignored.
This is the reason why the above OPEs (\ref{lastopeope})
do not preserve the covariance in the $SO(3)$ indices.  
In particular, the nonlinear terms 
between the higher spin currents appear in the last three OPEs of 
(\ref{lastopeope}).
It is easy to see that 
these come from the $\Psi$ dependent terms in the three places 
of the third and the fourth equations of (\ref{relation}).

\section{Conclusions and outlook }

We have constructed the eight higher spin currents, 
(\ref{spin3halfexpression}), Appendix (\ref{2expression}), 
Appendix (\ref{5halfexpression})
and Appendix (\ref{3expression}). 
We have found their OPEs given in (\ref{w3halfw3half}), 
Appendix (\ref{w3halfw2}) 
and Appendix (\ref{w2w2}) or its component results in 
Appendix (\ref{Fresult1}), Appendix (\ref{Fresult2}) and 
Appendix 
(\ref{compresult})
or its ${\cal N}=3$ version in (\ref{finalPhiPhi}).

Several comments are in order. The Jacobi identities used in 
section $6$ are not used completely because the OPEs between the 
higher spin currents are not known. We emphasize that 
those Jacobi identities are exactly zero. As we further study the OPEs 
between the lowest higher spin-$\frac{3}{2}$ current and the next 
higher spin currents, we expect that the Jacobi identities 
are satisfied up to null fields as in \cite{Ahn1604}. 
Non-freely generated algebra described in \cite{dFH}
due to the presence of the null fields
has been checked in \cite{CV1312} by applying to the Kazama-Suzuki model. 
See also the previous works \cite{Blumenhagenetal1,Blumenhagenetal2}
in the other specific  examples
how to use the vacuum character to check the null fields at a given spin.

One might ask 
whether there is a possibility to have the additional higher spin currents
in the right hand side of (\ref{finalPhiPhi}). 
In Appendix $H$, we present the $(N,M)=(3,2)$ case 
where there is NO extra higher spin current appearing in 
the right hand side of (\ref{finalPhiPhi}).
Although the vacuum character in the ${\cal N}=3$ Kazama-Suzuki model
will be found explicitly (and therefore 
the spin contents will be known up to a given spin), 
it will not be easy to 
see the extra higher spin currents completely 
as we add them in the right hand side of
(\ref{finalPhiPhi}). In order to determine the structure constants appearing 
in these extra higher spin currents, we should use the other Jacobi 
identities between them we do not know at the moment. Note that
the structure constants in (\ref{finalPhiPhi}) are not determined.

If the extra higher spin currents are present in the right hand side of
(\ref{finalPhiPhi}) in the large $(N,M)$ values, we expect 
that they will appear 
linearly \cite{BS} without spoiling the precise expression in the OPE 
(\ref{finalPhiPhi}). Of course, they can appear in the combinations with 
the ${\cal N}=3$ stress energy tensor as usual. 
The structure constants in them should contain the factors 
$(c-6)$ corresponding to $(N,M)=(2,2)$ case, 
$(c-9)$ corresponding to $(N,M)=(3,2)$ case and so on.
We may try to calculate the OPEs from Appendix $D$ manually
but this is beyond the scope of this paper because it is rather 
involved to obtain the higher spin currents ${\bf \Phi}^{(2)}(Z)$,
${\bf \Phi}^{(\frac{5}{2})}(Z)$ and ${\bf \Phi}^{(3),\alpha}(Z)$ 
explicitly with arbitrary 
$(N,M)$ dependence.

We list some open problems. 

$\bullet$
Marginal operator 

As in the large $c$ limit, we would like to 
construct the marginal operator which breaks the higher spin symmetry
but keeps the ${\cal N}=3$ superconformal symmetry.
It is an open problem to obtain the eigenvalue equation for the various 
zero modes acting on the corresponding states
and to obtain the mass terms for generic $c$.

$\bullet$
Orthogonal Kazama-Suzuki model

For the coset in \cite{CHR1306}
\bea
\frac{\hat{SO}(2N+M)_k \oplus \hat{SO}(2NM)_1}{\hat{SO}(2N)_{k+M} \oplus 
\hat{SO}(M)_{k+2N}}, \qquad c =\frac{ 3 k N M}{(M+2N+k-2)},
\label{othercoset}
\eea
the $M=2$ case in (\ref{othercoset}) 
corresponds to the other type of Kazama-Suzuki model.
What happens for the level $k=2N$ which is the dual Coxeter number 
of $SO(2N+2)$?
It would be interesting to see whether this coset model 
makes an enhancement of the original ${\cal N}=2$ supersymmetry
and to obtain the higher spin currents (if they exist). 
The relevant works \cite{Ahn1106,AP1410,AP1310,AP1301} 
on this direction will be useful to
construct the higher spin currents.
The relevant work can be found in \cite{KS} where the 
${\cal N}=1$ supersymmetry enhancement in the orthogonal coset exists 
at the 
critical level.

$\bullet$
Large ${\cal N}=4$ coset theory at the critical level

As described in the introduction, the $M=2$ case in the 
coset model (\ref{KScoset}) is similar to the large ${\cal N}=4$
coset theory. It is an open problem to 
check whether there exists an enhancement of the supersymmetry 
at the particular critical level or not. 
It is not known what is the algebra which has an ${\cal N}$ supersymmetry
with ${\cal N} > 4$. Therefore, it is better to study 
the fermion model having the ${\cal N}$ nonlinear superconformal algebra
\cite{Knizhnik,Bershadsky}. 
Then we should add the appropriate spin-$1$ and spin-$\frac{1}{2}$
currents in order to obtain the linear algebra. 
It is natural to ask the possibility for the higher spin extension.

$\bullet$
The OPEs between the next higher spin multiplets

So far, we have considered the OPE between the lowest 
${\cal N}=3$ higher spin multiplet. It is an open problem 
to calculate the other OPEs between the next 
${\cal N}=3$ higher spin multiplets.

\vspace{.7cm}

\centerline{\bf Acknowledgments}

We would like to thank Y. Hikida, M. Kim, and J. Paeng for discussions. 
This research was supported by Basic Science Research Program through
the National Research Foundation of Korea (NRF)  
funded by the Ministry of Education  
(No. 2016R1D1A1B03931786).
CA acknowledges warm hospitality from 
the School of  Liberal Arts (and Institute of Convergence Fundamental
Studies), Seoul National University of Science and Technology.

\newpage

\appendix

\renewcommand{\theequation}{\Alph{section}\mbox{.}\arabic{equation}}

\section{ The ${\cal N}=3$ superconformal algebra in section $2$}

We present the ${\cal N}=3$ superconformal algebra in component approach
corresponding to (\ref{j3j3})
as follows \cite{Ademolloetal,Ademolloetal1}:
{\small
\bea
\Psi(z) \, \Psi(w) &= & \frac{1}{(z-w)} \, \frac{c}{3} + \cdots,
\nonu \\
\Psi(z) \, G^i(w) & = & \frac{1}{(z-w)} \, J^i(w) +\cdots, \nonu \\
J^i(z) \, J^j(w) & = & \frac{1}{(z-w)^2} \, \frac{c}{3} \delta^{ij} +
\frac{1}{(z-w)} \, i \epsilon^{ijk} J^k(w) +\cdots,
\nonu \\
 J^i(z) \, G^j(w) & = & \frac{1}{(z-w)^2} \,  \delta^{ij} \Psi(w) +
\frac{1}{(z-w)} \, i \epsilon^{ijk} G^k(w) +\cdots,
\nonu \\
G^i(z) \, G^j(w) & = & \frac{1}{(z-w)^3} \, \frac{2c}{3} \delta^{ij}
+ \frac{1}{(z-w)^2} \, 2 i \epsilon^{ijk} J^k(w) + \frac{1}{(z-w)}
\left[ 2 \delta^{ij} T + i \epsilon^{ijk} \pa J^k \right](w) +\cdots,
\nonu \\
T(z) \, \Psi(w) & = &  
\frac{1}{(z-w)^2} \, \frac{1}{2} \Psi(w) + \frac{1}{(z-w)} \pa \Psi(w) 
+ \cdots,
\nonu \\
T(z) \, J^i(w) & = &  
\frac{1}{(z-w)^2} \,  J^i(w) + \frac{1}{(z-w)} \pa J^i(w) + \cdots,
\nonu \\
T(z) \, G^i(w) & = &  
\frac{1}{(z-w)^2} \, \frac{3}{2} G^i(w) + \frac{1}{(z-w)} \pa G^i(w) + \cdots,
\nonu \\
T(z) \, T(w) & = & \frac{1}{(z-w)^4} \, \frac{c}{2} + 
\frac{1}{(z-w)^2} \, 2 T(w) + \frac{1}{(z-w)} \pa T(w) + \cdots.
\label{compope}
\eea}
Note that there are no singular terms in the OPE between 
the spin-$\frac{1}{2}$ current $\Psi(z)$ and the spin-$1$ current $J^i(w)$.
One also obtains the OPE 
$G^i(z) \, \Psi(w) = \frac{1}{(z-w)} \, J^i(w) +\cdots$.
The nonlinear version of above ${\cal N}=3$ superconformal algebra
can be obtained by factoring out the spin-$\frac{1}{2}$ current 
\cite{GS,Schoutens}
and becomes the result of (\ref{nonKB}). 

\section{ The ${\cal N}=3$ primary conditions in the component 
approach  in section $2$}

From the  ${\cal N}=3$ primary higher spin multiplet
in (\ref{jphi}), we 
write down them in component approach as follows \cite{CK}:
{\small
\bea
\Psi(z) \, \psi_{\Delta}^{\alpha}(w) & = & + \cdots,
\nonu \\
\Psi(z) \, \phi_{\Delta+\frac{1}{2}}^{i, \alpha}(w) & = & -\frac{1}{(z-w)}
\, ({\bf T}^i)^{\alpha \beta} \psi_{\Delta}^{\beta}(w)+ \cdots,
\nonu \\
 \Psi(z) \, \psi_{\Delta+1}^{i, \alpha}(w) & = &
\frac{1}{(z-w)} \, \left[  \phi^{i,\alpha}_{\Delta+\frac{1}{2}} +
i \epsilon^{ijk} ({\bf T}^j)^{\alpha \beta} \phi^{k,\beta}_{\Delta+\frac{1}{2}}
\right](w) +\cdots,
\nonu \\
\Psi(z) \, \phi_{\Delta+\frac{3}{2}}^{ \alpha}(w) & = & 
\frac{1}{(z-w)^2} \, \Delta \, \psi^{\alpha}_{\Delta}(w)
-\frac{1}{(z-w)}
\, \left[ \frac{1}{2} \pa \psi^{\alpha}_{\Delta} + \frac{1}{2} ({\bf T}^i)^{\alpha \beta} 
\psi_{\Delta+1}^{i, \beta} \right](w)+ \cdots,
\nonu \\ 
J^i(z) \, \psi_{\Delta}^{\alpha}(w) & = & -\frac{1}{(z-w)} \, ({\bf T}^i)^{\alpha \beta} 
\psi_{\Delta}^{\beta}(w)+ \cdots,
\nonu \\
J^i(z) \, \phi_{\Delta+\frac{1}{2}}^{j, \alpha}(w) & = & \frac{1}{(z-w)}
\, \left[ i \epsilon^{ijk} \phi_{\Delta+\frac{1}{2}}^{k,\alpha}- 
({\bf T}^i)^{\alpha \beta} \phi_{\Delta+\frac{1}{2}}^{j, \beta} \right](w)+ \cdots,
\nonu \\
J^i(z) \, \psi_{\Delta+1}^{j, \alpha}(w) & = &
\frac{1}{(z-w)^2} \, \left[ 2 \Delta \delta^{ij} \psi_{\Delta}^{\alpha} -
i \epsilon^{ijk} ({\bf T}^k)^{\alpha \beta} \psi_{\Delta}^{\beta} \right](w)
\nonu \\
& + & 
\frac{1}{(z-w)} \, \left[   i \epsilon^{ijk} \psi^{k,\alpha}_{\Delta+1} -
({\bf T}^i)^{\alpha \beta} \psi^{j,\beta}_{\Delta+1}
\right](w) +\cdots,
\nonu \\
J^i(z) \, \phi_{\Delta+\frac{3}{2}}^{ \alpha}(w) & = & 
\frac{1}{(z-w)^2} \, \left[ (\Delta+\frac{1}{2}) \, 
\phi^{i,\alpha}_{\Delta+\frac{1}{2}} + \frac{i}{2} \epsilon^{ijk} 
({\bf T}^j)^{\alpha \beta} 
\phi^{k,\beta}_{\Delta+\frac{1}{2}}\right](w)
\nonu \\
& - & \frac{1}{(z-w)}
\,  ({\bf T}^i)^{\alpha \beta} \phi^{\beta}_{\Delta+\frac{3}{2}}(w)+ \cdots,
\nonu \\ 
G^i(z) \, \psi_{\Delta}^{\alpha}(w) & = & \frac{1}{(z-w)} \,  
\phi_{\Delta+\frac{1}{2}}^{i, \alpha}(w)+ \cdots,
\nonu \\
G^i(z) \, \phi_{\Delta+\frac{1}{2}}^{j, \alpha}(w) & = & 
\frac{1}{(z-w)^2} \, \left[ 2 \delta^{ij} \Delta \psi_{\Delta}^{\alpha} -
i \epsilon^{ijk} ({\bf T}^k)^{\alpha \beta} \psi_{\Delta}^{\beta} \right](w) 
\nonu \\
&+& \frac{1}{(z-w)}
\, \left[ \delta^{ij} \pa \psi_{\Delta}^{\alpha} + i 
\epsilon^{ijk} \psi_{\Delta+1}^{k,\alpha} \right](w)+ \cdots,
\nonu \\
G^i(z) \, \psi_{\Delta+1}^{j, \alpha}(w) & = &
\frac{1}{(z-w)^2} \, \left[ 2 (\Delta  +\frac{1}{2}) i \epsilon^{ijk} 
\phi_{\Delta+\frac{1}{2}}^{k,\alpha} +
({\bf T}^j)^{\alpha \beta} \phi_{\Delta+\frac{1}{2}}^{i, \beta} -\delta^{ij} 
({\bf T}^k)^{\alpha \beta} \phi_{\Delta+\frac{1}{2}}^{k,\beta} \right](w)
\nonu \\
& + & 
\frac{1}{(z-w)} \, \left[   2 \delta^{ij}  \phi^{\alpha}_{\Delta+\frac{3}{2}} +
i \epsilon^{ijk} \pa \phi^{k,\alpha}_{\Delta+\frac{1}{2}}
\right](w) +\cdots,
\nonu \\
G^i(z) \, \phi_{\Delta+\frac{3}{2}}^{ \alpha}(w) & = & 
-\frac{1}{(z-w)^3} \, ({\bf T}^i)^{\alpha \beta} \psi_{\Delta}^{\beta}(w)  
\nonu \\
&+&
\frac{1}{(z-w)^2} \, \left[ (\Delta+ 1) \, 
\psi^{i,\alpha}_{\Delta+1} + \frac{i}{2} \epsilon^{ijk} ({\bf T}^j)^{\alpha \beta} 
\psi^{k,\beta}_{\Delta+1}\right](w)
\nonu \\
& + & \frac{1}{(z-w)}
\,  \frac{1}{2} \pa \psi^{i, \alpha}_{\Delta+1}(w)+ \cdots,
\nonu \\ 
T(z) \, \psi_{\Delta}^{\alpha}(w) & = & \frac{1}{(z-w)^2}\,
\Delta  \psi_{\Delta}^{\alpha}(w) + \frac{1}{(z-w)} \, \pa 
\psi_{\Delta}^{\alpha}(w) + \cdots,
\nonu \\
T(z) \, \phi_{\Delta+\frac{1}{2}}^{i, \alpha}(w) & = & 
\frac{1}{(z-w)^2} \, (\Delta +\frac{1}{2})   
\phi_{\Delta+\frac{1}{2}}^{i, \alpha}(w)
+ \frac{1}{(z-w)}
\pa  \phi_{\Delta+\frac{1}{2}}^{i, \alpha}(w)
+ \cdots,
\nonu \\
T(z) \, \psi_{\Delta+1}^{i, \alpha}(w) & = &
- \frac{1}{(z-w)^3} \, ({\bf T}^i)^{\alpha \beta} \psi_{\Delta}^{\beta}(w)
+\frac{1}{(z-w)^2} \, (\Delta+1)  \psi_{\Delta+1}^{i, \alpha}(w)
+\frac{1}{(z-w)} \, 
\pa
\psi_{\Delta+1}^{i, \alpha}(w) +\cdots,
\nonu \\
T(z) \, \phi_{\Delta+\frac{3}{2}}^{ \alpha}(w) & = & 
-\frac{1}{(z-w)^3} \frac{1}{2} ({\bf T}^i)^{\alpha \beta} 
\phi_{\Delta+\frac{1}{2}}^{i,\beta}(w) 
+\frac{1}{(z-w)^2} \, (\Delta+\frac{3}{2}) \phi_{\Delta+\frac{3}{2}}^{ \alpha}(w)
+ \frac{1}{(z-w)}
\pa
\phi_{\Delta+\frac{3}{2}}^{ \alpha}(w)
\nonu \\
& + &  \cdots.
\label{jphicomp}
\eea}
For the $SO(3)$ singlet ${\cal N}=3$ higher spin multiplet, we ignore
the $SO(3)$ generator terms with boldface notation 
in (\ref{jphicomp}). 
Note that the higher spin currents 
$\psi_{\Delta+1}^{i, \alpha}(w)$ and 
$ \phi_{\Delta+\frac{3}{2}}^{ \alpha}(w)$
are not primary currents under the stress energy tensor
$T(z)$ according to the last two equations in (\ref{jphicomp}). 
We obtain the primary higher spin currents 
$(\psi_{\Delta+1}^{i, \alpha} + \frac{1}{2\Delta} 
({\bf T}^i)^{\alpha \beta} \pa \psi_{\Delta}^{\beta})(w)
$ and 
$ (\phi_{\Delta+\frac{3}{2}}^{ \alpha} +\frac{1}{2(2\Delta +1)} 
({\bf T}^i)^{\alpha \beta} \pa \phi_{\Delta+\frac{1}{2}}^{i,\beta})(w)$.

\section{ The fundamental OPEs between the spin-$\frac{1}{2}$ and 
the spin-$1$ currents in section $2$}

We present the various OPEs between the 
spin-$\frac{1}{2}$ currents and the spin-$1$ currents.

\subsection{The OPEs between the spin-$\frac{1}{2}$ currents}

There are five nontrivial OPEs as follows:
{\small
\bea
\Psi^{\al}(z) \, \Psi^{\beta}(w) & = & \frac{1}{(z-w)} \, \delta^{\al \beta}+
\cdots,
\nonu \\
\Psi^{\rho}(z) \, \Psi^{\si}(w) & = & \frac{1}{(z-w)} \, \delta^{\rho \si}+
\cdots,
\nonu \\
\Psi^{u(1)}(z) \, \Psi^{u(1)}(w) & = & \frac{1}{(z-w)}+
\cdots,
\nonu \\
\Psi^{a \bar{i}}(z) \, \Psi^{\bar{b} j}(w) & = & 
\frac{1}{(z-w)} \, \delta^{a \bar{b}} \, \delta^{\bar{i} j}+
\cdots,
\nonu \\
\psi^{a \bar{i}}(z) \, \psi^{\bar{b} j}(w) & = & \frac{1}{(z-w)} \, 
\delta^{a \bar{b}} \, \delta^{\bar{i} j}+
\cdots,
\label{1half1half}
\eea}
where the first four OPEs correspond to $(3.18)$ of \cite{CHR1406}.

\subsection{The OPEs between the spin-$\frac{1}{2}$ currents and the 
spin-$1$ currents}

By specifying the structure constants explicitly, we 
have the following OPEs between the spin-$\frac{1}{2}$ currents and 
the spin-$1$ currents
{\small
\bea
\Psi^{\alpha}(z) \, J_1^{\beta}(w) & = & 
\frac{1}{(z-w)} \, i f^{\al \beta \ga} 
\Psi^{\ga}(w) + \cdots,
\nonu \\
\Psi^{\alpha}(z) \, J^{a \bar{i}}(w)  & = & 
\frac{1}{(z-w)} \,  
\Psi^{b \bar{i}} \, t^{\al}_{b \bar{a}}(w) + \cdots,
\nonu \\
\Psi^{\alpha}(z) \, J^{\bar{a} i}(w)  & = & 
-\frac{1}{(z-w)} \,  
 t^{\al}_{a \bar{b}} \, \Psi^{\bar{b} i}(w) + \cdots,
\nonu \\
\Psi^{\rho}(z) \, J_1^{\si}(w) & = & 
\frac{1}{(z-w)} \, i f^{\rho \si \tau } 
\Psi^{\tau}(w) + \cdots,
\nonu \\
\Psi^{\rho}(z) \, J^{a \bar{i}}(w)  & = & 
-\frac{1}{(z-w)} \,   t^{\rho}_{i \bar{j}} \,
\Psi^{a \bar{j}} \,(w) + \cdots,
\nonu \\
\Psi^{\rho}(z) \, J^{\bar{a} i}(w)  & = & 
\frac{1}{(z-w)} \,  
 \Psi^{\bar{a} j} \,  t^{\rho}_{j \bar{i}}(w) + \cdots,
\nonu \\
\Psi^{u(1)}(z) \, J^{a \bar{i}}(w)  & = & 
\frac{1}{(z-w)} \, \sqrt{\frac{N+M}{N M}}  
 \Psi^{a \bar{i}}(w) + \cdots,
\nonu \\
\Psi^{u(1)}(z) \, J^{ \bar{a} i}(w)  & = & - 
\frac{1}{(z-w)} \, \sqrt{\frac{N+M}{N M}}  
 \Psi^{ \bar{a} i}(w) + \cdots,
\nonu \\
\Psi^{a \bar{i}}(z) \, J_2^{\al}(w) & = & 
-\frac{1}{(z-w)} \, 
\Psi^{b \bar{i}} \,  t^{\al}_{b \bar{a}}(w) + \cdots,
\nonu \\
\Psi^{a \bar{i}}(z) \, J_2^{\rho}(w) & = & 
\frac{1}{(z-w)} \,  t^{\rho}_{i \bar{j}}
\Psi^{a \bar{j}}(w) + \cdots,
\nonu \\
\Psi^{a \bar{i}}(z) \, J^{u(1)}(w) & = & 
-\frac{1}{(z-w)} \,  \sqrt{\frac{N+M}{N M}}
\Psi^{a \bar{i}}(w) + \cdots,
\nonu \\
\Psi^{b \bar{j}}(z) \, J^{\bar{a} i}(w) & = & 
\frac{1}{(z-w)} \Bigg(  \sqrt{\frac{N+M}{N M}}
\delta^{b \bar{a}} \, \delta^{\bar{j} i} \Psi^{u(1)} +
\delta^{\bar{j} i} \,   t^{\al}_{a \bar{b}} \, \Psi^{\al}-
\delta^{b \bar{a} } \,   \Psi^{\rho}  \, t^{\rho}_{ j \bar{i}} \Bigg)(w) + \cdots,
\nonu \\
\Psi^{ \bar{a} i }(z) \, J_2^{\al}(w) & = & 
\frac{1}{(z-w)} \,  t^{\al}_{a \bar{b}} \,
\Psi^{ \bar{b} i} \, (w) + \cdots,
\nonu \\
\Psi^{ \bar{a} i}(z) \, J_2^{\rho}(w) & = & 
-\frac{1}{(z-w)} \, 
\Psi^{ \bar{a} j} \,  t^{\rho}_{j \bar{i}}(w) + \cdots,
\nonu \\
\Psi^{ \bar{a} i}(z) \, J^{u(1)}(w) & = & 
\frac{1}{(z-w)} \,  \sqrt{\frac{N+M}{N M}}
\Psi^{ \bar{a} i}(w) + \cdots,
\nonu \\
\Psi^{\bar{b} j}(z) \, J^{a \bar{i} }(w) & = & 
\frac{1}{(z-w)} \Bigg(  -\sqrt{\frac{N+M}{N M}}
\delta^{\bar{b} a} \, \delta^{j \bar{i} } \Psi^{u(1)} -
\delta^{j \bar{i} } \,   t^{\al}_{b \bar{a}} \, \Psi^{\al}+
\delta^{ \bar{b} a } \,   \Psi^{\rho}  \, t^{\rho}_{ i \bar{j} } \Bigg)(w) + \cdots,
\nonu \\
\psi^{a \bar{i} }(z) \, j^{\al}(w) & = & 
-\frac{1}{(z-w)} \,  
\psi^{ b \bar{i} } \, t^{\al}_{b \bar{a}}(w) + \cdots,
\nonu \\
\psi^{a \bar{i} }(z) \, j^{\rho}(w) & = & 
\frac{1}{(z-w)} \,  t^{\rho}_{i \bar{j}}\,  
\psi^{ a \bar{j} }(w) + \cdots,
\nonu \\
\psi^{a \bar{i} }(z) \, j^{u(1)}(w) & = & 
-\frac{1}{(z-w)} \, 
\psi^{ a \bar{i} }(w) + \cdots,
\nonu  \\
\psi^{ \bar{a} i }(z) \, j^{\al}(w) & = & 
\frac{1}{(z-w)} \, t^{\al}_{a \bar{b}} 
\psi^{  \bar{b} i } \, (w) + \cdots,
\nonu \\
\psi^{ \bar{a} i }(z) \, j^{\rho}(w) & = & 
-\frac{1}{(z-w)} \,   
\psi^{  \bar{a} j }  \, t^{\rho}_{j \bar{i}}(w) + \cdots,
\nonu \\
\psi^{ \bar{a} i}(z) \, j^{u(1)}(w) & = & 
\frac{1}{(z-w)} \, 
\psi^{  \bar{a} i }(w) + \cdots,
\label{1half1}
\eea}
corresponding to $(3.20)$ of \cite{CHR1406}.
The spin-$1$ currents are
given by
{
\small
\bea
J_1^{\al}(z) & \equiv & -\frac{i}{2} f^{\al \beta \ga} \Psi^{\beta} \Psi^{\ga}(z),
\nonu \\
J_2^{\al}(z) & \equiv & \delta_{\bar{i} j}
\Psi^{b \bar{i}} t^{\al}_{b \bar{a}} \Psi^{\bar{a} j} 
(z),
\nonu \\
J_1^{\rho}(z) & \equiv & -\frac{i}{2} f^{\rho \si \tau} \Psi^{\si} \Psi^{\tau}(z),
\nonu \\
J_2^{\rho}(z) & = &  \delta_{\bar{a} b}
\Psi^{ \bar{a} i} t^{\rho}_{i \bar{j}} \Psi^{b \bar{j} } 
(z),
\nonu \\
J^{u(1)}(z) & \equiv & \sqrt{\frac{N+M}{N M}} \delta_{a\bar{b}} \delta_{\bar{i} j}
\Psi^{a\bar{i}} \Psi^{\bar{b}j}(z),
\nonu \\
J^{a\bar{i}}(z) & \equiv &  \Bigg(
\sqrt{\frac{N+M}{N M}} \Psi^{u(1)} \Psi^{a \bar{i}}
+ \delta^{\bar{a} a} t^{\al}_{b \bar{a}} 
\Psi^{\al} \Psi^{b \bar{i}} - \Psi^{\rho} \delta^{\bar{i} i} t^{\rho}_{i \bar{j}}
\Psi^{a \bar{j}} \Bigg)(z), 
\nonu  \\
J^{\bar{a} i }(z) & \equiv &  \Bigg( -
\sqrt{\frac{N+M}{N M}} \Psi^{u(1)} \Psi^{ \bar{a} i}
- \delta^{\bar{a} a} t^{\al}_{a \bar{b}} 
\Psi^{\al} \Psi^{ \bar{b} i} + \Psi^{\rho} \delta^{\bar{i} i} t^{\rho}_{j \bar{i}}
\Psi^{\bar{a} j} \Bigg)(z), 
\nonu  \\
j^{\al}(z) & \equiv & \delta_{\bar{i} j} \psi^{b\bar{i}} t^{\al}_{b \bar{a}} 
\psi^{\bar{a} j}(z),
\nonu \\
j^{\rho}(z) & \equiv &  \delta_{\bar{a} b} \psi^{\bar{a} i} t^{\rho}_{i \bar{j}} 
\psi^{b \bar{j} }(z),
\nonu \\
j^{u(1)}(z) & \equiv &  \delta_{a\bar{b} } \delta_{\bar{i} j} 
\psi^{a\bar{i} } \psi^{\bar{b} j}(z).
\label{spin1intermsof1half}
\eea}
Note that $\hat{J}^{u(1)}(z) \equiv \sqrt{\frac{N M}{ (N+M)}} \, 
J^{u(1)}(z)$.
Then the currents in the denominator of the coset (\ref{KScoset}) 
are given by 
$(J^{\al} + j^{\al})(z)$ for $SU(N)$ factor,
$(J^{\rho} + j^{\rho})(z)$ for $SU(M)$ factor,
and $(N+M)(\hat{J}^{u(1)} +  j^{u(1)})(z)$ for 
$U(1)$ factor where $J^{\al}(z) \equiv (J_1^{\al}+J_2^{\al})(z)$
and $J^{\rho}(z) \equiv (J_1^{\rho}+J_2^{\rho})(z)$.

\subsection{The OPEs between the 
spin-$1$ currents and itself}

Furthermore, one
obtains the following OPEs 
by expanding the structure constants explicitly 
as before 
where the spin-$1$ currents are given in Appendix
(\ref{spin1intermsof1half})
{\small
\bea
J_1^{\al}(z) \, J_1^{\beta}(w) & = & \frac{1}{(z-w)^2} \, N \delta^{\al \beta}
+
\frac{1}{(z-w)} \, i f^{\al \beta \ga} \, J_1^{\ga}(w) + \cdots,
\nonu \\
J_1^{\al}(z) \, J^{a \bar{i}}(w) & = & 
\frac{1}{(z-w)} \, i f^{\al \beta \ga} \, \Psi^{\ga} \, \Psi^{b \bar{i}}
\, t^{\beta}_{b \bar{a}}(w) + \cdots,
\nonu \\
J_1^{\al}(z) \, J^{ \bar{a} i}(w) & = & 
-\frac{1}{(z-w)} \, i f^{\al \beta \ga} \, \Psi^{\ga} \, \Psi^{ \bar{b} i}
\, t^{\beta}_{a \bar{b}}(w) + \cdots,
\nonu \\
J_2^{\al}(z) \, J_2^{\beta}(w) & = & \frac{1}{(z-w)^2} \, M \delta^{\al \beta}
+
\frac{1}{(z-w)} \, i f^{\al \beta \ga} \, J_2^{\ga}(w) + \cdots,
\nonu \\
J_2^{\al}(z) \, J^{a \bar{i}}(w) & = & 
\frac{1}{(z-w)} \Bigg( 
t^{\al}_{b \bar{a}} J^{b \bar{i}} -
i f^{\al \beta \ga} \, \Psi^{\ga} \, \Psi^{b \bar{i}}
\, t^{\beta}_{b \bar{a}} \Bigg)(w) + \cdots,
\nonu \\
J_2^{\al}(z) \, J^{ \bar{a} i}(w) & = & 
\frac{1}{(z-w)} \Bigg(
- J^{\bar{b} i} \, t^{\al}_{a \bar{b}}+ 
i f^{\al \beta \ga} \, \Psi^{\ga} \,  t^{\beta}_{a \bar{b}} \,
\Psi^{ \bar{b} i} \Bigg)(w) + \cdots,
\nonu \\
J_1^{\rho}(z) \, J^{a \bar{i}}(w) & = & - 
\frac{1}{(z-w)} \, i f^{\rho \si \tau} \, \Psi^{\tau} \, t^{\si}_{i \bar{j}}\,
\Psi^{a \bar{j}}(w) + \cdots,
\nonu \\
J_1^{\rho}(z) \, J^{ \bar{a} i}(w) & = & 
\frac{1}{(z-w)} \, i f^{\rho \si \tau} \, \Psi^{\tau} \,
\Psi^{ \bar{a} j}  \, t^{\si}_{j \bar{i}}(w) + \cdots,
\nonu \\
J_1^{\rho}(z) \, J_1^{ \si}(w) & = & 
\frac{1}{(z-w)^2} \, M \, \delta^{\rho \si} 
+ \frac{1}{(z-w)} \, i
f^{\rho \si \tau} J_1^{\tau}(w) + \cdots,
\nonu \\
J_2^{\rho}(z) \, J_2^{ \si}(w) & = & 
\frac{1}{(z-w)^2} \, N \, \delta^{\rho \si} 
+ \frac{1}{(z-w)} \, i
f^{\rho \si \tau} J_2^{\tau}(w) + \cdots,
\nonu \\
J_2^{\rho}(z) \, J^{a \bar{i}}(w) & = & 
\frac{1}{(z-w)} \Bigg( - J^{a \bar{j}} \, t^{\rho}_{i \bar{j}}
 +i f^{\rho \si \tau} \, \Psi^{\tau} \, t^{\si}_{i \bar{j}}\,
\Psi^{a \bar{j}} \Bigg)(w) + \cdots,
\nonu \\
J_2^{\rho}(z) \, J^{ \bar{a} i}(w) & = &  
\frac{1}{(z-w)} \Bigg( 
J^{\bar{a} j} \, t^{\rho}_{j \bar{i}}
-i f^{\rho \si \tau} \, \Psi^{\tau} \,
\Psi^{ \bar{a} j}  \, t^{\si}_{j \bar{i}} \Bigg)(w) + \cdots,
\nonu \\
J^{u(1)}(z) \, J^{u(1)}(w) & = & \frac{1}{(z-w)^2} \, (N+M)
+ \cdots,
\nonu \\
J^{u(1)}(z) \, J^{a \bar{i}}(w) & = & \frac{1}{(z-w)} \, \sqrt{\frac{N+M}{N M}}
 J^{a \bar{i}}(w)+ \cdots,
\nonu \\
J^{u(1)}(z) \, J^{ \bar{a} i}(w) & = & -\frac{1}{(z-w)} \, \sqrt{\frac{N+M}{N M}}
 J^{ \bar{a} i}(w)+ \cdots,
\nonu \\
J^{a\bar{i}}(z) \, J^{\bar{b} j}(w) & = &
\frac{1}{(z-w)^2} \, (N+M) \delta^{a \bar{b}} \, \delta^{\bar{i} j}
\nonu \\
& + & \frac{1}{(z-w)}  \Bigg( \delta^{\bar{i} j} \, t^{\al}_{b \bar{a}} \, 
(J_1^{\al} +J_2^{\al}) -\delta^{a\bar{b}} \, t^{\rho}_{i \bar{j}} (J_1^{\rho}
+ J_2^{\rho}) + \sqrt{\frac{N+M}{N M}} \, \delta^{a\bar{b}} \, 
\delta^{\bar{i} j} \, J^{u(1) }\Bigg)(w) \nonu \\
& + &  \cdots, 
\nonu \\
j^{\al}(z) \, j^{\beta}(w) & =& 
\frac{1}{(z-w)^2} \, M \delta^{\al \beta} + \frac{1}{(z-w)} \, i f^{\al \beta \ga}
\, j^{\ga}(w) + \cdots,
\nonu \\
j^{\rho}(z) \, j^{\si}(w) & =& 
\frac{1}{(z-w)^2} \, N \delta^{\rho \si} + \frac{1}{(z-w)} \, i f^{\rho \si 
\tau}
\, j^{\tau}(w) + \cdots,
\nonu \\
j^{u(1)}(z) \, j^{u(1)}(w) & = & \frac{1}{(z-w)^2} \, N M + \cdots,
\label{11}
\eea}
corresponding to $(3.21)$ of \cite{CHR1406}.
We observe that the levels are given by 
$(N+2M)$, $(2N+M)$ and $2N M (N+M)^2$
corresponding to the $SU(N)$, $SU(M)$ and $U(1)$ factors
appearing on the denominator of the coset (\ref{KScoset}) 
respectively.
This can be seen from the second-order pole of each OPE 
between the corresponding current we described in previous subsection.
The first level is obtained from 
the first, the fourth and the seventeenth equations
of (\ref{11}).
The second level is given by the ninth, the tenth and the eighteenth 
equations
of (\ref{11}).
Finally the third level can be obtained from the thirteenth and the last 
equations of (\ref{11}).

\section{ The remaining higher spin currents  in section $3$}

Let us introduce the following $(N,M)$ dependent coefficient functions
{\small
\bea
A(N,M) & \equiv
&\sqrt{\frac{3N^2 M}{2(M+N)(2M+3N)(MN-1)}}, 
\nonu \\
 B(N,M) & \equiv & A(M,N) = \sqrt{\frac{3M^2 N}{2(M+N)(2N+3M)(MN-1)}},
\nonu \\
a(N,M)& \equiv &-\frac{2M}{3N}, 
\nonu \\
 b(N,M) & \equiv & a(M,N) = -\frac{2N}{3M}.
\label{nmdependent}
\eea}
We present the remaining $7$ higher spin currents 
in terms of 
various fermions as follows with the help of 
(\ref{1half1half}), (\ref{1half1}) and (\ref{11}).

\subsection{The higher spin-$2$ currents}

By using (\ref{g3halfope}) and (\ref{Gphis}),
the three higher spin-$2$ currents, together with (\ref{nmdependent}), 
can be described as
{\small
\bea
\phi^{(2),+}(z) &=& -\frac{A(N,M)}{\sqrt{N+M}}    
t^{\alpha}_{b \bar{a}} (\psi^{\bar{a} i} \Psi^{b \bar{i}})
\Bigg( (3a(N,M)-2) J_1^{\alpha}+J_2^{\alpha}+j^{\alpha} \Bigg) (z)
\nonu \\
 &+&  \frac{B(N,M)}{\sqrt{N+M}}   
t^{\rho}_{i \bar{j}} (\psi^{\bar{a} i} \Psi^{a \bar{j}})
\Bigg( (3b(N,M)-2) J_1^{\rho}+J_2^{\rho}+j^{\rho} \Bigg) (z),
\nonu \\
\phi^{(2),-}(z) &=&\frac{A(N,M)}{\sqrt{N+M}}     
t^{\alpha}_{a \bar{b}} (\psi^{a \bar{i}} \Psi^{\bar{b} i})
\Bigg( (3a(N,M)-2) J_1^{\alpha}+J_2^{\alpha}+j^{\alpha}\Bigg) (z)
\nonu \\
&-& \frac{B(N,M)}{\sqrt{N+M}}    
t^{\rho}_{j \bar{i}} (\psi^{a \bar{i}} \Psi^{\bar{a} j})
\Bigg( (3b(N,M)-2) J_1^{\rho}+J_2^{\rho}+j^{\rho} \Bigg) (z),
\nonu \\
\phi^{(2),3}(z) &=&\frac{A(N,M)}{\sqrt{2(N+M)}}  
(J_2^{\alpha}-j^{\alpha}) \Bigg((3a(N,M)-2) 
J_1^{\alpha}+J_2^{\alpha}+j^{\alpha}\Bigg)(z)
\nonu \\
&+&  \frac{B(N,M)}{\sqrt{2(N+M)}}  
(J_2^{\rho}-j^{\rho}) \Bigg((3b(N,M)-2) J_1^{\rho}+J_2^{\rho}+j^{\rho}\Bigg) (z).
\label{2expression}
\eea}
The first two of (\ref{2expression})
are not fully normal ordered products in the spirit of \cite{BBSS1,BBSS2,BS}.

\subsection{The higher spin-$\frac{5}{2}$ currents}

The three higher spin-$\frac{5}{2}$ currents by using (\ref{5halfdefinition}) 
are summarized by
{\small
\bea
\psi^{(\frac{5}{2}),+}(z)&=&\frac{A(N,M)}{\sqrt{2}(N+M)} \Bigg[ 
  2  t^{\alpha}_{b \bar{a}}  (\psi^{\bar{a} i} J^{b \bar{i}} )
\Bigg((3a(N,M)-2) J_1^{\alpha}+J_2^{\alpha}+j^{\alpha} \Bigg)  
+   i f^{\alpha \beta \gamma} (\psi^{ \bar{a} i} \Psi^{\beta} \Psi^{b \bar{i}} )  
\nonu \\
& \times & 
\Bigg((3a(N,M)-2) J_1^{\alpha}+(4-3a(N,M))J_2^{\alpha}+(3a(N,M)-2)j^{\alpha} \Bigg)
 t^{\gamma}_{b \bar{a}}  \Bigg](z)
 \nonu \\
 &-& \frac{B(N,M)}{\sqrt{2}(N+M)}  \Bigg[ 
2 t^{\rho}_{i \bar{j}} (\psi^{\bar{a} i } J^{a \bar{j}}) \Bigg((3b(N,M)-2) J_1^{\rho}
+J_2^{\rho}+j^{\rho} \Bigg) 
 +    i f^{\rho \sigma \tau} (\psi^{\bar{a} i} \Psi^{\sigma} \Psi^{a \bar{j}} ) 
\nonu \\ 
& \times & \Bigg((3b(N,M)-2) J_1^{\rho}+(4-3b(N,M))J_2^{\rho}+(3b(N,M)-2)j^{\rho}
\Bigg)
 t^{\tau}_{i \bar{j}} \Bigg] (z),
 \nonu \\
 \psi^{(\frac{5}{2}),-}(z)&=&\frac{A(N,M)}{\sqrt{2}(N+M)} \Bigg[ 
 2  t^{\alpha}_{a \bar{b}}  (\psi^{a \bar{i}} J^{ \bar{b} i } ) \Bigg((3a(N,M)-2) 
J_1^{\alpha}+J_2^{\alpha}+j^{\alpha} \Bigg)
+ i f^{\alpha \beta \gamma} (\psi^{a \bar{i}} \Psi^{\beta} \Psi^{ \bar{b}i } ) 
\nonu \\
& \times & 
\Bigg((3a(N,M)-2) J_1^{\alpha}+(4-3a(N,M))J_2^{\alpha}+(3a(N,M)-2)j^{\alpha}\Bigg)
 t^{\gamma}_{a \bar{b}} \Bigg](z)
 \nonu \\
 &-&\frac{B(N,M)}{\sqrt{2}(N+M)}  
\Bigg[ 2 t^{\rho}_{j \bar{i}} (\psi^{a \bar{i}  } J^{\bar{a} j})
\Bigg((3b(N,M)-2) J_1^{\rho}+J_2^{\rho}+j^{\rho} \Bigg) 
 +   i f^{\rho \sigma \tau} (\psi^{a \bar{i} } \Psi^{\sigma} \Psi^{ \bar{a} j} ) 
\nonu \\
& \times &  
\Bigg((3b(N,M)-2) J_1^{\rho}+(4-3b(N,M))J_2^{\rho}+(3b(N,M)-2)j^{\rho} \Bigg)
 t^{\tau}_{j \bar{i}}  \Bigg](z),
 \nonu \\
  \psi^{(\frac{5}{2}),3}(z)&=& \frac{A(N,M)}{(N+M)} 
\Bigg[ t^{\alpha}_{a \bar{b}} \left(J^{a \bar{i}} \Psi^{\bar{b} i}
 -\sqrt{\frac{N+M}{N M}} \psi^{a \bar{i}} \psi^{\bar{b} i} \Psi^{u(1)}
 - \psi^{a \bar{i}} \psi^{\bar{c} i} \Psi^{\beta} t^{\beta}_{b \bar{c}}
 + \psi^{a \bar{i}} \psi^{\bar{b} j} \Psi^{\rho} t^{\rho}_{j \bar{i}} \right) 
 \nonu \\
& \times&  ((3a(N,M)-2) J_1^{\alpha}+J_2^{\alpha}+j^{\alpha})
+ 3i (1-a(N,M)) f^{\alpha \beta \gamma} t^{\alpha}_{a \bar{b}} t^{\beta}_{d \bar{c}}
(\psi^{a \bar{i}}  \Psi^{ \bar{b}i } ) 
(\psi^{ \bar{c} j} \Psi^{\gamma} \Psi^{ d\bar{j} })  \Bigg](z)
\nonu \\
&+&  \frac{B(N,M)}{(N+M)} 
\left[ t^{\rho}_{j \bar{i}} \left(-J^{a \bar{i}} \Psi^{\bar{a} j}
 +\sqrt{\frac{N+M}{N M}} \psi^{a \bar{i}} \psi^{ \bar{a} j} \Psi^{u(1)}
 + \psi^{a \bar{i}} \psi^{\bar{b} j} \Psi^{\beta} t^{\beta}_{a \bar{b}}
 - \psi^{a \bar{i}} \psi^{\bar{a} k} \Psi^{\sigma} t^{\sigma}_{k \bar{j}} \right) \right.
 \nonu \\
& \times& \left. ((3b(N,M)-2) J_1^{\rho}+J_2^{\rho}+j^{\rho})
+ 3i (1-b(N,M)) f^{\rho \sigma \tau } t^{\rho}_{j \bar{i}} t^{\sigma}_{k \bar{l}}
(\psi^{a \bar{i}}  \Psi^{ \bar{a} j } ) 
(\psi^{ \bar{b} k} \Psi^{\tau} \Psi^{ b \bar{l} })  \right](z) 
\nonu \\
& - &  \pa \psi^{(\frac{3}{2})}(z),
\label{5halfexpression} 
\eea}
with (\ref{nmdependent}).
The lowest higher spin-$\frac{3}{2}$ current
appearing in Appendix (\ref{5halfexpression})
is given by (\ref{spin3halfexpression}).
In order to express the above in the form of fully normal ordered products
we should further arrange them carefully.

\subsection{The higher spin-$3$ current}

By using (\ref{spin3inter}) and (\ref{spin3defintion}),
the higher spin-$3$ current, together with (\ref{nmdependent}),  
can be written as
{\small
\bea
\phi^{(3)}(z)&=& \frac{A(N,M)}{ \sqrt{8(N+M)^3}}  \Bigg[ 
2 t^{\alpha}_{a \bar{b}} \Bigg(
J^{a \bar{i}} J^{\bar{b} i} - \psi^{a \bar{i}} \psi^{ \bar{c} i} 
t^{\beta}_{b \bar{c}} (J_1^{\beta}+J_2^{\beta})
+\psi^{a \bar{i}} \psi^{ \bar{b} j} 
t^{\rho}_{j \bar{i}} (J_1^{\rho}+J_2^{\rho}) 
\nonu \\
&-&  \sqrt{\frac{(N+M)}{N M}} \psi^{a \bar{i}} \psi^{ \bar{b} i } J^{u(1)}
- (N+M) \psi^{a \bar{i}} \pa \psi^{ \bar{b} i} \Bigg)
((3a(N,M)-2) J_1^{\alpha}+J_2^{\alpha}+j^{\alpha} )
\nonu \\
&+& 6i (a(N,M)-1) f^{\alpha \beta \gamma} t^{\alpha}_{a \bar{b}} t^{\beta}_{d \bar{c}}
(\psi^{a \bar{i}}  J^{ \bar{b}i } ) 
(\psi^{ \bar{c} j} \Psi^{\gamma} \Psi^{ d\bar{j} }) 
\nonu \\
&+& i   f^{\alpha \beta \gamma} \Bigg( J^{a \bar{i}} \Psi^{\beta} \Psi^{\bar{b} i}
+\psi^{a \bar{i}} (\psi^{\bar{c} j} \Psi^{d \bar{j}})  \Psi^{ \bar{b} i} t^{\beta}_{d \bar{c}} 
- \sqrt{\frac{(N+M)}{N M}} 
\psi^{a \bar{i}} \psi^{ \bar{b} i } \Psi^{\beta} \Psi^{u(1)} 
\nonu \\
&-&  \psi^{a \bar{i}} \psi^{ \bar{c} i } \Psi^{\beta} \Psi^{\delta} t^{\delta}_{b \bar{c}} 
+ \psi^{a \bar{i}} \psi^{ \bar{b} j } \Psi^{\beta} \Psi^{\rho} t^{\rho}_{j \bar{i}} 
\Bigg)
\Bigg((3a(N,M)-2) J_1^{\alpha}+(4-3a(N,M)) J_2^{\alpha} \nonu \\
& + & ( 3a(N,M)-2) 
j^{\alpha} \Bigg) t^{\gamma}_{a \bar{b}}
+  6i (1-a(N,M))  
f^{\alpha \beta \gamma} (\psi^{a \bar{i}} \Psi^{\beta} \Psi^{ \bar{b}i } ) 
\nonu \\
& \times & 
\Bigg(
i f^{\alpha \delta \epsilon} \psi^{\bar{c} j } \Psi^{\delta} \Psi^{d \bar{j}} 
t^{\epsilon}_{d \bar{c}}
+ J^{c \bar{j}} \psi^{\bar{d} j} t^{\alpha}_{c \bar{d}} \Bigg)
  t^{\gamma}_{a \bar{b}}  \Bigg](z)
  \nonu \\
&+&   \frac{B(N,M)}{ \sqrt{8(N+M)^3}}  \Bigg[ 
2 t^{\rho}_{j \bar{i}} \Bigg(-J^{a \bar{i}} J^{\bar{a} j} + \psi^{a \bar{i}} \psi^{ \bar{b} j} 
t^{\alpha}_{a \bar{b}} (J_1^{\alpha}+J_2^{\alpha})
-\psi^{a \bar{i}} \psi^{ \bar{a} k} 
t^{\sigma}_{k \bar{j}} (J_1^{\sigma}+J_2^{\sigma}) 
\nonu \\
&+& \sqrt{\frac{N+M}{N M}} \psi^{a \bar{i}} \psi^{ \bar{a} j } J^{u(1)}
+ (N+M) \psi^{a \bar{i}} \pa \psi^{ \bar{a} j} \Bigg)
\Bigg((3b(N,M)-2) J_1^{\rho}+J_2^{\rho}+j^{\rho} \Bigg)
\nonu \\
&+& 6i (b(N,M)-1) f^{\rho \sigma \tau} t^{\rho}_{j \bar{i}} t^{\sigma}_{k \bar{l}}
(\psi^{a \bar{i}}  J^{ \bar{a} j } ) 
(\psi^{ \bar{b} k} \Psi^{\tau} \Psi^{ b \bar{l} }) 
\nonu \\
&-& i   f^{\rho \sigma \tau} \Bigg( J^{a \bar{i}} \Psi^{\sigma} \Psi^{\bar{a} j}
-\psi^{a \bar{i}} (\psi^{\bar{b} k} \Psi^{b \bar{l}})  \Psi^{ \bar{a} j} t^{\sigma}_{k \bar{l}} 
- \sqrt{\frac{N+M}{N M}} \psi^{a \bar{i}} \psi^{ \bar{a} j } \Psi^{\sigma} \Psi^{u(1)} 
\nonu \\
&-&  
\psi^{a \bar{i}} \psi^{ \bar{b} j } \Psi^{\sigma} \Psi^{\gamma} t^{\gamma}_{a \bar{b}} 
+ \psi^{a \bar{i}} \psi^{ \bar{a} k } \Psi^{\sigma} \Psi^{\upsilon} t^{\upsilon }_{k \bar{j}} \Bigg)
\nonu \\
& \times & \Bigg((3b(N,M)-2) 
J_1^{\rho}+(4-3b(N,M)) J_2^{\rho}+(3b(N,M)-2) j^{\rho} \Bigg) t^{\tau}_{j \bar{i}}
\nonu \\
&+&  
6i (1-b(N,M))  
f^{\rho \sigma \tau} (\psi^{a \bar{i}} \Psi^{\sigma} \Psi^{ \bar{a} j } ) 
\Bigg(
i f^{\rho \upsilon \phi} \psi^{\bar{b} k } \Psi^{\upsilon} \Psi^{b \bar{l}} t^{\phi}_{k \bar{l}}
+  \psi^{\bar{b} k} J^{b \bar{l}} t^{\rho}_{k \bar{l}} \Bigg)
  t^{\tau}_{j \bar{i}}  \Bigg](z) \nonu \\
& - &  \frac{1}{2} \pa \phi^{(2),3}(z),
\label{3expression}
\eea}
where 
one of the higher spin-$2$ current
appearing in Appendix (\ref{3expression})
is given by Appendix (\ref{2expression}).
Further rearrangement of the multiple products
should be done in order to write this in terms of 
fully normal ordered product \cite{BBSS1,BBSS2,BS}.  

\section{ The OPEs in the ${\cal N}=2$ superspace  in section $6$}

In this Appendix, we present the remaining  ${\cal N}=2$ OPEs
discussed in section $6$.

\subsection{ The OPEs in ${\cal N}=2$ superspace 
corresponding to Appendix $A$}

In order to describe the ${\cal N}=2$ description for the 
${\cal N}=2$ OPEs between the higher spin currents,
we should 
write down the standard ${\cal N}=3$ superconformal algebra 
in ${\cal N}=2$ superspace.
By writing down each component current with an appropriate combinations 
correctly, the following 
${\cal N}=2$ stress energy tensor satisfies the standard 
${\cal N}=2$ OPE   
{\small
\bea
{\bf T}(Z) & = &  J^3(z)- \theta \,\frac{1}{2} (G^1 + i G^2)(z) 
+ \bar{\theta} \, \frac{1}{2} (-G^1 + i G^2)(z) + 
\theta \bar{\theta} \,
T(z)
\nonu \\
& \equiv & \Bigg( J^3, \quad -\frac{1}{2} (G^1 + i G^2), \quad
  \frac{1}{2} (-G^1 + i G^2), \quad T \Bigg).
\label{stressn2}
\eea}
Then the remaining four currents 
of ${\cal N}=3$ superconformal algebra can combine 
as the following ${\cal N}=2$ multiplet
{\small
\bea
{\bf T}^{(\frac{1}{2})}(Z) & = &  2 \Psi(z)+ \theta \,
 (J^1 + i J^2)(z) 
+ \bar{\theta} \,  (-J^1 + i J^2)(z) + 
\theta \bar{\theta} \,
G^3(z)
\nonu \\
& \equiv & \Bigg( 2 \Psi, \quad  (J^1 + i J^2), \quad
   (-J^1 + i J^2), \quad G^3 \Bigg).
\label{multipletn2}
\eea}

Then we identify the previous component results 
presented in Appendix (\ref{compope}) can be interpreted as 
the following three ${\cal N}=2$ OPEs between the ${\cal N}=2$
stress energy tensor (\ref{stressn2}) and 
the ${\cal N}=2$ multiplet (\ref{multipletn2})
{\small
\bea
{\bf T}(Z_1) \, {\bf T}(Z_2) & = &  \frac{1}{z_{12}^2} \, \frac{c}{3}+
\frac{\theta_{12} \bar{\theta}_{12}}{z_{12}^2} \, {\bf T}(Z_2)
+ \frac{\bar{\theta}_{12}}{z_{12}} \, \overline{D} {\bf T}(Z_2)
-\frac{{\theta}_{12}}{z_{12}} \, D {\bf T}(Z_2)
\nonu \\
& + & \frac{\theta_{12} \bar{\theta}_{12}}{z_{12}} \, \pa {\bf T}(Z_2)
+ \cdots, 
\nonu \\ 
{\bf T}(Z_1) \, {\bf T}^{(\frac{1}{2})}(Z_2) & = &  
\frac{\theta_{12} \bar{\theta}_{12}}{z_{12}^2} \,\frac{1}{2} 
{\bf T}^{(\frac{1}{2})}(Z_2)
+ \frac{\bar{\theta}_{12}}{z_{12}} \, \overline{D} {\bf T}^{(\frac{1}{2})}(Z_2)
-\frac{{\theta}_{12}}{z_{12}} \, D {\bf T}^{(\frac{1}{2})}(Z_2)
\nonu \\
& + & 
\frac{\theta_{12} \bar{\theta}_{12}}{z_{12}} \, \pa {\bf T}^{(\frac{1}{2})}(Z_2)
+ \cdots, 
\nonu \\
{\bf T}^{(\frac{1}{2})}(Z_1) \, {\bf T}^{(\frac{1}{2})}(Z_2) & = &  
 \frac{1}{z_{12}} \, \frac{4c}{3}
+ \frac{\theta_{12} \bar{\theta}_{12}}{z_{12}} \,  2 {\bf T}(Z_2)
+ \cdots. 
\label{threen2}
\eea}
Note that this ${\cal N}=2$ OPEs (\ref{threen2})
can be obtained from the large ${\cal N}=4$ superconformal algebra 
observed in \cite{BO,RASS}. 
The complex spinor covariant derivatives 
are given by $D = \frac{\pa}{\pa \theta} -\frac{1}{2} \bar{\theta} 
\frac{\pa}{\pa z}$ and $ \overline{D} = 
\frac{\pa}{\pa \bar{\theta}} -\frac{1}{2} \theta 
\frac{\pa}{\pa z}$ \cite{KT}.

\subsection{ The OPEs in ${\cal N}=2$ superspace 
corresponding to Appendix $B$ }

Furthermore, the component results presented in Appendix $B$
can be described in ${\cal N}=2$ superspace as follows:
{\small
\bea
{\bf T}(Z_1) \, {\bf W}^{(\frac{3}{2})}(Z_2) & = &  
\frac{\theta_{12} \bar{\theta}_{12}}{z_{12}^2} \,\frac{3}{2} 
{\bf W}^{(\frac{3}{2})}(Z_2)
+ \frac{\bar{\theta}_{12}}{z_{12}} \, \overline{D} {\bf W}^{(\frac{3}{2})}(Z_2)
-\frac{{\theta}_{12}}{z_{12}} \, D {\bf W}^{(\frac{3}{2})}(Z_2)
\nonu \\
& + &
 \frac{\theta_{12} \bar{\theta}_{12}}{z_{12}} \, \pa {\bf W}^{(\frac{3}{2})}(Z_2)
+ \cdots, 
\nonu \\
{\bf T}(Z_1) \, {\bf W}^{(2)}(Z_2) & = &  
\frac{\theta_{12} \bar{\theta}_{12}}{z_{12}^2} \,2
{\bf W}^{(2)}(Z_2)
+ \frac{\bar{\theta}_{12}}{z_{12}} \, \overline{D} {\bf W}^{(2)}(Z_2)
-\frac{{\theta}_{12}}{z_{12}} \, D {\bf W}^{(2)}(Z_2)
\nonu \\
& + & \frac{\theta_{12} \bar{\theta}_{12}}{z_{12}} \, \pa {\bf W}^{(2)}(Z_2)
+ \cdots, 
\nonu \\
{\bf T}(Z_1) \, {\bf W}^{(2')}(Z_2) & = &  
\frac{\theta_{12} \bar{\theta}_{12}}{z_{12}^2} \,2
{\bf W}^{(2')}(Z_2)
+ \frac{\bar{\theta}_{12}}{z_{12}} \, \overline{D} {\bf W}^{(2')}(Z_2)
-\frac{{\theta}_{12}}{z_{12}} \, D {\bf W}^{(2')}(Z_2)
\nonu \\
& + & \frac{\theta_{12} \bar{\theta}_{12}}{z_{12}} \, \pa {\bf W}^{(2')}(Z_2)
+ \cdots, 
\nonu \\
{\bf T}(Z_1) \, {\bf W}^{(\frac{5}{2})}(Z_2) & = &  
\frac{\theta_{12} \bar{\theta}_{12}}{z_{12}^2} \, \frac{5}{2}
{\bf W}^{(\frac{5}{2})}(Z_2)
+ \frac{\bar{\theta}_{12}}{z_{12}} \, \overline{D} {\bf W}^{(\frac{5}{2})}(Z_2)
-\frac{{\theta}_{12}}{z_{12}} \, D {\bf W}^{(\frac{5}{2})}(Z_2)
\nonu \\
& + & 
\frac{\theta_{12} \bar{\theta}_{12}}{z_{12}} \, \pa {\bf W}^{(\frac{5}{2})}(Z_2)
+ \cdots, 
\nonu \\
{\bf T}(Z_1) \, {\bf W}^{(\frac{5}{2}')}(Z_2) & = &  
\frac{\theta_{12} \bar{\theta}_{12}}{z_{12}^2} \, \frac{5}{2}
{\bf W}^{(\frac{5}{2}')}(Z_2)
+ \frac{\bar{\theta}_{12}}{z_{12}} \, \overline{D} {\bf W}^{(\frac{5}{2}')}(Z_2)
-\frac{{\theta}_{12}}{z_{12}} \, D {\bf W}^{(\frac{5}{2}')}(Z_2)
\nonu \\
& + &
 \frac{\theta_{12} \bar{\theta}_{12}}{z_{12}} \, \pa {\bf W}^{(\frac{5}{2}')}(Z_2)
+ \cdots, 
\nonu \\
{\bf T}(Z_1) \, {\bf W}^{(3)}(Z_2) & = &  
\frac{\theta_{12} \bar{\theta}_{12}}{z_{12}^2} \, 3
{\bf W}^{(3)}(Z_2)
+ \frac{\bar{\theta}_{12}}{z_{12}} \, \overline{D} {\bf W}^{(3)}(Z_2)
-\frac{{\theta}_{12}}{z_{12}} \, D {\bf W}^{(3)}(Z_2)
\nonu \\
& + & \frac{\theta_{12} \bar{\theta}_{12}}{z_{12}} \, \pa {\bf W}^{(3)}(Z_2)
+ \cdots, 
\nonu \\
{\bf T}(Z_1) \, {\bf W}^{(3),1}(Z_2) & = &  
\frac{\theta_{12} \bar{\theta}_{12}}{z_{12}^2} \, 3
{\bf W}^{(3),1}(Z_2)
+ \frac{\bar{\theta}_{12}}{z_{12}} \, \overline{D} {\bf W}^{(3),1}(Z_2)
-\frac{{\theta}_{12}}{z_{12}} \, D {\bf W}^{(3),1}(Z_2)
\nonu \\
& + & \frac{\theta_{12} \bar{\theta}_{12}}{z_{12}} \, \pa {\bf W}^{(3),1}(Z_2)
+  \frac{1}{z_{12}} \, i \,  {\bf W}^{(3),2}(Z_2) + \cdots, 
\nonu \\
{\bf T}(Z_1) \, {\bf W}^{(\frac{7}{2}),1}(Z_2) & = &  
\frac{\theta_{12}}{z_{12}^2} \,  \frac{1}{2} {\bf W}^{(3),3}(Z_2)
+ \frac{\bar{\theta}_{12}}{z_{12}^2} \,  \frac{1}{2} {\bf W}^{(3),3}(Z_2)
+ \frac{\theta_{12} \bar{\theta}_{12}}{z_{12}^2} \, \frac{7}{2}
{\bf W}^{(\frac{7}{2}),1}(Z_2)
\nonu \\
& + & 
 \frac{\bar{\theta}_{12}}{z_{12}} \, \overline{D} {\bf W}^{(\frac{7}{2}),1}(Z_2)
-\frac{{\theta}_{12}}{z_{12}} \, D {\bf W}^{(\frac{7}{2}),1}(Z_2)
+ \frac{\theta_{12} \bar{\theta}_{12}}{z_{12}} \, \pa {\bf W}^{(\frac{7}{2}),1}(Z_2)
\nonu \\
& + & \frac{1}{z_{12}} \, i \,  {\bf W}^{(\frac{7}{2}),2}(Z_2) + \cdots, 
\nonu \\
{\bf T}(Z_1) \, {\bf W}^{(3),2}(Z_2) & = &  
\frac{\theta_{12} \bar{\theta}_{12}}{z_{12}^2} \, 3
{\bf W}^{(3),2}(Z_2)
+ \frac{\bar{\theta}_{12}}{z_{12}} \, \overline{D} {\bf W}^{(3),2}(Z_2)
-\frac{{\theta}_{12}}{z_{12}} \, D {\bf W}^{(3),2}(Z_2)
\nonu \\
& + & \frac{\theta_{12} \bar{\theta}_{12}}{z_{12}} \, \pa {\bf W}^{(3),2}(Z_2)
-  \frac{1}{z_{12}} \, i \,  {\bf W}^{(3),1}(Z_2) + \cdots, 
\nonu \\
{\bf T}(Z_1) \, {\bf W}^{(\frac{7}{2}),2}(Z_2) & = &  
\frac{\theta_{12}}{z_{12}^2} \,  \frac{i}{2} {\bf W}^{(3),3}(Z_2)
- \frac{\bar{\theta}_{12}}{z_{12}^2} \,  \frac{i}{2} {\bf W}^{(3),3}(Z_2)
+ \frac{\theta_{12} \bar{\theta}_{12}}{z_{12}^2} \, \frac{7}{2}
{\bf W}^{(\frac{7}{2}),2}(Z_2)
\nonu \\
& + & 
 \frac{\bar{\theta}_{12}}{z_{12}} \, \overline{D} {\bf W}^{(\frac{7}{2}),2}(Z_2)
-\frac{{\theta}_{12}}{z_{12}} \, D {\bf W}^{(\frac{7}{2}),2}(Z_2)
+ \frac{\theta_{12} \bar{\theta}_{12}}{z_{12}} \, \pa {\bf W}^{(\frac{7}{2}),2}(Z_2)
\nonu \\
& - & \frac{1}{z_{12}} \, i \,  {\bf W}^{(\frac{7}{2}),1}(Z_2) + \cdots, 
\nonu \\
{\bf T}(Z_1) \, {\bf W}^{(3),3}(Z_2) & = &  
\frac{\theta_{12} \bar{\theta}_{12}}{z_{12}^2} \, 3
{\bf W}^{(3),3}(Z_2)
+ \frac{\bar{\theta}_{12}}{z_{12}} \, \overline{D} {\bf W}^{(3),3}(Z_2)
-\frac{{\theta}_{12}}{z_{12}} \, D {\bf W}^{(3),3}(Z_2)
\nonu \\
& + &  \frac{\theta_{12} \bar{\theta}_{12}}{z_{12}} \, \pa {\bf W}^{(3),3}(Z_2)
+   \cdots, 
\nonu \\
{\bf T}(Z_1) \, {\bf W}^{(\frac{7}{2}),3}(Z_2) & = &  
-\frac{\theta_{12}}{z_{12}^2} \,  \frac{1}{2} {\bf W}^{(3),1}(Z_2)
- \frac{\bar{\theta}_{12}}{z_{12}^2} \,  \frac{1}{2} {\bf W}^{(3),1}(Z_2)
-\frac{\theta_{12}}{z_{12}^2} \,  \frac{i}{2} {\bf W}^{(3),2}(Z_2)
\nonu \\
& + &  \frac{\bar{\theta}_{12}}{z_{12}^2} \,  \frac{i}{2} {\bf W}^{(3),2}(Z_2)
+   \frac{\theta_{12} \bar{\theta}_{12}}{z_{12}^2} \, \frac{7}{2}
{\bf W}^{(\frac{7}{2}),3}(Z_2)
\nonu \\
& + & 
 \frac{\bar{\theta}_{12}}{z_{12}} \, \overline{D} {\bf W}^{(\frac{7}{2}),3}(Z_2)
-\frac{{\theta}_{12}}{z_{12}} \, D {\bf W}^{(\frac{7}{2}),3}(Z_2)
+ \frac{\theta_{12} \bar{\theta}_{12}}{z_{12}} \, \pa {\bf W}^{(\frac{7}{2}),3}(Z_2)
 + \cdots, 
\nonu \\
{\bf T}^{(\frac{1}{2})}(Z_1) \, {\bf W}^{(\frac{3}{2})}(Z_2) & = &  
\frac{\theta_{12} \bar{\theta}_{12}}{z_{12}} \, {\bf W}^{(2)}(Z_2)
+   \cdots, 
\nonu \\
{\bf T}^{(\frac{1}{2})}(Z_1) \, {\bf W}^{(2)}(Z_2) & = &  
  \frac{\theta_{12} \bar{\theta}_{12}}{z_{12}^2} \, 3
{\bf W}^{(\frac{3}{2})}(Z_2)
\nonu \\
& + & 
 \frac{\bar{\theta}_{12}}{z_{12}} \, 2 \overline{D} {\bf W}^{(\frac{3}{2})}(Z_2)
-\frac{{\theta}_{12}}{z_{12}} \, 2 D {\bf W}^{(\frac{3}{2})}(Z_2)
+ \frac{\theta_{12} \bar{\theta}_{12}}{z_{12}} \, \pa {\bf W}^{(\frac{3}{2})}(Z_2)
 + \cdots, 
\nonu \\
{\bf T}^{(\frac{1}{2})}(Z_1) \, {\bf W}^{(2')}(Z_2) & = &  
\frac{\theta_{12} \bar{\theta}_{12}}{z_{12}} \, {\bf W}^{(\frac{5}{2})}(Z_2)
+   \cdots, 
\nonu \\
{\bf T}^{(\frac{1}{2})}(Z_1) \, {\bf W}^{(\frac{5}{2})}(Z_2) & = &  
  \frac{\theta_{12} \bar{\theta}_{12}}{z_{12}^2} \, 4
{\bf W}^{(2')}(Z_2)
\nonu \\
& + & 
 \frac{\bar{\theta}_{12}}{z_{12}} \, 2 \overline{D} {\bf W}^{(2')}(Z_2)
-\frac{{\theta}_{12}}{z_{12}} \, 2 D {\bf W}^{(2')}(Z_2)
+ \frac{\theta_{12} \bar{\theta}_{12}}{z_{12}} \, \pa {\bf W}^{(2')}(Z_2)
 + \cdots, 
\nonu \\
{\bf T}^{(\frac{1}{2})}(Z_1) \, {\bf W}^{(\frac{5}{2}')}(Z_2) & = &  
\frac{\theta_{12} \bar{\theta}_{12}}{z_{12}} \, {\bf W}^{(3)}(Z_2)
+   \cdots, 
\nonu \\
{\bf T}^{(\frac{1}{2})}(Z_1) \, {\bf W}^{(3)}(Z_2) & = &  
  \frac{\theta_{12} \bar{\theta}_{12}}{z_{12}^2} \, 5
{\bf W}^{(\frac{5}{2}')}(Z_2)
\nonu \\
& + & 
 \frac{\bar{\theta}_{12}}{z_{12}} \, 2 \overline{D} {\bf W}^{(\frac{5}{2}')}(Z_2)
-\frac{{\theta}_{12}}{z_{12}} \, 2 D {\bf W}^{(\frac{5}{2}')}(Z_2)
+ \frac{\theta_{12} \bar{\theta}_{12}}{z_{12}} \, \pa {\bf W}^{(\frac{5}{2}')}(Z_2)
 + \cdots, 
\nonu \\
{\bf T}^{(\frac{1}{2})}(Z_1) \, {\bf W}^{(3),1}(Z_2) & = &  
 \frac{\bar{\theta}_{12}}{z_{12}} \,  {\bf W}^{(3),3}(Z_2)
+\frac{{\theta}_{12}}{z_{12}} \,  {\bf W}^{(3),3}(Z_2)
+ \frac{\theta_{12} \bar{\theta}_{12}}{z_{12}} \,  {\bf W}^{(\frac{7}{2}),1}(Z_2)
 + \cdots, 
\nonu \\
{\bf T}^{(\frac{1}{2})}(Z_1) \, {\bf W}^{(\frac{7}{2}),1}(Z_2) & = &  
  \frac{\theta_{12} \bar{\theta}_{12}}{z_{12}^2} \, 6
{\bf W}^{(3),1}(Z_2)
\nonu \\
& + & 
 \frac{\bar{\theta}_{12}}{z_{12}} \, \left[
2 \overline{D} {\bf W}^{(3),1}  +{\bf W}^{(\frac{7}{2}),3} \right](Z_2)
+\frac{{\theta}_{12}}{z_{12}} \, \left[ -2 D {\bf W}^{(3),1} 
+{\bf W}^{(\frac{7}{2}),3}
\right](Z_2)
\nonu \\
& + & \frac{\theta_{12} \bar{\theta}_{12}}{z_{12}} \, 
\pa {\bf W}^{(3),1}(Z_2)
+ \frac{1}{z_{12}} \, 2 i  {\bf W}^{(3),2}(Z_2) 
+ \cdots, 
\nonu \\
{\bf T}^{(\frac{1}{2})}(Z_1) \, {\bf W}^{(3),2}(Z_2) & = &  
 -\frac{\bar{\theta}_{12}}{z_{12}} \,  i {\bf W}^{(3),3}(Z_2)
+\frac{{\theta}_{12}}{z_{12}} \,  i {\bf W}^{(3),3}(Z_2)
+ \frac{\theta_{12} \bar{\theta}_{12}}{z_{12}} \,  {\bf W}^{(\frac{7}{2}),2}(Z_2)
 + \cdots, 
\nonu \\
{\bf T}^{(\frac{1}{2})}(Z_1) \, {\bf W}^{(\frac{7}{2}),2}(Z_2) & = &  
  \frac{\theta_{12} \bar{\theta}_{12}}{z_{12}^2} \, 6
{\bf W}^{(3),2}(Z_2)
\nonu \\
& + & 
 \frac{\bar{\theta}_{12}}{z_{12}} \, \left[
2 \overline{D} {\bf W}^{(3),2}  - i {\bf W}^{(\frac{7}{2}),3} \right](Z_2)
+\frac{{\theta}_{12}}{z_{12}} \, \left[ -2 D {\bf W}^{(3),2} 
+ i {\bf W}^{(\frac{7}{2}),3}
\right](Z_2)
\nonu \\
& + & \frac{\theta_{12} \bar{\theta}_{12}}{z_{12}} \, 
\pa {\bf W}^{(3),2}(Z_2)
-\frac{1}{z_{12}} \, 2 i  {\bf W}^{(3),1}(Z_2) 
+ \cdots, 
\nonu \\
{\bf T}^{(\frac{1}{2})}(Z_1) \, {\bf W}^{(3),3}(Z_2) & = &  
 \frac{\bar{\theta}_{12}}{z_{12}} \,  \left[ -{\bf W}^{(3),1} + i{\bf W}^{(3),2} 
\right](Z_2)
-\frac{{\theta}_{12}}{z_{12}} \, \left[  {\bf W}^{(3),1} + i {\bf W}^{(3),2}
\right](Z_2)
\nonu \\
& + & \frac{\theta_{12} \bar{\theta}_{12}}{z_{12}} \,  {\bf W}^{(\frac{7}{2}),3}(Z_2)
 + \cdots, 
\label{primaryopes}
\\
{\bf T}^{(\frac{1}{2})}(Z_1) \, {\bf W}^{(\frac{7}{2}),3}(Z_2) & = &  
  \frac{\theta_{12} \bar{\theta}_{12}}{z_{12}^2} \, 6
{\bf W}^{(3),3}(Z_2) +
  \frac{\bar{\theta}_{12}}{z_{12}} \, \left[
2 \overline{D} {\bf W}^{(3),3}-  {\bf W}^{(\frac{7}{2}),1}  
+ i {\bf W}^{(\frac{7}{2}),2} \right](Z_2)
\nonu \\
& - & \frac{{\theta}_{12}}{z_{12}} \, \left[ 2 D {\bf W}^{(3),3} 
+  {\bf W}^{(\frac{7}{2}),1}
+ i {\bf W}^{(\frac{7}{2}),2}
\right](Z_2)
+  \frac{\theta_{12} \bar{\theta}_{12}}{z_{12}} \, 
\pa {\bf W}^{(3),3}(Z_2)
+ \cdots. 
\nonu
\eea}
We see that the six higher spin currents, ${\bf W}^{(3),\al}(Z_2)$
and ${\bf W}^{(\frac{7}{2}),\al}(Z_2)$ are not primary under the 
${\cal N}=2$ stress energy tensor ${\bf T}(Z_1)$.
We observe that 
the combination 
of $({\bf W}^{(3), 1} \pm i {\bf W}^{(3),2})(Z_2)$ has  $U(1)$ charge
$\pm 1$ by calculating the OPE
with ${\bf T}(Z_1)$
{\small
\bea
{\bf T}(Z_1) \, ({\bf W}^{(3), 1} \pm i {\bf W}^{(3),2})(Z_2) &= & 
\pm \frac{1}{z_{12}} \,  ({\bf W}^{(3), 1} \pm i {\bf W}^{(3),2})(Z_2)  + 
\mbox{sing. terms}.
\label{3u1charge}
\eea}
Similarly,
we can see
{\small
\bea
{\bf T}(Z_1) \, ({\bf W}^{(\frac{7}{2}), 1} \pm i {\bf W}^{(\frac{7}{2}),2})(Z_2) &= & 
\pm \frac{1}{z_{12}} \,  ({\bf W}^{(\frac{7}{2}), 1} \pm i 
{\bf W}^{(\frac{7}{2}),2})(Z_2)  + 
\mbox{sing. terms}.
\label{7halfu1charge}
\eea}
Therefore, 
 $({\bf W}^{(\frac{7}{2}), 1} \pm i {\bf W}^{(\frac{7}{2}),2})(Z_2)$
have the $U(1)$ charges (of ${\cal N}=2$ superconformal algebra) 
$\pm 1$ respectively. 

\subsection{ The remaining two OPEs in ${\cal N}=2$ superspace }

The ${\cal N}=2$ OPE between the ${\cal N}=2$ higher spin-$\frac{3}{2}$
current and the ${\cal N}=2$ higher spin-$2$ current can be summarized
by  
{\small
\bea
{\bf W}^{(\frac{3}{2})}(Z_1) \, {\bf W}^{(2)}(Z_2) & = & 
\frac{\theta_{12} \bar{\theta}_{12}}{z_{12}^4} \, 6 
{\bf T}^{(\frac{1}{2})}(Z_2) -
\frac{\theta_{12}}{z_{12}^3} \, 12 D {\bf T}^{(\frac{1}{2})}(Z_2)
+\frac{\bar{\theta}_{12}}{z_{12}^3} \, 12 \overline{D} {\bf T}^{(\frac{1}{2})}(Z_2)
\nonu \\
& + &
\frac{\theta_{12} \bar{\theta}_{12}}{z_{12}^3} \, 6 \pa {\bf T}^{(\frac{1}{2})}(Z_2)
\nonu \\
&+& \frac{1}{z_{12}^2}  \frac{1}{(2 c-3)} \Bigg[ -36 
{\bf T} {\bf T}^{(\frac{1}{2})}
-12 c  [D, \overline{D}]   {\bf T}^{(\frac{1}{2})} \Bigg](Z_2) 
\nonu \\
&+& \frac{\theta_{12}}{z_{12}^2}  \frac{1}{(2 c-3)} \Bigg[ 
- 12 (c-3)  \pa D  {\bf T}^{(\frac{1}{2})} - 36
{\bf T} D  {\bf T}^{(\frac{1}{2})} 
\Bigg](Z_2) \nonu \\
&+& 
\frac{\bar{\theta}_{12}}{z_{12}^2}  \frac{1}{(2 c-3)} 
\Bigg[ 12 (c-3)
\pa \overline{D}  {\bf T}^{(\frac{1}{2})}
-36  {\bf T} \overline{D}  {\bf T}^{(\frac{1}{2})} \Bigg](Z_2) 
\nonu \\
&+& \frac{\theta_{12} \bar{\theta}_{12}}{z_{12}^2}
\Bigg[ 
\frac{1}{(c+1) (c+6) (2 c-3)}
\Bigg(
-18 (13 c+18)
{\bf T} {\bf T}  {\bf T}^{(\frac{1}{2})} \nonu \\
& + & 3 (2 c^3-25 c^2-123 c-126) 
\pa^2  {\bf T}^{(\frac{1}{2})}
+  144 c (c+1) 
D {\bf T} \overline{D}  {\bf T}^{(\frac{1}{2})}
\nonu \\
& - &   144 c (c+1)
\overline{D} {\bf T} D  {\bf T}^{(\frac{1}{2})}
+  18 (13 c+18) 
 {\bf T}^{(\frac{1}{2})} \overline{D} {\bf T}^{(\frac{1}{2})}  D {\bf T}^{(\frac{1}{2})} 
\nonu \\
& - & 12 (2 c+3) (c+6) 
[D, \overline{D}] {\bf T} {\bf T}^{(\frac{1}{2})}
-  54 (c-2) (c+1) 
{\bf T} [D, \overline{D}]  {\bf T}^{(\frac{1}{2})}
\nonu \\
& - & 18 (13 c+18) 
\pa {\bf T}  {\bf T}^{(\frac{1}{2})}
\Bigg)
+  
\frac{3  }{2 c}  C_{(\frac{3}{2})(\frac{3}{2})}^{(2)}  {\bf T}^{(\frac{1}{2})} 
{\bf W}^{(2')}  +
2   C_{(\frac{3}{2})(3)}^{(\frac{5}{2})} {\bf W}^{(\frac{5}{2}')}
\Bigg](Z_2)
\nonu \\
&+& 
\frac{1}{z_{12}} \Bigg[
\frac{1 }{(c+1) (2 c-3)} \Bigg(
-12 (c+3) 
{\bf T} \pa  {\bf T}^{(\frac{1}{2})} 
-12 c 
\pa {\bf T}  {\bf T}^{(\frac{1}{2})} 
\nonu \\
& - & 4 (c-3) c 
\pa [D, \overline{D}]  {\bf T}^{(\frac{1}{2})} 
+  24 c
\overline{D} {\bf T} D {\bf T}^{(\frac{1}{2})} 
+  24 c 
D {\bf T} \overline{D}  {\bf T}^{(\frac{1}{2})}
\Bigg)
-      C_{(\frac{3}{2})(\frac{3}{2})}^{(2)}  {\bf W}^{(\frac{5}{2})}
 \Bigg](Z_2) \nonu \\
&+ &
\frac{\theta_{12}}{z_{12}} \Bigg[ 
\frac{1 }{(c+1) (c+6) (2 c-3)}
\Bigg(
-12 (c^2+24 c+36)
{\bf T} \pa D   {\bf T}^{(\frac{1}{2})}
\nonu \\
& - & 12 (2 c^2-27 c-54)
\pa {\bf T} D  {\bf T}^{(\frac{1}{2})}
+  36 (5 c+6) 
D {\bf T} \pa  {\bf T}^{(\frac{1}{2})}
\nonu \\
& + & 36 c 
\pa {\bf T}^{(\frac{1}{2})}  {\bf T}^{(\frac{1}{2})} D  {\bf T}^{(\frac{1}{2})} 
-  2 (2 c^3-27 c^2-198 c-216) 
\pa^2 D  {\bf T}^{(\frac{1}{2})}
\nonu \\
& + & 12 (4 c^2+15 c+18) 
[D, \overline{D}] {\bf T} D {\bf T}^{(\frac{1}{2})}
-  108 (c+2)
\overline{D}  {\bf T}^{(\frac{1}{2})} D  {\bf T}^{(\frac{1}{2})} D  
{\bf T}^{(\frac{1}{2})}
\nonu \\
& + & 12 c (5 c+6)
D {\bf T} [D, \overline{D}]  {\bf T}^{(\frac{1}{2})}
+9 (5 c+6)
 {\bf T}^{(\frac{1}{2})} [D, \overline{D}]  {\bf T}^{(\frac{1}{2})}  
D {\bf T}^{(\frac{1}{2})}
\nonu \\
& + & 108 (c+2) 
{\bf T} {\bf T} D  {\bf T}^{(\frac{1}{2})}
+  36 (5 c+6) 
{\bf T} D {\bf T}  {\bf T}^{(\frac{1}{2})}
\Bigg)
\nonu \\
& + & 
C_{(\frac{3}{2})(\frac{3}{2})}^{(2)} \Bigg(
-\frac{3 (c+3)  }{5 (c-3) c} 
 {\bf T}^{(\frac{1}{2})} D {\bf W}^{(2)'}
-  \frac{9 (3 c-1)  }{5 (c-3) c}  
D {\bf T}^{(\frac{1}{2})}  {\bf W}^{(2')}
\nonu \\
& - & \frac{2 (c+3)  }{5 (c-3)}
D {\bf W}^{(\frac{5}{2})} \Bigg)
-\frac{4}{5}   
C_{(\frac{3}{2})(3)}^{(\frac{5}{2})} D  {\bf W}^{(\frac{5}{2}')}
-   C_{(\frac{3}{2})(\frac{5}{2})}^{(3)} \Bigg( {\bf W}^{(3),1}
+ i    {\bf W}^{(3),2} \Bigg)
\Bigg](Z_2) \nonu \\
& + & 
\frac{\bar{\theta}_{12}}{z_{12}} \Bigg[ 
\frac{1 }{(c+1) (c+6) (2 c-3)}
\Bigg(
-12 (c^2+24 c+36) 
{\bf T} \pa \overline{D} {\bf T}^{(\frac{1}{2})}
\nonu \\
& - & 12 (2 c^2+9 c+18)
\pa {\bf T} \overline{D} {\bf T}^{(\frac{1}{2})}
+  36 (5 c+6) 
\overline{D} {\bf T} \pa   {\bf T}^{(\frac{1}{2})}
\nonu \\
& - & 36 (5 c+6) 
\pa \overline{D} {\bf T}   {\bf T}^{(\frac{1}{2})}
-  36 c 
\pa {\bf T}^{(\frac{1}{2})} {\bf T}^{(\frac{1}{2})} \overline{D}
{\bf T}^{(\frac{1}{2})}
\nonu \\
& + & 2 (2 c^3-27 c^2-207 c-270) 
\pa^2 \overline{D} {\bf T}^{(\frac{1}{2})}
\nonu \\
& - & 12 (4 c^2+15 c+18)
[D, \overline{D}] {\bf T} \overline{D}   {\bf T}^{(\frac{1}{2})}
-  12 c (5 c+6) 
\overline{D} {\bf T}  [D, \overline{D}]
 {\bf T}^{(\frac{1}{2})}
\nonu \\
& + & 9 (5 c+6)
 {\bf T}^{(\frac{1}{2})} \overline{D}  {\bf T}^{(\frac{1}{2})} [D, \overline{D}]
 {\bf T}^{(\frac{1}{2})}
\nonu \\
& + & 108 (c+2)
\overline{D}  {\bf T}^{(\frac{1}{2})} \overline{D}  {\bf T}^{(\frac{1}{2})} 
D   {\bf T}^{(\frac{1}{2})} 
\nonu \\
& - & 108 (c+2)
{\bf T} {\bf T} \overline{D} {\bf T}^{(\frac{1}{2})}
-  36 (5 c+6) 
{\bf T} \overline{D} {\bf T} {\bf T}^{(\frac{1}{2})}
\Bigg)
\nonu \\
& + & 
C_{(\frac{3}{2})(\frac{3}{2})}^{(2)} \Bigg(
\frac{3 (c+3)  }{5 (c-3) c}  
{\bf T}^{(\frac{1}{2})} \overline{D} {\bf W}^{(2')}
+  \frac{9 (3 c-1)  }{5 (c-3) c}
 \overline{D}  {\bf T}^{(\frac{1}{2})}
{\bf W}^{(2')}
\nonu \\
& - & \frac{2 (c+3)  }{5 (c-3)}
\overline{D} {\bf W}^{(\frac{5}{2})} \Bigg)
+  
\frac{4}{5}  C_{(\frac{3}{2})(3)}^{(\frac{5}{2})} \overline{D} {\bf W}^{(\frac{5}{2}')} 
-  C_{(\frac{3}{2})(\frac{5}{2})}^{(3)} \Bigg( {\bf W}^{(3),1}
- i     {\bf W}^{(3),2} \Bigg)
\Bigg](Z_2) 
\nonu \\
& + & 
\frac{\theta_{12} \bar{\theta}_{12}}{z_{12}} \Bigg[ 
\frac{1 }{(c+1) (c+6) (2 c-3)} \Bigg(
72 c^2 
D {\bf T} \pa \overline{D} {\bf T}^{(\frac{1}{2})}
\nonu \\
& - & 72 c^2 
\overline{D} {\bf T} \pa D  {\bf T}^{(\frac{1}{2})}
- 18 (c^2-9 c-18) 
{\bf T} \pa [D, \overline{D}] 
 {\bf T}^{(\frac{1}{2})}
\nonu \\
& - & 18 (3 c^2+7 c+6) 
\pa {\bf T} [D, \overline{D}]  {\bf T}^{(\frac{1}{2})}
 +  48 c (2 c+3) 
\pa D {\bf T} \overline{D}  {\bf T}^{(\frac{1}{2})}
\nonu \\
& - & 48 c (2 c+3) 
\pa \overline{D} {\bf T} D  {\bf T}^{(\frac{1}{2})}
+  27 (5 c+6) 
\pa \overline{D}  {\bf T}^{(\frac{1}{2})}  {\bf T}^{(\frac{1}{2})} 
D {\bf T}^{(\frac{1}{2})}
\nonu \\
& + & 27 (5 c+6) 
\pa D {\bf T}^{(\frac{1}{2})} {\bf T}^{(\frac{1}{2})} \overline{D}
{\bf T}^{(\frac{1}{2})}
+  162 (c+2)
\pa {\bf T}^{(\frac{1}{2})} \overline{D} {\bf T}^{(\frac{1}{2})} D 
{\bf T}^{(\frac{1}{2})}
\nonu \\
& - & 12 (2 c+3)(c+6)
[D, \overline{D}] {\bf T} \pa {\bf T}^{(\frac{1}{2})}
-  6 (2 c+3)(c+6)
\pa [D, \overline{D}] {\bf T}  {\bf T}^{(\frac{1}{2})}
\nonu \\
& - & 18 (7 c+6)
{\bf T} {\bf T} \pa 
 {\bf T}^{(\frac{1}{2})}
-  144 (2 c+3) 
\pa {\bf T} {\bf T}  {\bf T}^{(\frac{1}{2})}
\nonu \\
& - & 162 (c+2) 
\pa {\bf T} \pa  {\bf T}^{(\frac{1}{2})}
+  (2 c^3-57 c^2-243 c-270)
\pa^3  {\bf T}^{(\frac{1}{2})} 
\nonu \\
& + & 36 (c+6)
{\bf T} \overline{D} {\bf T} D  {\bf T}^{(\frac{1}{2})}
+  36 (c+6)
{\bf T} D {\bf T} \overline{D} {\bf T}^{(\frac{1}{2})}
\Bigg)
\nonu \\
& + & 
C_{(\frac{3}{2})(\frac{3}{2})}^{(2)} \Bigg(
- \frac{3  }{(c-3)}  
{\bf T} {\bf W}^{(\frac{5}{2})}
+  \frac{3 (2 c-9)  }{10 (c-3) c}
 {\bf T}^{(\frac{1}{2})} \pa {\bf W}^{(2')}
\nonu \\
& + & \frac{3 (7 c-9)  }{10 (c-3) c}
  \pa  {\bf T}^{(\frac{1}{2})}
{\bf W}^{(2')} +\frac{(c+3)  }{10 (c-3)}  
[D, \overline{D}] {\bf W}^{(\frac{5}{2})}
\nonu \\
& - & 
 \frac{3  }{(c-3)}  
\overline{D} {\bf T}^{(\frac{1}{2})} D {\bf W}^{(2')}
-  \frac{3 }{(c-3)}  
D {\bf T}^{(\frac{1}{2})} \overline{D} {\bf W}^{(2')} \Bigg)
+  \frac{6}{5}   C_{(\frac{3}{2})(3)}^{(\frac{5}{2})}
\pa {\bf W}^{(\frac{5}{2}')}
\nonu \\
& + &  C_{(\frac{3}{2})(\frac{5}{2})}^{(3)} \Bigg(
-\frac{1}{2}  
\overline{D} ( {\bf W}^{(3),1} +i  {\bf W}^{(3),2}) 
+  \frac{1}{2}  
D ( {\bf W}^{(3),1} -i {\bf W}^{(3),2})
-  \frac{1}{2}    {\bf W}^{(\frac{7}{2}),3} \Bigg)
\Bigg](Z_2) 
\nonu \\
& + &
\cdots.
\label{w3halfw2}
\eea}
Note that the higher spin multiplets
${\bf W}^{(3),1}$ and ${\bf W}^{(3),2}$ appear together in the above OPE. 
According to (\ref{3u1charge}),
we see that 
$U(1)$ charge in (\ref{w3halfw2}) is preserved.

Furthermore,
the ${\cal N}=2$ OPE between the ${\cal N}=2$ higher spin-$2$
current and itself can be summarized
by  
{\small
\bea
{\bf W}^{(2)}(Z_1) \, {\bf W}^{(2)}(Z_2) & = & 
\frac{1}{z_{12}^4} \, 8c + \frac{\theta_{12} \bar{\theta}_{12}}{z_{12}^4}\, 48
{\bf T}(Z_2) \nonu \\
& + & 
\frac{\theta_{12}}{z_{12}^3} \frac{1 }{(2 c-3)} \Bigg[ 
-36 
 {\bf T}^{(\frac{1}{2})} D  {\bf T}^{(\frac{1}{2})}
-24 (4 c-3) D {\bf T}
\Bigg](Z_2) 
\nonu \\
&+& \frac{\bar{\theta}_{12}}{z_{12}^3}  \frac{1 }{(2 c-3)}
\Bigg[ 
24 (4 c-3) 
\overline{D} {\bf T}
-36  {\bf T}^{(\frac{1}{2})} \overline{D}  {\bf T}^{(\frac{1}{2})}
\Bigg](Z_2)
\nonu \\
&+& \frac{\theta_{12} \bar{\theta}_{12}}{z_{12}^3}
\Bigg[ 
48  \pa {\bf T} -\frac{18}{(2 c-3)}   {\bf T}^{(\frac{1}{2})} [D, \overline{D}]
  {\bf T}^{(\frac{1}{2})}
\Bigg](Z_2)
\nonu \\
&+& 
\frac{1}{z_{12}^2} \Bigg[ 
\frac{1 }{(c+1) (2 c-3)} \Bigg(
24 c 
  \overline{D} {\bf T}^{(\frac{1}{2})} D  {\bf T}^{(\frac{1}{2})}
-24 (4 c+3)
{\bf T} {\bf T}
\nonu \\
& - & 12 (2 c+3)
\pa {\bf T}^{(\frac{1}{2})} {\bf T}^{(\frac{1}{2})}
- 16 c (2 c+3)
[D, \overline{D}] {\bf T}
- 
24 c  \pa {\bf T}  \Bigg) 
\nonu \\
& + &
 2  C_{(\frac{3}{2})(\frac{3}{2})}^{(2)} {\bf W}^{(2')}
\Bigg](Z_2)
\nonu \\
&+& 
\frac{\theta_{12}}{z_{12}^2}
\Bigg[ \frac{1}{(c+1) (2 c-3)} \Bigg(
+6 c
[D, \overline{D}] {\bf T}^{(\frac{1}{2})} D {\bf T}^{(\frac{1}{2})}
-24 (4 c+3) 
{\bf T} D {\bf T}
\nonu \\
& - & 6 (5 c+6)
\pa D  {\bf T}^{(\frac{1}{2})}  {\bf T}^{(\frac{1}{2})}
-12 c 
\pa  {\bf T}^{(\frac{1}{2})} D  {\bf T}^{(\frac{1}{2})}
\nonu \\
& - & 8 (8 c^2-9) 
\pa D {\bf T} \Bigg)
+
 C_{(\frac{3}{2})(\frac{3}{2})}^{(2)} D {\bf W}^{(2')}
\Bigg](Z_2)
\nonu \\
&+& 
\frac{\bar{\theta}_{12}}{z_{12}^2}
\Bigg[ \frac{1 }{(c+1) (2 c-3)} \Bigg(
-6 c 
\overline{D}   {\bf T}^{(\frac{1}{2})} [D,\overline{D}]  {\bf T}^{(\frac{1}{2})}
-24 (4 c+3)
{\bf T} \overline{D} {\bf T}
\nonu \\
& - & 6 (5 c+6) 
\pa \overline{D}   {\bf T}^{(\frac{1}{2})}   {\bf T}^{(\frac{1}{2})}
-12 c 
\pa   {\bf T}^{(\frac{1}{2})} \overline{D}   {\bf T}^{(\frac{1}{2})}
\nonu \\
& + & 8 (8 c^2-3 c-9) 
\pa \overline{D} {\bf T} \Bigg)
+  C_{(\frac{3}{2})(\frac{3}{2})}^{(2)}  \overline{D} {\bf W}^{(2')}
\Bigg](Z_2)
\nonu \\
& + & 
\frac{\theta_{12} \bar{\theta}_{12}}{z_{12}^2}
\Bigg[\frac{1}{(c+1) (c+6) (2 c-3)} \Bigg(
-24 (8 c^2+15 c+18)
{\bf T} [D, \overline{D}] {\bf T}
\nonu \\
& + & 3 (c^2-87 c-126) 
\pa [D, \overline{D}]  {\bf T}^{(\frac{1}{2})}  {\bf T}^{(\frac{1}{2})}
\nonu \\
& - & 9 (5 c^2+5 c-6)
\pa  {\bf T}^{(\frac{1}{2})}  [D, \overline{D}]  {\bf T}^{(\frac{1}{2})} 
+  288 (2 c+3) 
{\bf T} \overline{D}  {\bf T}^{(\frac{1}{2})} D  {\bf T}^{(\frac{1}{2})}
\nonu \\
& - & 18 (31 c+42) 
\overline{D} {\bf T} {\bf T}^{(\frac{1}{2})} D {\bf T}^{(\frac{1}{2})}
-  18 (31 c+42)
D {\bf T} {\bf T}^{(\frac{1}{2})} \overline{D} {\bf T}^{(\frac{1}{2})}
\nonu \\
& - & 144 c (5 c+6) 
\overline{D} {\bf T} D {\bf T}
+18 (4 c+7) (c+6) 
\pa \overline{D} {\bf T}^{(\frac{1}{2})} D {\bf T}^{(\frac{1}{2})}
\nonu \\
& - & 18 (4 c+7)(c+6)
\pa D {\bf T}^{(\frac{1}{2})} \overline{D} {\bf T}^{(\frac{1}{2})}
- 288 (2 c+3) 
{\bf T} {\bf T} {\bf T}
\nonu \\
& - & 54 (5 c+6)
{\bf T} \pa  {\bf T}^{(\frac{1}{2})}  {\bf T}^{(\frac{1}{2})}
-  288 (2 c+3)
\pa {\bf T} {\bf T}
\nonu \\
& + & 6 (8 c^3+17 c^2-147 c-198)
\pa^2 {\bf T}
-36 c (5 c+6) 
\pa [D, \overline{D}] {\bf T} \Bigg)
\nonu \\
& + &
 C_{(\frac{3}{2})(\frac{3}{2})}^{(2)} \Bigg(
 \frac{6 (7 c-9)  }{5 (c-3) c}
{\bf T} {\bf W}^{(2')}
-\frac{3 (2 c-9)  }{5 (c-3) c} 
 {\bf T}^{(\frac{1}{2})}  {\bf W}^{(\frac{5}{2})}
\nonu \\
& - & \frac{3 (c-7)  }{10 (c-3)}  
[D, \overline{D}] {\bf W}^{(2')} \Bigg)
+\frac{12}{5}  C_{(\frac{3}{2})(3)}^{(\frac{5}{2})}  {\bf W}^{(3)}
+  C_{(\frac{3}{2})(\frac{5}{2})}^{(3)} {\bf W}^{(3),3} \Bigg](Z_2)
\nonu \\
&+& 
\frac{1}{z_{12}} \Bigg( \frac{1 }{(c+1) (2 c-3)} \Bigg[ 
12 c 
\pa \overline{D}  {\bf T}^{(\frac{1}{2})} D  {\bf T}^{(\frac{1}{2})}
+12 c  \pa D  {\bf T}^{(\frac{1}{2})} \overline{D}
 {\bf T}^{(\frac{1}{2})} 
\nonu \\
& - & 
24 (4 c+3)  \pa {\bf T} {\bf T}
-6 (2 c+3) \pa^2  {\bf T}^{(\frac{1}{2})}  
{\bf T}^{(\frac{1}{2})}
\nonu \\
& - & 8 c (2 c+3)
\pa [D, \overline{D}] {\bf T} \Bigg]
+  C_{(\frac{3}{2})(\frac{3}{2})}^{(2)} \pa  {\bf W}^{(2')}
\Bigg)(Z_2)
\nonu \\
&+& 
\frac{\theta_{12}}{z_{12}} \Bigg[ 
\frac{1 }{(c+1) (c+6) (2 c-3)} \Bigg(
72 (3 c^2+7 c+6)
[D,\overline{D}] {\bf T} D {\bf T}
\nonu \\
& + & 18 (2 c^2+9 c+6)
\pa D  {\bf T}^{(\frac{1}{2})}   [D, \overline{D}]  
{\bf T}^{(\frac{1}{2})}
\nonu \\
& - & 18 (c^2+15 c+30)
\pa [D,\overline{D}] {\bf T}^{(\frac{1}{2})}
D {\bf T}^{(\frac{1}{2})}
-24 (2 c^2+45 c+54)
\pa D {\bf T} {\bf T}
\nonu \\
& - & 72 (c^2-12 c-24) 
\pa {\bf T} D {\bf T}
-3 (4 c^2+3 c+18)
\pa^2 D  {\bf T}^{(\frac{1}{2})}  {\bf T}^{(\frac{1}{2})}
\nonu \\
& - & 18 (c^2-12)
\pa D {\bf T}^{(\frac{1}{2})} \pa {\bf T}^{(\frac{1}{2})}
-  216 (c+2) 
\overline{D} {\bf T} D  {\bf T}^{(\frac{1}{2})}
D  {\bf T}^{(\frac{1}{2})}
\nonu \\
& - & 432 (c+2) 
D {\bf T} \overline{D}  {\bf T}^{(\frac{1}{2})} D  {\bf T}^{(\frac{1}{2})}
+  18 (5 c+6) 
[D, \overline{D}] {\bf T}  {\bf T}^{(\frac{1}{2})} D  {\bf T}^{(\frac{1}{2})}
\nonu \\
& + & 
18 (5 c+6) 
{\bf T } \pa D {\bf T}^{(\frac{1}{2})}
{\bf T}^{(\frac{1}{2})}
-   18 (5 c+6)
{\bf T} [D, \overline{D}]  {\bf T}^{(\frac{1}{2})}  D
 {\bf T}^{(\frac{1}{2})}  
\nonu \\
& - & 72 c 
\pa {\bf T}  {\bf T}^{(\frac{1}{2})}  D  {\bf T}^{(\frac{1}{2})} 
+18 (5 c+6) 
D {\bf T}  {\bf T}^{(\frac{1}{2})}  [D, \overline{D}]  {\bf T}^{(\frac{1}{2})}
\nonu \\
& + & 54 (3 c+2) 
D {\bf T} \pa  {\bf T}^{(\frac{1}{2})}  {\bf T}^{(\frac{1}{2})}
-36 (c+6) 
{\bf T} \pa {\bf T}^{(\frac{1}{2})} D {\bf T}^{(\frac{1}{2})} 
\nonu \\
& + & 288 (2 c+3)
{\bf T} {\bf T} D {\bf T}
+9 (c+6)
\pa^2  {\bf T}^{(\frac{1}{2})}  D {\bf T}^{(\frac{1}{2})}
\nonu \\
& - & 12 c (2 c^2-3 c-6)
\pa^2 D {\bf T} \Bigg)
\nonu \\
& + & 
 C_{(\frac{3}{2})(\frac{3}{2})}^{(2)} \Bigg(
\frac{6 (c+3)  }{5 (c-3) c}  
{\bf T} D {\bf W}^{(2')}
-\frac{18 (3 c-1)  }{5 (c-3) c}   
D {\bf T} {\bf W}^{(2')}
\nonu \\
& + & \frac{3 (c+3)  }{5 (c-3) c}  
{\bf T}^{(\frac{1}{2})} D {\bf W}^{(\frac{5}{2})} 
+\frac{9 (3 c-1)  }{5 (c-3) c}   
D {\bf T}^{(\frac{1}{2})}  {\bf W}^{(\frac{5}{2})} 
\nonu \\
&
+ & \frac{3 (c-7)  }{5 (c-3)}   
\pa D {\bf W}^{(2')} \Bigg)
\nonu \\
& - & \frac{4}{5}  C_{(\frac{3}{2})(3)}^{(\frac{5}{2})}  
D {\bf W}^{(3)}+
  C_{(\frac{3}{2})(\frac{5}{2})}^{(3)} ( {\bf W}^{(\frac{7}{2}),1} 
+i   {\bf W}^{(\frac{7}{2}),2} ) 
\Bigg](Z_2)
\nonu \\
&+& 
\frac{\bar{\theta}_{12}}{z_{12}} \Bigg[ 
\frac{1 }{(c+1) (c+6) (2 c-3)} \Bigg(
-72 (3 c^2+7 c+6)
\overline{D} {\bf T} [D, \overline{D}] {\bf T}
\nonu \\
& + & 18 (c^2+15 c+30) 
\pa  [D, \overline{D}]  
 {\bf T}^{(\frac{1}{2})} \overline{D} {\bf T}^{(\frac{1}{2})}
\nonu \\
& - & 18 (2 c^2+9 c+6)
\pa \overline{D}  {\bf T}^{(\frac{1}{2})}   [D, \overline{D}]  
 {\bf T}^{(\frac{1}{2})} 
-  72 c^2 
\pa {\bf T} \overline{D} {\bf T}
\nonu \\
& - & 
48 (c^2+30 c+36)
\pa \overline{D} {\bf T} {\bf T}
-3 (4 c^2+3 c+18) 
\pa^2 \overline{D} {\bf T}^{(\frac{1}{2})}  {\bf T}^{(\frac{1}{2})} 
\nonu \\
& - & 
18 (c^2-12) 
\pa \overline{D}  {\bf T}^{(\frac{1}{2})} \pa  {\bf T}^{(\frac{1}{2})}
+18 (5 c+6)
[D, \overline{D}]  {\bf T}  {\bf T}^{(\frac{1}{2})} \overline{D}  
{\bf T}^{(\frac{1}{2})}
\nonu \\
& + & 18 (5 c+6) 
\overline{D} {\bf T}  {\bf T}^{(\frac{1}{2})}  [D, \overline{D}]  
{\bf T}^{(\frac{1}{2})} 
-  18 (5 c+6)
{\bf T} \pa \overline{D}  {\bf T}^{(\frac{1}{2})}  {\bf T}^{(\frac{1}{2})} 
\nonu \\
& + & 432 (c+2)
\overline{D} {\bf T} \overline{D}  {\bf T}^{(\frac{1}{2})}  D 
 {\bf T}^{(\frac{1}{2})} 
+  216 (c+2)
D {\bf T} \overline{D} {\bf T}^{(\frac{1}{2})} \overline{D} {\bf T}^{(\frac{1}{2})} 
\nonu \\
& + & 72 c 
\pa {\bf T}  {\bf T}^{(\frac{1}{2})} \overline{D} {\bf T}^{(\frac{1}{2})}
+36(c+6)
{\bf T} \pa   {\bf T}^{(\frac{1}{2})}  \overline{D}   {\bf T}^{(\frac{1}{2})} 
\nonu \\
& - & 18 (5 c+6) 
{\bf T} \overline{D} {\bf T}^{(\frac{1}{2})} [D, \overline{D}]
{\bf T}^{(\frac{1}{2})}
-  54 (3 c+2)
\overline{D} {\bf T} \pa {\bf T}^{(\frac{1}{2})} {\bf T}^{(\frac{1}{2})}
\nonu \\
& - & 288 (2 c+3)
{\bf T} {\bf T} \overline{D} {\bf T}
+9(c+6)
\pa^2 {\bf T}^{(\frac{1}{2})} \overline{D} {\bf T}^{(\frac{1}{2})}
\nonu \\
& + & 
24 (c^3+3 c^2-3 c-18) 
\pa^2 \overline{D} {\bf T} \Bigg)
\nonu \\
& + & 
 C_{(\frac{3}{2})(\frac{3}{2})}^{(2)} \Bigg(
\frac{18 (3 c-1)  }{5 (c-3) c}  
\overline{D} {\bf T} {\bf W}^{(2')}
-\frac{6 (c+3)  }{5 (c-3) c}  
{\bf T} \overline{D} {\bf W}^{(2')}
\nonu \\
& - & \frac{3 (c+3)  }{5 (c-3) c}  
{\bf T}^{(\frac{1}{2})} \overline{D} {\bf W}^{(\frac{5}{2})}
-\frac{9 (3 c-1)  }{5 (c-3) c}  
\overline{D} {\bf T}^{(\frac{1}{2})} {\bf W}^{(\frac{5}{2})}
\nonu \\
&
+ & \frac{3 (c-7)  }{5 (c-3)}
 \pa \overline{D} {\bf W}^{(2')} \Bigg)
\nonu \\
& + & \frac{4}{5}  C_{(\frac{3}{2})(3)}^{(\frac{5}{2})} \overline{D} {\bf W}^{(3)}
+
  C_{(\frac{3}{2})(\frac{5}{2})}^{(3)} ( {\bf W}^{(\frac{7}{2}),1} 
-i     {\bf W}^{(\frac{7}{2}),2}) 
\Bigg](Z_2)
\nonu \\
&+& 
\frac{\theta_{12} \bar{\theta}_{12}}{z_{12}} \Bigg[ 
 \frac{1}{(c+1) (c+6) (2 c-3)} \Bigg( 72 (5 c+6)
{\bf T} \pa \overline{D}  {\bf T}^{(\frac{1}{2})} D  {\bf T}^{(\frac{1}{2})}
\nonu \\
& + & 72 (5 c+6) 
{\bf T} \pa D  {\bf T}^{(\frac{1}{2})} \overline{D}  {\bf T}^{(\frac{1}{2})}
\nonu \\
&
+ & 12 (4 c+9) (c+6)
\pa^2 \overline{D}  {\bf T}^{(\frac{1}{2})} D  {\bf T}^{(\frac{1}{2})}
+432 (c+2)
\pa {\bf T} \overline{D} {\bf T}^{(\frac{1}{2})} D {\bf T}^{(\frac{1}{2})}
\nonu \\
& + & 12 (c^2-27 c-54) 
\pa [D, \overline{D}]  {\bf T}^{(\frac{1}{2})} 
\pa  {\bf T}^{(\frac{1}{2})} 
\nonu \\
& + & 
96 c (5 c+6)
\pa D {\bf T} \overline{D} {\bf T}
+  8 (2 c^3+c^2-57 c-90)
\pa^3 {\bf T}
\nonu \\
&
- & 12 (4 c+9)(c+6) 
\pa^2 D  {\bf T}^{(\frac{1}{2})} \overline{D}  {\bf T}^{(\frac{1}{2})}
\nonu \\
& - & 36 (5 c+6)
{\bf T} \pa^2  {\bf T}^{(\frac{1}{2})}
 {\bf T}^{(\frac{1}{2})}
-72 (5 c+6)
\pa \overline{D} {\bf T} {\bf T}^{(\frac{1}{2})} D {\bf T}^{(\frac{1}{2})}
\nonu \\
& - & 432 (c+2) 
\pa {\bf T} \pa {\bf T}
-96 c (5 c+6)
\pa \overline{D} {\bf T} D {\bf T}
\nonu \\
& - & 72 (5 c+6)
\overline{D} {\bf T} \pa D {\bf T}^{(\frac{1}{2})} {\bf T}^{(\frac{1}{2})}
-  96 c^2 
\pa [D, \overline{D}] {\bf T} {\bf T}
\nonu \\
& - & 36 (5 c+6)
\pa {\bf T} \pa  {\bf T}^{(\frac{1}{2})}  {\bf T}^{(\frac{1}{2})}
-  72 (5 c+6)
D {\bf T} \pa \overline{D}  {\bf T}^{(\frac{1}{2})}   {\bf T}^{(\frac{1}{2})} 
\nonu \\
& - & 432 (c+2)
\overline{D} {\bf T} \pa  {\bf T}^{(\frac{1}{2})} D  {\bf T}^{(\frac{1}{2})}
-  72 (5 c+6)
\pa D {\bf T}   {\bf T}^{(\frac{1}{2})} \overline{D}   {\bf T}^{(\frac{1}{2})} 
\nonu \\
& - & 48 (4 c^2+15 c+18)
\pa {\bf T} [D, \overline{D}] {\bf T}
-576 (2 c+3) 
\pa {\bf T} {\bf T} {\bf T}
\nonu \\
& - & 36 (5 c+6)
\pa^2 [D, \overline{D}]  {\bf T}^{(\frac{1}{2})}  {\bf T}^{(\frac{1}{2})}
-  432 (c+2)
D {\bf T} \pa   {\bf T}^{(\frac{1}{2})} \overline{D}  {\bf T}^{(\frac{1}{2})}
\nonu \\
&- & 
36 c (c+2)
\pa^2  {\bf T}^{(\frac{1}{2})}  [D, \overline{D}]  {\bf T}^{(\frac{1}{2})} 
\Bigg)
\nonu \\
& + & 
 C_{(\frac{3}{2})(\frac{3}{2})}^{(2)}  \Bigg(
\frac{18 (c-2)  }{5 (c-3) c} 
{\bf T} \pa {\bf W}^{(2')}
+\frac{6  }{(c-3)}   
D {\bf T} \overline{D} {\bf W}^{(2')}
\nonu \\
& + & \frac{3 }{(c-3)}  
D  {\bf T}^{(\frac{1}{2})} \overline{D}  {\bf W}^{(\frac{5}{2})}
+\frac{6  }{(c-3)}   
\overline{D} {\bf T} D {\bf W}^{(2')}
\nonu \\
&+& \frac{3  }{(c-3)}   
\overline{D} {\bf T}^{(\frac{1}{2})} D {\bf W}^{(\frac{5}{2})}
+  \frac{12 (4 c-3)  }{5 (c-3) c}   
\pa {\bf T} {\bf W}^{(2')}
\nonu \\
& - &
\frac{3 (c-12)  }{10 (c-3) c}   
 {\bf T}^{(\frac{1}{2})} \pa  {\bf W}^{(\frac{5}{2})}
-  \frac{3 (11 c-12)  }{10 (c-3) c}   
\pa  {\bf T}^{(\frac{1}{2})}  {\bf W}^{(\frac{5}{2})}
\nonu \\
& - & 
\frac{(c-12)  }{5 (c-3)}   
\pa [D, \overline{D}] {\bf W}^{(2')} \Bigg)
+  \frac{8}{5} 
 C_{(\frac{3}{2})(3)}^{(\frac{5}{2})} \pa {\bf W}^{(3)}
\nonu \\
& + &
 C_{(\frac{3}{2})(\frac{5}{2})}^{(3)}  \Bigg(
\frac{1}{2} 
 \pa {\bf W}^{(3),3}
+ 
\frac{1}{2}   
\overline{D}  ( {\bf W}^{(\frac{7}{2}),1} 
+ i   {\bf W}^{(\frac{7}{2}),2} )  
\nonu \\
& - &  \frac{1}{2} 
  D  ( {\bf W}^{(\frac{7}{2}),1} 
- i 
  {\bf W}^{(\frac{7}{2}),2} ) \Bigg)
\Bigg](Z_2) +\cdots.
\label{w2w2}
\eea}
Note that the higher spin multiplets
${\bf W}^{(\frac{7}{2}),1}(Z_2)$ and ${\bf W}^{(\frac{7}{2}),2}(Z_2)$ 
appear together in the above OPE. 
According to (\ref{7halfu1charge}),
we see that 
$U(1)$ charge in (\ref{w2w2}) is preserved.

\section{ The OPEs between the lowest eight higher spin currents 
in the component approach corresponding to the section $7$ }

From the ${\cal N}=2$ OPE results in section $6$,
we obtain its component results. 

\subsection{ The complete $36$ OPEs (between the lowest eight higher
spin currents) in the component approach  }

We present the following complete OPEs by 
reading off the appropriate OPEs from the ${\cal N}=2$ superspace
description. The first four types of OPEs 
are given by
{\small
\bea
\psi^{(\frac{3}{2})}(z) \, \psi^{(\frac{3}{2})}(w) & = & 
\frac{1}{(z-w)^3} \, \frac{2c}{3} \nonu \\
& + & \frac{1}{(z-w)} \, \Bigg[ \frac{1 }{(c+1) (2 c-3)} \Bigg(
-6 c  J^i J^i - 
6 (c+3) \pa \Psi \Psi +
4 c (c+3) T  \Bigg) \nonu \\
& + &  C_{(\frac{3}{2}) (\frac{3}{2})}^{(2)} \psi^{(2)} 
\Bigg](w) + 
\cdots,
\nonu \\
\psi^{(\frac{3}{2})}(z) \, \phi^{(2),i}(w) & = &
\frac{1}{(z-w)^2} \, \frac{1}{(2c-3)} \Bigg[ 6 c G^i 
-18  \Psi J^i
\Bigg](w) 
\nonu \\
&+& \frac{1}{(z-w)} \Bigg[ \frac{1}{3} \pa (\mbox{pole two})
+ \frac{1}{(c+1)(2c-3)} \Bigg( 6 i c  ( \epsilon^{ijk} J^j G^k
-\frac{2}{3} i \pa G^i)
\nonu \\
& - & 18 ( \pa \Psi  J^i -\frac{1}{3} \pa (\Psi J^i)) \Bigg)
- 
\frac{1}{2}  C_{(\frac{3}{2}) (\frac{3}{2})}^{(2)} \phi^{(\frac{5}{2}),i}
\Bigg](w) + \cdots,
\nonu \\
\psi^{(\frac{3}{2})}(z) \, \psi^{(\frac{5}{2}),i}(w) & = &
\frac{1}{(z-w)^3} \,  6 J^i(w)
-\frac{1}{(z-w)^2} \,  \frac{18 }{(2 c-3)} \Psi G^i(w)
\nonu \\
& + & \frac{1}{(z-w)} \Bigg[ 
 \frac{1}{4} \pa (\mbox{pole two}) \nonu \\
& + & 
 \frac{1}{(c+1) (c+6) (2 c-3)} \Bigg(
18 i (5 c+6) (  \epsilon^{ijk} \Psi
J^j G^k -\frac{2}{3} i \Psi \pa G^i) 
\nonu \\
& - & 18 i c (c+2)
(\epsilon^{ijk} G^j G^k -\frac{2}{3} i \pa^2 J^i)
+  36 (c^2+3 c+6)
(T J^i -\frac{1}{2} \pa^2 J^i)
\nonu \\  
& - & 54 (c+2) J^i J^j J^j
+  72 i c
(\epsilon^{ijk} \pa J^j J^k -\frac{1}{3} i \pa^2 J^i)
\nonu \\
& + &  
6 (c^2-17 c-42)
(\pa \Psi G^i -\frac{1}{4} \pa (\Psi G^i)) 
-72 c \pa \Psi \Psi J^i \Bigg)
\nonu \\
&+ &  C_{(\frac{3}{2}) (\frac{3}{2})}^{(2)} \Bigg(
\frac{3 (c+3)  }{5 (c-3) c} 
\Psi \phi^{(\frac{5}{2}), i} 
+ \frac{9 (3 c-1)  }{5 (c-3) c} 
J^i \psi^{(2)}
+ \frac{3 (c-7) }{10 (c-3)}
  \psi^{(3), i} \Bigg)
 \nonu \\
&+& \frac{2}{5}   C_{(\frac{3}{2}) (3)}^{(\frac{5}{2})}  \phi^{(3),i} +  
 C_{(\frac{3}{2}) (\frac{5}{2})}^{(3)} \psi^{(3),\alpha=i}
\Bigg](w) +\cdots,
\nonu \\
\psi^{(\frac{3}{2})}(z) \, \phi^{(3)}(w) & = &
\frac{1}{(z-w)^4} \, 3 \Psi(w) -
\frac{1}{(z-w)^3} \, 3 \pa \Psi(w) 
\nonu \\
& + & \frac{1}{(z-w)^2} \Bigg[ \frac{1}{(c+1) (2 c-3)}
\Bigg(
6 (c+3) (T \Psi - \frac{3}{4} \pa^2 \Psi)
+ 
36 \frac{c(c+1)}{(c+6)}
J^i G^i
\nonu \\
& - & \frac{9 (13 c+18)}{(c+6)} \Psi J^i J^i \Bigg)
+\frac{3  }{2 c} C_{(\frac{3}{2}) (\frac{3}{2})}^{(2)} \Psi \psi^{(2)}
+  C_{(\frac{3}{2}) (3)}^{(\frac{5}{2})} \psi^{(\frac{5}{2})}
\Bigg](w) \nonu \\
& + & \frac{1}{(z-w)}
\Bigg[ 
\frac{1}{5} \pa (\mbox{pole two})
+  \frac{1}{(c+1) (2 c-3)} \Bigg(
-24 (c+3) (\pa \Psi T -\frac{1}{5} \pa (\Psi T))
\nonu \\
& - & 12 c (\pa J^i G^i -\frac{2}{5} \pa (J^i G^i))
+ 
18
(\pa \Psi J^i J^i -\frac{1}{5}\pa (\Psi J^i J^i)) \Bigg)
\nonu \\
& + & 
C_{(\frac{3}{2}) (\frac{3}{2})}^{(2)} \Bigg(-
\frac{9 }{2 (c-3)} 
(\pa \Psi \psi^{(2)} -\frac{1}{5} \pa (\Psi \psi^{(2)}))
-\frac{3 }{2 (c-3)}  J^i \phi^{(\frac{5}{2}),i}
\nonu \\
&- & 
\frac{(c-12) }{5 (c-3)}   \phi^{(\frac{7}{2})} \Bigg)
-\frac{1 }{4}  C_{(\frac{3}{2}) (\frac{5}{2})}^{(3)} \phi^{(\frac{7}{2}),i,\alpha=i} 
\Bigg](w) 
+ \cdots.
\label{Fresult1}
\eea
}
These OPEs appeared in the section $4$.

The next three types of OPEs can be written as
{\small
\bea
\phi^{(2), i}(z) \, \phi^{(2),j}(w)  & = &
\frac{1}{(z-w)^4} \, 2 \delta^{ij} c 
+ \frac{1}{(z-w)^3} \, 6 i \epsilon^{ijk} J^k(w)
\nonu \\
& + &  \frac{1}{(z-w)^2} \Bigg[ 
\frac{8 c (2 c+3) }{(c+1) (2 c-3)} \delta^{ij} T 
-\frac{18}{ (2 c-3)} J^i J^j 
-\frac{6 c }{(c+1) (2 c-3)} \delta^{ij} J^k J^k
\nonu \\
& - & \frac{12 (2 c+3) }{(c+1) (2 c-3)} \delta^{ij} \pa \Psi \Psi
 + \frac{6 i c}{(2c-3)} \epsilon^{ijk} \pa J^k + \delta^{ij} 
 C_{(\frac{3}{2}) (\frac{3}{2})}^{(2)}
\psi^{(2)}
\Bigg](w)
\nonu \\
&+& \frac{1}{(z-w)} \Bigg[ 
\frac{4 c (2 c+3)}{(c+1) (2 c-3)}  
\delta^{ij} \pa T 
-  \frac{6 c }{(c+1) (2 c-3)} \delta^{ij} \pa J^k J^k
\nonu \\
& - & \frac{6 (2 c+3) }{(c+1) (2 c-3)} \delta^{ij} \pa^2 \Psi \Psi
\nonu \\
&+& 
\frac{1 }{(c+1) (c+6) (2 c-3)} \Bigg(
3 c (5 c+6)  ( G^i G^j -G^j G^i) 
-54 i (c+2)  \epsilon^{ijk} J^l J^l J^k
\nonu \\
& - & 18 (5 c+6)  \Psi (J^i G^j-J^j G^i)
+18 i (5 c+6) \epsilon^{ijk} \Psi \pa G^k 
\nonu \\
&+& 12 i (4 c^2+15 c+18)
\epsilon^{ijk} T J^k
-18 i (5 c+6)  \epsilon^{ijk} \pa \Psi G^k
\nonu \\
& - & 
72 i c  \epsilon^{ijk} \pa \Psi \Psi J^k 
-12 (c^2-9 c-18) J^i \pa J^j
\nonu \\
& - & 6 (c^2+39 c+54)  \pa J^i J^j
+ 2 i (c^3-9 c^2-45 c-54) 
\epsilon^{ijk} \pa^2 J^k
\Bigg)
\nonu \\
&  + & C_{(\frac{3}{2}) (\frac{3}{2})}^{(2)} \Bigg(
\frac{3 i (c+3)  }{5 (c-3) c} 
\epsilon^{ijk} \Psi \phi^{(\frac{5}{2}),k}
+\frac{9 i (3 c-1)  }{5 (c-3) c} 
 \epsilon^{ijk} J^k \psi^{(2)} +  
 \frac{1}{2} \delta^{ij} 
\pa \psi^{(2)}
\nonu \\
&  - & \frac{i (c+3)  }{5 (c-3)}  
\epsilon^{ijk} \psi^{(3),k} \Bigg)
+\frac{2}{5} i  C_{(\frac{3}{2}) (3)}^{(\frac{5}{2})} \epsilon^{ijk} \phi^{(3),k}
+ i  C_{(\frac{3}{2}) (\frac{5}{2})}^{(3)} \epsilon^{ijk} \psi^{(3),\alpha=k}
\Bigg](w) + \cdots,
\nonu \\
\phi^{(2),i}(z) \, \psi^{(\frac{5}{2}),j} & = &
\frac{1}{(z-w)^4} \, 6 \delta^{ij} \Psi(w) +
\frac{1}{(z-w)^3} \, \frac{1}{(2c-3)} \Bigg[ 6 i (4 c-3)  \epsilon^{ijk}
G^k  - 36 i  \epsilon^{ijk} \Psi J^k \Bigg](w)
\nonu \\
&+ & \frac{1}{(z-w)^2} \Bigg[ 
 \frac{1}{(c+1) (2 c-3)}
\Bigg(
\frac{72 c(c+1) }{(c+6)} \delta^{ij} J^k G^k
-18(c+1) J^i G^j
\nonu \\
& - & \frac{18 (13 c+18) }{ (c+6) } \delta^{ij} \Psi J^k J^k
+  24 (2 c+3)  \delta^{ij} T \Psi
\nonu \\
& - & 18 (2 c+3) \delta^{ij} \pa^2 \Psi
- 6 c  (J^i G^j-J^j G^i)
\nonu \\
& - & 6 i c  \epsilon^{ijk} \Psi \pa J^k
- 12 i (2 c+3) 
\epsilon^{ijk} \pa \Psi  J^k
+  4 i c (2 c+3) \epsilon^{ijk} \pa G^k \Bigg)
\nonu \\
&+ & \frac{3}{c} C_{(\frac{3}{2}) (\frac{3}{2})}^{(2)} \delta^{ij}
\Psi \psi^{(2)}
+ 2  C_{(\frac{3}{2}) (3)}^{(\frac{5}{2})} \delta^{ij} \psi^{(\frac{5}{2})}
-\frac{i}{2}  C_{(\frac{3}{2}) (\frac{3}{2})}^{(2)} \epsilon^{ijk} 
\phi^{(\frac{5}{2}),k}
\Bigg](w) \nonu \\
& + & \frac{1}{(z-w)} \Bigg[ 
\frac{1 }{(c+1) (c+6) (2 c-3)}
\Bigg(
-18 i(c+6) 
 J^i J^{i+1} G^{i+2} 
\nonu \\
& + &  18 i(c+6) 
 J^i J^{i+2} G^{i+1}
+ 12 c(c+6)  J^i \pa G^i
+ 24 c^2 \delta^{ij} \sum_{k} J^k \pa G^k
\nonu \\
& + &  18(c+6) \Psi \pa J^i J^i
-  18(11c+18) \delta^{ij} \sum_{k} \Psi \pa J^k J^k
\nonu \\
&- & 36(c+1)(c+6) \pa J^i G^i
+ 36c(c+2) \delta^{ij} \sum_{k} \pa J^k G^k
 -36(c+6) \pa \Psi  J^i J^i
\nonu \\
& - & 72c \delta^{ij} \sum_{k} \pa \Psi J^k J^k 
- 4 (c+6)(2c+3) \delta^{ij} \pa^3 \Psi
+   12 (2 c+3)(c+6)  \delta^{ij} \pa T \Psi
\Bigg)_{j=i}
\nonu \\
&+& \frac{1 }{(c+1) (c+6) (2 c-3)}
\Bigg(  
-72 i (2c+3) J^i J^i G^{i+2}
+ 18 i (5c+6) J^i J^{i+2} G^i
\nonu \\
&-& 6 (c-18) (2 c+3) J^i \pa G^{i+1}
-162 i (c+2) J^{i+1} J^{i+1} G^{i+2} 
\nonu \\
&+& 108 i (c+2) J^{i+1} J^{i+2} G^{i+1}
+ 6 (4 c^2-21 c-54) J^{i+1} \pa G^i
\nonu \\
& - &
54 i (c+2) J^{i+2} J^{i+2} G^{i+2}
+
18 (5 c+6) \Psi G^i G^{i+1}
\nonu \\
&+& 
18 (5c+6) \Psi J^i \pa J^{i+1}
-72 c \Psi \pa J^i J^{i+1}
+36 i (3 c^2+7 c+6) 
 T G^{i+2}
\nonu \\
&  - & 36 i (5 c+6) 
 T \Psi J^{i+2} 
-6 (c^2+36 c+36) \pa J^i G^{i+1} 
- 6 c (5c+6) \pa J^{i+1} G^i
\nonu \\
&
- & 
36(c+6) \pa \Psi J^i J^{i+1}
-
54i (3c+2)  \pa \Psi \Psi G^{i+2}
-6i (c-18)(2c+3) \pa \Psi \pa J^{i+2}
\nonu \\
& + & i (2 c^3-21 c^2-117 c-162)  \pa^2 G^{i+2}
-  3 i (c+6)(2c+3)  \pa^2 \Psi J^{i+2} 
\Bigg)_{j=i+1}
 \nonu \\
&+& 
 \frac{1 }{(c+1) (c+6) (2 c-3)}
\Bigg(
162i (c+2) J^j J^j G^{j+2}-
108 i (c+2) J^j J^{j+2} G^j
\nonu \\
& 
+ & 6 (4 c^2-21 c-54) J^j \pa G^{j+1}
\nonu \\
&+& 72 i (2c+3) J^{j+1} J^{j+1} G^{j+2}
-18 i (5c+6) J^{j+1} J^{j+2} G^{j+1}
\nonu \\
&-& 6 (c-18) (2 c+3) J^{j+1} \pa G^{j}
+ 54 i (c+2) J^{j+2} J^{j+2} G^{j+2}
\nonu \\
&-& 18(5c+6) \Psi G^j G^{j+1}
-72 c \Psi J^j \pa J^{j+1} + 
18 (5c+6) \Psi \pa J^j J^{j+1}
\nonu \\
& - & 9 i (c+6) \Psi \pa^2 J^{j+2}
-36 i (3 c^2+7 c+6) T G^{j+2}
+ 36 i (5c+6) T \Psi J^{j+2}
\nonu \\
&-& 6 c (5c+6) \pa J^j G^{j+1}
- 6 (c^2+36 c+36) \pa J^{j+1} G^j
-36 (c+6) \pa \Psi J^j J^{j+1}
\nonu \\
&+& 54 i (3c+2) \pa \Psi \Psi G^{j+2}
+ 6 i (2 c^2-27 c-18) \pa \Psi \pa J^{j+2}
+ 3 i (c+6) (2 c+3) \pa^2 \Psi J^{j+2}
\nonu \\
& - & i (2 c^3-21 c^2-117 c-162 ) \pa^2 G^{j+2} 
\Bigg)_{i=j+1}
\nonu \\
& + & 
 C_{(\frac{3}{2}) (\frac{3}{2})}^{(2)} \Bigg(-
\frac{3 }{(c-3)} 
\delta^{ij} J^k \phi^{(\frac{5}{2}),k}
+ \frac{6 }{(c-3)}  
 J^i \phi^{(\frac{5}{2}),j} 
+   \frac{(c+3)  }{5 (c-3)}    
  \delta^{ij} \phi^{(\frac{7}{2})}
\nonu \\
& - & \frac{9 (3c-1)  }{5 (c-3) c}   
(J^i \phi^{(\frac{5}{2}),j}-J^j \phi^{(\frac{5}{2}),i})
-\frac{3 i (c+3)  }{5 (c-3) c} 
  \epsilon^{ijk} \Psi \psi^{(3),k}
 +  \frac{9 i (3 c-1)  }{5 (c-3) c}  
\epsilon^{ijk} G^k \psi^{(2)} 
\nonu \\
&+& \frac{9 (c-2) }{5 (c-3) c} \delta^{ij} \Psi \pa \psi^{(2)}
-\frac{6 (c+3) }{5 (c-3) c} \delta^{ij} \pa \Psi \psi^{(2)}
-\frac{i(c+3)}{5(c-3)} \epsilon^{ijk} \pa \psi^{(\frac{5}{2}),k}
\Bigg)
\nonu \\
& + &  C_{(\frac{3}{2}) (\frac{5}{2})}^{(3)}
\Bigg(-
\frac{1}{2}  \delta^{ij} \phi^{(\frac{7}{2}),k,
\alpha=k}+     \phi^{(\frac{7}{2}),i,
\alpha=j} \Bigg) 
\nonu \\
&+ &  C_{(\frac{3}{2}) (3)}^{(\frac{5}{2})} \Bigg(
\frac{4}{5}   \delta^{ij} 
\pa \psi^{(\frac{5}{2})} 
+ \frac{2}{5} i    \epsilon^{ijk}
\psi^{(\frac{7}{2}),k} \Bigg)
\Bigg](w) +  \cdots,
\nonu \\
\phi^{(2),i}(z) \, \phi^{(3)}(w) &= & 
\frac{1}{(z-w)^4} \, 12 J^i(w) 
-\frac{1}{(z-w)^3} \,  \frac{18 }{(2 c-3)} \, \Psi G^i(w)
+\frac{1}{(z-w)^2} \, \Bigg[ 
\nonu \\
&  
+ & \frac{1 }{(c+1) (c+6) (2 c-3)}
\Bigg(-
9 i c (5 c+6)  
\epsilon^{ijk} G^j G^k 
-72 (2 c+3) 
J^i J^j J^j
\nonu \\
& + & 36 i c^2  \epsilon^{ijk}  J^j \pa J^k
+9 i (31 c+42) \epsilon^{ijk} \Psi J^j G^k 
\nonu \\
& - & 
3 (5 c^2-57 c-90)  \Psi \pa G^i
+12 (8 c^2+15 c+18) 
T J^i
\nonu \\
& + & 
27 (c^2-3 c-6)  \pa \Psi G^i
- 54 (5 c+6)  \pa \Psi \Psi J^i
\nonu \\
& - & 
18 (3 c^2+7 c+6)  \pa^2 J^i \Bigg)
\nonu \\
&+&  C_{(\frac{3}{2}) (\frac{3}{2})}^{(2)} \Bigg(
\frac{3 (c-7)  }{20 (c-3)} 
\psi^{(3),i}
+ \frac{3 (7 c-9)  }{5 (c-3) c}  
J^i \psi^{(2)} -\frac{3 (2 c-9)  }{5 (c-3) c} 
 \Psi \phi^{(\frac{5}{2}),i} \Bigg)
\nonu \\
& + & 
\frac{6}{5}  C_{(\frac{3}{2}) (3)}^{(\frac{5}{2})} \phi^{(3),i}
+\frac{1}{2}  C_{(\frac{3}{2}) (\frac{5}{2})}^{(3)} \psi^{(3),\alpha=i}
\Bigg](w)
\nonu \\
&+& \frac{1}{(z-w)} \Bigg[ 
\frac{1}{(c+1) (c+6) (2 c-3)} 
\Bigg(
-6 i c (5 c+6)  \epsilon^{ijk} G^j \pa G^k
\nonu \\
& + & 12 i c^2  \epsilon^{ijk} J^j \pa^2 J^k
+ 9 i (11 c+18)  \epsilon^{ijk} \Psi J^j \pa G^k 
\nonu \\
&+& 9 i (11 c+18) \epsilon^{ijk} \Psi \pa J^j G^k
-6 (c^2-9 c-18)  \Psi \pa^2 G^i
\nonu \\
& - & 36 (5 c+6) 
T \pa J^i +
9 i (7 c-6)
\epsilon^{ijk} \pa \Psi J^j G^k
\nonu \\
&  - & 18 (5 c+6) \pa \Psi \Psi \pa J^i
+ 18 c(c+6)  \pa \Psi \pa G^i
\nonu \\
& + & 12 (4 c^2+15 c+18)  \pa T J^i
-18 (5 c+6)  \pa^2 \Psi G^i
\nonu \\
& - & 18 (5 c+6)  \pa^2 \Psi \Psi J^i
+ (-17 c^2+12 c+36)  \pa^3 J^i
\nonu \\
& - & 108 (c+2)  J^i \pa J^j J^j
-36 c  \pa J^i J^j J^j \Bigg)
\nonu \\
&+&  C_{(\frac{3}{2}) (\frac{3}{2})}^{(2)}  \Bigg(
\frac{(c+3)  }{20 (c-3)} \pa \psi^{(3),i}
+ \frac{3 i  }{2 (c-3)}  \epsilon^{ijk}
G^j \phi^{(\frac{5}{2}),k}
+\frac{3 (4 c-3)  }{5 (c-3) c}  
 J^i \pa \psi^{(2)}
\nonu \\
& - & \frac{3 i }{2 (c-3)} 
\epsilon^{ijk} J^j \psi^{(3),k}
-\frac{9 (c-2)  }{10 (c-3) c}  
\Psi \pa \phi^{(\frac{5}{2}),i}
\nonu \\
& - & \frac{3 (c+3)  }{5 (c-3) c} 
\pa J^i \psi^{(2)}
+ \frac{3 (7 c+6)  }{10 (c-3) c} 
\pa \Psi \phi^{(\frac{5}{2}),i} \Bigg)
+\frac{2}{5}   C_{(\frac{3}{2}) (3)}^{(\frac{5}{2})} \pa \phi^{(3),i}
\nonu \\
&+&  C_{(\frac{3}{2}) (\frac{5}{2})}^{(3)} \Bigg(
\frac{1}{4}  \pa \psi^{(3),\alpha=i}
 +
\frac{i}{4}   \epsilon^{ijk} \psi^{(4),j,\alpha=k}
\Bigg)
\Bigg](w) + \cdots.
\label{Fresult2}
\eea
}
Note that $(
\frac{1}{3} \frac{1}{4}  \pa \psi^{(3),\alpha=i}
 +
\frac{i}{4}   \epsilon^{ijk} \psi^{(4),j,\alpha=k})(w)
$ is a primary current under the stress energy tensor.
In the second OPE, we use the notation for the 
index as follows: $i+2$ becomes $1$ for $i=2$, and $2$ for $i=3$.

The remaining three types of OPEs are 
{\small
\bea
\psi^{(\frac{5}{2}),i}(z) \, \psi^{(\frac{5}{2}),j}(w) & = &
\frac{1}{(z-w)^5} \, 8 \delta^{ij} c 
+\frac{1}{(z-w)^4} \, 24 i \epsilon^{ijk} J^k(w)
\nonu \\
& + & \frac{1}{(z-w)^3} \Bigg[ \frac{1 }{(c+1) (2 c-3)} 
\Bigg(
4 (20 c^2+15 c-9)  \delta^{ij} T
-12 (4c+3)  \delta^{ij} J^k J^k
\nonu \\
& - & 12 (7 c+9)  \delta^{ij} \pa \Psi \Psi
+ 36(c+1)  J^i J^j + 6 i (4 c-9)(c+1)
\epsilon^{ijk} \pa J^k \Bigg)
\nonu \\
&+& 2  C_{(\frac{3}{2}) (\frac{3}{2})}^{(2)} \delta^{ij} \psi^{(2)} 
\Bigg](w)
\nonu \\
&+& \frac{1}{(z-w)^2} \Bigg[ 
\frac{1 }{(c+1) (c+6) (2 c-3)} \Bigg(
2c (20 c^2+57 c+54)  \delta^{ij} \pa T
\nonu \\
& - & 12 (4c+3)(c+6)  \delta^{ij} \pa J^k J^k
\nonu \\
& - & 6 (7 c+9)(c+6)  \delta^{ij} \pa^2 \Psi \Psi
+ 6 (26 c^2+9 c-18) 
G^i G^j
\nonu \\
&-& 144 i (2 c+3) 
\epsilon^{ijk} J^l J^l J^k
-12 (4 c^2-63 c-90)
J^i \pa J^j
\nonu \\
& -& 18 (31 c+42) 
\Psi (J^i G^j-J^j G^i)
- 6 i (4 c^2-63 c-90) 
\epsilon^{ijk} \Psi \pa G^k
\nonu \\
&+& 12 i (20 c^2+57 c+54) 
\epsilon^{ijk} T J^k
+ 12 (7 c^2-42 c-72) 
\pa J^i J^j
\nonu \\
&+& 18 i (4 c^2-c-6)
\epsilon^{ijk} \pa \Psi G^k
- 108 i (5 c+6) 
\epsilon^{ijk} \pa \Psi \Psi J^k
\nonu \\
&+& 2 i (4 c^3-63 c^2-234 c-216) 
\epsilon^{ijk} \pa^2 J^k \Bigg)
\nonu \\
&+&    C_{(\frac{3}{2}) (\frac{3}{2})}^{(2)} 
\Bigg( \delta^{ij} \pa \psi^{(2)}
-\frac{i (c+3) }{5 (c-3)}  \epsilon^{ijk} 
 \psi^{(3),k}
-\frac{6 i (2 c-9)  }{5 (c-3) c}   
\epsilon^{ijk} \Psi \phi^{(\frac{5}{2}),k}
\nonu \\
& + &  \frac{6 i (7 c-9)  }{5 (c-3) c}
  \epsilon^{ijk} J^k \psi^{(2)} \Bigg)
+  
 \frac{12}{5} i  C_{(\frac{3}{2}) (3)}^{(\frac{5}{2})}
\epsilon^{ijk}  \phi^{(3),k}
+ i  \epsilon^{ijk} C_{(\frac{3}{2}) (\frac{5}{2})}^{(3)} \psi^{(3),\alpha=k}
\Bigg](w)
\nonu \\
& + & \frac{1}{(z-w)} \Bigg[ \frac{1 }{(c+1) (c+6) (2 c-3)}
\Bigg(
-324 i (c+2) 
\delta^{ij}  J^i G^{i+1} G^{i+2} 
\nonu \\
& - & 36 i(c+6)   J^i J^{i+1} \pa J^{i+2}
+36i(c+6)  J^i   \pa J^{i+1} J^{i+2}
+ 144 i (2c+3) J^{i+1} G^{i} G^{i+2}
\nonu \\
& + & 
18 (5 c+6) \delta^{ij} \sum_k \Psi J^k \pa G^k
-54(3c+2) \delta^{ij} \sum_k \Psi \pa J^k G^k
-144i (2c+3) J^{i+2} G^{i} G^{i+1}
\nonu \\
&-& 18(c+6)  \Psi J^i \pa G^i - 18(c+6) \Psi \pa J^i G^i
\nonu \\
&+& 72 (3 c^2+7 c+6) 
\delta^{ij} T T
-36(c+6)  T J^i J^i
\nonu \\
& - & 144 (2 c+3)  \delta^{ij} \sum_k T J^k J^k
- 72 (7 c+6)  \delta^{ij} T \pa \Psi \Psi
\nonu \\
& - & 6 (8 c+9)(c+6)
\pa G^i G^i
+ 18 (3 c^2-10 c-24) 
\delta^{ij} \sum_k \pa G^k G^k
\nonu \\
& - & 18 c(c+6) 
\pa J^i \pa J^i
-18 (c^2-7 c-6) 
\delta^{ij} \sum_k \pa J^k \pa J^k
\nonu \\
&+& 90(c+6)
\pa \Psi J^i G^i
+ 54 (c-2)  \delta^{ij} \sum_{k} \pa \Psi J^k G^k
\nonu \\
&+ & 6 (5 c+3)(c+6)  \pa^2 J^i J^i
-6(c+6)(2c-3)  \delta^{ij} \sum_{k} \pa^2 J^k J^k
-18 c(c+6) \delta^{ij} \pa^2 \Psi \pa \Psi
\nonu \\
& - & 6 (2 c^2-13 c-6) 
\delta^{ij} \pa^3 \Psi \Psi
+ 6 (2 c+3) (c^2-3 c+18) 
\delta^{ij} \pa^2 T \Bigg)_{j=i}
\nonu \\
&  +  &  
\frac{1 }{(c+1) (c+6) (2 c-3)}
\Bigg(
6 (17 c^2+33 c+18) 
G^i \pa G^{i+1}
\nonu \\
& + & 18 i(c+6)   J^i G^i G^{i+2} 
- 18 i(c+6)   J^{i+1} G^{i+1} G^{i+2} 
\nonu \\
&+& 6 i (20 c^2+57 c+54)
 \pa T J^{i+2}
-36(c+6)  T J^i J^{i+1}
\nonu \\
&+& 24 i (5 c^2+15 c+18) 
 T \pa J^{i+2}
+ 18 (3 c^2-8 c-12) 
\pa G^i G^{i+1}
\nonu \\
&-& 18 i (7 c+6) 
 J^i J^i \pa J^{i+2} 
-9 (2 c^2-47 c-66) 
J^i \pa^2 J^{i+1}
\nonu \\
&-& 
 18 i (17 c+30) 
 \pa J^i J^i J^{i+2} 
-162 i (c+2) J^{i+1} J^{i+1} \pa J^{i+2}
\nonu \\
& - & 18 c(c+6)
\pa J^i \pa J^{i+1}
-  54 i (5 c+6)
 \pa J^{i+1} J^{i+1} J^{i+2}
\nonu \\
&-& 216 i (2 c+3) 
 \pa J^{i+2} J^{i+2} J^{i+2}
\nonu \\
&+& 3 (16 c^2-69 c-126) 
\pa^2 J^i J^{i+1}
+ i (2 c^3-77 c^2-249 c-234) 
 \pa^3 J^{i+2}
\nonu \\
&- & 144 (2 c+3)
\Psi J^i \pa G^{i+1}
+ 54 (5 c+6) 
\Psi J^{i+1} \pa G^i
\nonu \\
& - & 144 (2 c+3) 
\Psi \pa J^i G^{i+1}
+54 (5 c+6) 
\Psi \pa J^{i+1} G^i
\nonu \\
&-&  3 i (4 c^2-63 c-90) 
 \Psi \pa^2 G^{i+2}
-18 (13 c+6) 
\pa \Psi J^i G^{i+1}
\nonu \\
&+& 324 (c+2) 
\pa \Psi J^{i+1} G^i
- 54 i (5 c+6)   \pa \Psi
\Psi \pa J^{i+2}
\nonu \\
&+& 12 i (2 c+3)(c+6) 
 \pa \Psi \pa G^{i+2}
+9 i (4 c^2-c-6) 
 \pa^2 \Psi G^{i+2}
\nonu \\
& - & 54 i (5 c+6) 
 \pa^2 \Psi \Psi J^{i+2} \Bigg)_{j=i+1}
\nonu \\
&+&  C_{(\frac{3}{2}) (\frac{3}{2})}^{(2)}
\Bigg(
-  \frac{9 (3 c-1) }{5 (c-3) c}  
\delta^{ij} G^k \phi^{(\frac{5}{2}),k}
\nonu \\
& - & \frac{3 (c+3)  }{5 (c-3) c}  
\delta^{ij} J^k \psi^{(3),k}
+\frac{6 (c+3)  }{5 (c-3) c} 
\delta^{ij} \Psi \phi^{(\frac{7}{2})}
+ \frac{18 (3 c-1)  }{5 (c-3) c} 
 \delta^{ij} T \psi^{(2)} 
\nonu \\
& + & \frac{3 (c-7)  }{10 (c-3)} 
\delta^{ij} \pa^2 \psi^{(2)}
+ \frac{3  }{(c-3)} 
G^i \phi^{(\frac{5}{2}),j}
+\frac{3  }{(c-3)} 
G^j \phi^{(\frac{5}{2}),i}
\nonu \\
& + & \frac{3  }{(c-3)} 
J^i \psi^{(3),j} +
\frac{3  }{(c-3)} 
J^j \psi^{(3),i}
+  \frac{3 i (7 c-9)  }{5 (c-3) c}  
\epsilon^{ijk} \pa (J^k  \psi^{(2)})
\nonu \\
& - & \frac{3 i (2 c-9)  }{5 (c-3) c} 
\epsilon^{ijk} \pa (\Psi  \phi^{(\frac{5}{2}),k})
-  \frac{i (c+3) }{10 (c-3)}  
\epsilon^{ijk} \pa \psi^{(3),k} \Bigg)
\nonu \\
&
+ & C_{(\frac{3}{2}) (3)}^{(\frac{5}{2})} \Bigg(
\frac{6}{5} i   \epsilon^{ijk} \pa \phi^{(3),k}
 + \frac{4}{5}   \delta^{ij}   \phi^{(4)} \Bigg) 
\nonu \\
&+ & 
C_{(\frac{3}{2}) (\frac{5}{2})}^{(3)} \Bigg(
\frac{1}{2} i   
\epsilon^{ijk} \pa \psi^{(3),\alpha=k} 
+\frac{1}{2}   \psi^{(4),i,\alpha=j}
+\frac{1}{2}  \psi^{(4),j,\alpha=i}
\Bigg)
\Bigg](w)
+ \cdots,
\nonu \\
\psi^{(\frac{5}{2}),i}(z) \, \phi^{(3)}(w) & = &
\frac{1}{(z-w)^4}  \frac{1}{(2c-3)} \Bigg[ 
3 (20 c-21)  G^i
-54  \Psi J^i
\Bigg](w) \nonu \\
& + & \frac{1}{(z-w)^3} \Bigg[ \frac{1 }{(c+1) (2 c-3)}
\Bigg(
6 i (4 c+3) 
\epsilon^{ijk} J^j G^k
-6 c  \Psi \pa J^i
\nonu \\
& - & 6 (7 c+9) 
\pa \Psi J^i
+ (20 c^2+15 c-9) 
\pa G^i \Bigg)
-\frac{1}{2}  C_{(\frac{3}{2}) (\frac{3}{2})}^{(2)} \phi^{(\frac{5}{2}),i}
\Bigg](w)
\nonu \\
&+& \frac{1}{(z-w)^2}  \Bigg[ \frac{1 }{(c+1) (c+6) (2 c-3)}
\Bigg(
9 (37 c+54)  J^i J^j G^j
-252 (2 c+3)  J^j J^j G^i
\nonu \\
&+ & 6 i (11 c^2-84 c-144) 
\epsilon^{ijk} J^j \pa G^k
-54 i (3 c+4) 
\epsilon^{ijk} \Psi G^j G^k
\nonu \\
&- & \frac{63}{2} i (5 c+6)  \epsilon^{ijk} 
\Psi J^j \pa J^k
+  \frac{63}{2} i (5 c+6)  \epsilon^{ijk} 
\Psi \pa J^j  J^k
\nonu \\
&+& \frac{3}{2} (8 c+15)(c+6)  \Psi \pa^2 J^i
+6 (55 c^2+87 c+54) 
T G^i
\nonu \\
& - & 216 (3 c+4) 
T \Psi J^i
-3 i (25 c^2-138 c-216)  \epsilon^{ijk}
\pa J^j G^k
\nonu \\
& - & 126 (5 c+6)  \pa \Psi \Psi G^i
-6 (14 c^2-57 c-90)  \pa \Psi \pa J^i
\nonu \\
& + & 3 (7 c+12)(c+6)  \pa^2 \Psi J^i
+ \frac{1}{2} (10 c^3-147 c^2-603 c-702) 
\pa^2 G^i \Bigg)
\nonu \\
& + & 
 C_{(\frac{3}{2}) (\frac{3}{2})}^{(2)} \Bigg(
\frac{3 (23 c-21)  } 
{10 (c-3) c}  G^i \psi^{(2)}
+  \frac{3 i (13 c-21)  }{10 (c-3) c}
 \epsilon^{ijk} J^j \phi^{(\frac{5}{2}),k}
\nonu \\
& + & \frac{9 (c-7)  }{10 (c-3) c} 
\Psi \psi^{(3),i}
-  \frac{(c+3)  }{5 (c-3)}  
\pa \phi^{(\frac{5}{2}),i} \Bigg)
\nonu \\
&+ & 
\frac{7  }{5}  C_{(\frac{3}{2}) (3)}^{(\frac{5}{2})} \psi^{(\frac{7}{2}),i}
-\frac{1}{2} i 
 C_{(\frac{3}{2}) (\frac{5}{2})}^{(3)} \epsilon^{ijk} \phi^{(\frac{7}{2}),j,\alpha=k}
\Bigg](w)
\nonu \\
& + & \frac{1}{(z-w)} \Bigg[ 
\frac{1}{(c+1) (c+6) (2 c-3)}
\Bigg(
9 (17 c+30)
J^i J^j \pa G^j
-108 (2 c+3)
J^j J^j \pa G^i
\nonu \\
&+& 9 (17 c+30)
\pa J^i J^j  G^j
-216 (2 c+3)
\pa J^j J^j  G^i
\nonu \\
&+ & 9 (13 c+6) 
J^i \pa J^j G^j
+ 3 i (2 c+3) (5 c-42) 
\epsilon^{ijk} J^j \pa^2 G^k
\nonu \\
&-& 36 i (2 c+3)
\epsilon^{ijk} \Psi \pa (G^j G^k)
-144 (2 c+3)  T \Psi \pa J^i
\nonu \\
&-& 
216 c  T \pa \Psi J^i
+ 6 (19 c^2+3 c-18) 
T \pa G^i
\nonu \\
&-& 3 i (7 c+6)(c+6)
\epsilon^{ijk} \pa J^j \pa G^k
-54 i c  \epsilon^{ijk} \pa \Psi G^j G^k
\nonu \\
& - & 27 i (5 c+6)
\epsilon^{ijk} \pa \Psi J^j \pa J^k
-54 (5 c+6)  \pa \Psi \Psi \pa G^i
\nonu \\
&+& 54 (3 c^2+7 c+6)  \pa T G^i
-144 (2 c+3)  \pa T \Psi J^i
\nonu \\
&-& 9 i (3 c^2-23 c-30)  \epsilon^{ijk} 
\pa^2 J^j G^k
-54 (5 c+6) 
\pa^2 \Psi \Psi G^i
\nonu \\
& + & \frac{1}{2} (2 c^3-67 c^2-183 c-198) 
\pa^3 G^i 
-27 i (5 c+6)  \epsilon^{ijk} 
\Psi J^j \pa^2 J^k
\nonu \\
&+& \frac{3}{2} (4 c^2+3 c+18)
\Psi \pa^3 J^i
-9 (4 c^2-7 c-6)  
\pa \Psi \pa^2 J^i
\nonu \\
&-& 9 (c^2-39 c-54) 
\pa^2 \Psi \pa J^i
+3 (3 c-2)(c+6) 
\pa^3 \Psi J^i \Bigg)
\nonu \\
&+&   C_{(\frac{3}{2}) (\frac{3}{2})}^{(2)}
\Bigg(
\frac{9 (4 c-3)  }{10 (c-3) c} 
G^i \pa \psi^{(2)}
-\frac{3 i  }{2 (c-3)}   \epsilon^{ijk}
G^j \psi^{(3),k}
\nonu \\
& - & \frac{3  }{(c-3)}  J^i \phi^{(\frac{7}{2})}
+\frac{3 (2 c-9)  }{10 (c-3) c} 
\Psi \pa \psi^{(3),i}
-\frac{3  }{(c-3)}  T \phi^{(\frac{5}{2}),i}
\nonu \\
&+& \frac{3 i (2 c-9)  }{10 (c-3) c}  
\epsilon^{ijk} \pa J^j \phi^{(\frac{5}{2}),k}
+\frac{3 (7 c-9)  }{10 (c-3) c}  
\pa G^i \psi^{(2)}
\nonu \\
& - & \frac{9 (c+3)  }{10 (c-3) c} 
\pa \Psi \psi^{(3),i} 
-\frac{(c+3)  }{20 (c-3)} 
\pa^2 \phi^{(\frac{5}{2}),i} +  \frac{3 i (7 c-9)  }{10 (c-3) c}  
\epsilon^{ijk} J^j \pa \phi^{(\frac{5}{2}),k}\Bigg)
\nonu \\
&+& \frac{3}{5}  C_{(\frac{3}{2}) (3)}^{(\frac{5}{2})} \pa \psi^{(\frac{7}{2}),i}
+ C_{(\frac{3}{2}) (\frac{5}{2})}^{(3)} \Bigg(
-\frac{1}{4} i   \epsilon^{ijk} 
\pa \phi^{(\frac{7}{2}),j,\alpha=k} -\frac{1}{2}  
\phi^{(\frac{9}{2}),\alpha=i} \Bigg)
\Bigg](w)
+\cdots,
\nonu \\
\phi^{(3)}(z) \, \phi^{(3)}(w) & = & 
\frac{1}{(z-w)^6} \, 10 c +\frac{1}{(z-w)^4} \Bigg[ 
\frac{1 }{(c+1) (2 c-3)} \Bigg(
6 (20 c^2+7 c-15)  T
\nonu \\
& - & 9 (4 c+3)  J^i J^i
-  
18 (5 c+6)  \pa \Psi \Psi \Bigg)
+
\frac{3}{2}   C_{(\frac{3}{2}) (\frac{3}{2})}^{(2)} \psi^{(2)}
\Bigg](w)
\nonu \\
&+& \frac{1}{(z-w)^3} \Bigg[ \frac{1}{2} \pa (\mbox{pole four})
\Bigg](w)
\nonu \\
& + & \frac{1}{(z-w)^2} \Bigg[ 
\frac{1 }{(c+1) (c+6) (2 c-3)} \Bigg(
-144 i (2 c+3)  \epsilon^{ijk} J^i G^j G^k
\nonu \\
& - & 216 (5 c+6)  T \pa \Psi \Psi
+36(5c+6) \Psi J^i \pa G^i
-72(5c+6) \Psi \pa J^i G^i
\nonu \\
&+ & 12 (7 c^2-42 c-72)  \pa G^i G^i
-\frac{9}{2} (12 c^2-35 c-66)   \pa J^i \pa J^i 
\nonu \\
&+& 36 (5 c+6) \pa \Psi J^i G^i
+\frac{9}{2} (4 c+11) (c+6) \pa^2 J^i J^i
+24 (16 c^2+27 c+18) T T 
\nonu \\
& - & \frac{27}{2} (5 c+6)(c+6)  \pa^2 \Psi \pa \Psi
-\frac{3}{2} (c-42) (5 c+6) 
\pa^3 \Psi \Psi
\nonu \\
& + & \frac{9}{2} (4 c^3-17 c^2+63 c+126)
\pa^2 T 
-288(2c+3) T J^i J^i
\Bigg)
\nonu \\
&+&  C_{(\frac{3}{2}) (\frac{3}{2})}^{(2)} \Bigg(
-\frac{3 (11 c-12)  }{10 (c-3) c} 
G^i \phi^{(\frac{5}{2}),i}
+  \frac{9 (c-2)  }{5 (c-3) c}   J^i 
\psi^{(3),i}
+\frac{9 (c-7)  }{40 (c-3)}   
\pa^2 \psi^{(2)} \nonu \\
&-& \frac{3 (c-12) }{5 (c-3) c} \Psi \phi^{(\frac{7}{2})} 
+ \frac{12 (4 c-3) }{5 (c-3) c} T \psi^{(2)}
\Bigg)
+  \frac{8}{5} C_{(\frac{3}{2}) (3)}^{(\frac{5}{2})} \phi^{(4)}
+ \frac{1}{4}    C_{(\frac{3}{2}) (\frac{5}{2})}^{(3)}  \psi^{(4),i,\alpha=i}
\Bigg](w) 
\nonu \\
&+& \frac{1}{(z-w)} \Bigg[  -\frac{1}{24} \pa^3 (\mbox{pole four}) + \frac{1}{2} \pa  (\mbox{pole two} )
\Bigg](w)
+\cdots.
\label{compresult}
\eea}
In the first OPE, we use the same convention for the index as before.
One sees that $(
-\frac{1}{7} \frac{1}{4} i   \epsilon^{ijk} 
\pa \phi^{(\frac{7}{2}),j,\alpha=k} -\frac{1}{2}  
\phi^{(\frac{9}{2}),\alpha=i} 
)(w)
$ is a primary current under the stress energy tensor.
We can also describe the above OPEs in the manifest way of  
$U(1)$ charge. Any three higher spin currents having $SO(3)$ index $i$
can be decomposed into the one with $U(1)$ charge $+1$,
the one with $U(1)$ charge $-1$, and the one with vanishing 
$U(1)$ charge.
In the large $c$ limit, all the nonlinear terms
on the right hand side  
in (\ref{compresult}) disappear.      

\subsection{ The $8$ OPEs between the eight higher spin currents 
and the lowest higher spin-$\frac{3}{2}$ current  }

As emphasized in section $7$, it is very useful to write down
the following eight OPEs (including the first OPE of
(\ref{Fresult1})), by taking the OPEs between the 
eight higher spin currents located at $z$ coordinate and the lowest higher 
spin current located at $w$ coordinate, 
{\small
\bea
 \phi^{(2),i}(z) \, \psi^{(\frac{3}{2})}(w)  & = &
\frac{1}{(z-w)^2} \, \frac{1}{(2c-3)} \Bigg[ 6 c G^i 
-18  \Psi J^i
\Bigg](w) 
\nonu \\
&+& \frac{1}{(z-w)} \Bigg[ \frac{2}{3} \pa (\mbox{pole two})
- \frac{1}{(c+1)(2c-3)} \Bigg( 6 i c  ( \epsilon^{ijk} J^j G^k
-\frac{2}{3} i \pa G^i)
\nonu \\
& - & 18 ( \pa \Psi  J^i -\frac{1}{3} \pa (\Psi J^i)) \Bigg)
+ 
\frac{1}{2}  C_{(\frac{3}{2}) (\frac{3}{2})}^{(2)} \phi^{(\frac{5}{2}),i}
\Bigg](w) + \cdots,
\nonu \\
\psi^{(\frac{5}{2}),i}(z) \psi^{(\frac{3}{2})}(w) & = &
\frac{1}{(z-w)^3} \,  6 J^i(w)
+\frac{1}{(z-w)^2} \,  \Bigg[ \frac{18 }{(2 c-3)} \Psi G^i 
+ 6 \pa J^i
\Bigg](w)
\nonu \\
& + & \frac{1}{(z-w)} \Bigg[ 
 \frac{3}{4} \pa (\mbox{pole two}) \nonu \\
& + & 
 \frac{1}{(c+1) (c+6) (2 c-3)} \Bigg(
18 i (5 c+6) (  \epsilon^{ijk} \Psi
J^j G^k -\frac{2}{3} i \Psi \pa G^i) 
\nonu \\
& - & 18 i c (c+2)
(\epsilon^{ijk} G^j G^k -\frac{2}{3} i \pa^2 J^i)
+  36 (c^2+3 c+6)
(T J^i -\frac{1}{2} \pa^2 J^i)
\nonu \\  
& - & 54 (c+2) J^i J^j J^j
+  72 i c
(\epsilon^{ijk} \pa J^j J^k -\frac{1}{3} i \pa^2 J^i)
\nonu \\
& + &  
6 (c^2-17 c-42)
(\pa \Psi G^i -\frac{1}{4} \pa (\Psi G^i)) 
-72 c \pa \Psi \Psi J^i \Bigg) -\frac{3}{2} \pa^2 J^i
\nonu \\
&+ & C_{(\frac{3}{2}) (\frac{3}{2})}^{(2)} \Bigg(
\frac{3 (c+3)  }{5 (c-3) c}  
\Psi \phi^{(\frac{5}{2}), i} 
+ \frac{9 (3 c-1)  }{5 (c-3) c} 
J^i \psi^{(2)}
+ \frac{3 (c-7) }{10 (c-3)}
  \psi^{(3), i} \Bigg)
 \nonu \\
&+& \frac{2}{5}   C_{(\frac{3}{2}) (3)}^{(\frac{5}{2})}  \phi^{(3),i} +  
 C_{(\frac{3}{2}) (\frac{5}{2})}^{(3)} \psi^{(3),\alpha=i}
\Bigg](w) +\cdots,
\nonu \\
 \phi^{(3)}(z) \, \psi^{(\frac{3}{2})}(w) & = &
\frac{1}{(z-w)^4} \, 3 \Psi(w) +
\frac{1}{(z-w)^3} \, 6 \pa \Psi(w) 
\nonu \\
& + & \frac{1}{(z-w)^2} \Bigg[ \frac{1}{(c+1) (2 c-3)}
\Bigg(
6 (c+3) (T \Psi - \frac{3}{4} \pa^2 \Psi)
+ 
36 \frac{c(c+1)}{(c+6)}
J^i G^i
\nonu \\
& - & \frac{9 (13 c+18)}{(c+6)} \Psi J^i J^i \Bigg)
+ \frac{9}{2} \pa^2 \Psi
+\frac{3  }{2 c} C_{(\frac{3}{2}) (\frac{3}{2})}^{(2)} \Psi \psi^{(2)}
+  C_{(\frac{3}{2}) (3)}^{(\frac{5}{2})} \psi^{(\frac{5}{2})}
\Bigg](w) \nonu \\
& + & \frac{1}{(z-w)}
\Bigg[ 
\frac{4}{5} \pa (\mbox{pole two})
+  \frac{1}{(c+1) (2 c-3)} \Bigg(
24 (c+3) (\pa \Psi T -\frac{1}{5} \pa (\Psi T))
\nonu \\
& + & 12 c (\pa J^i G^i -\frac{2}{5} \pa (J^i G^i))
- 
18
(\pa \Psi J^i J^i -\frac{1}{5}\pa (\Psi J^i J^i)) 
 \Bigg)
-\frac{8}{5} \pa^3 \Psi
\nonu \\
& + &  C_{(\frac{3}{2}) (\frac{3}{2})}^{(2)}
\Bigg(
\frac{9 }{2 (c-3)}
(\pa \Psi \psi^{(2)} -\frac{1}{5} \pa (\Psi \psi^{(2)}))
+\frac{3 }{2 (c-3)}  J^i \phi^{(\frac{5}{2}),i}
\nonu \\
&+ & 
\frac{(c-12) }{5 (c-3)}   \phi^{(\frac{7}{2})} \Bigg)
+\frac{1 }{4}  C_{(\frac{3}{2}) (\frac{5}{2})}^{(3)} \phi^{(\frac{7}{2}),i,\alpha=i} 
\Bigg](w) 
+ \cdots.
\label{reverseorder}
\eea}
That is, the OPE
(\ref{Fresult1}) and the above three OPEs (\ref{reverseorder})
will appear in the ${\cal N}=3$ version (\ref{finalPhiPhi}) in appropriate
supersymmetric way as in \cite{AK1509}.

\section{ The OPEs between the lowest eight higher spin currents 
in the component approach corresponding to the section $8$ }

Because the higher spin currents after decoupling the 
spin-$\frac{1}{2}$ current of ${\cal N}=3$ superconformal algebra
are determined via (\ref{relation}), we can calculate the OPEs between 
them.

\subsection{ The complete $36$ OPEs (between the lowest eight higher
spin currents) in the component approach  }

Based on the results of Appendix $F$, the OPEs between the 
composite fields appearing on the right hand sides of 
(\ref{relation}) are known, we calculate the OPEs between the 
higher spin currents after factoring out the spin-$\frac{1}{2}$ current
and they 
are described as follows.
The first four types of OPEs (totally eight OPEs) 
can be summarized as 
{\small
\bea
\hat{\psi}^{(\frac{3}{2})}(z) \, \hat{\psi}^{(\frac{3}{2})}(w) & = & 
\frac{1}{(z-w)^3} \, \frac{2c}{3} \nonu \\
& + & \frac{1}{(z-w)} \, \Bigg[ \frac{1 }{(c+1) (2 c-3)} \Bigg(
-6 c  \hat{J}^i \hat{J}^i  +
4 c (c+3) \hat{T}  \Bigg) +   
C_{(\frac{3}{2}) (\frac{3}{2})}^{(2)} \hat{\psi}^{(2)} 
\Bigg](w) + 
\cdots,
\nonu \\
\hat{\psi}^{(\frac{3}{2})}(z) \, \hat{\phi}^{(2),i}(w) & = &
\frac{1}{(z-w)^2} \, \frac{6c}{(2c-3)}  \hat{G}^i(w) 
\nonu \\
&+& \frac{1}{(z-w)} \Bigg[
\frac{1}{(c+1) (2 c-3)} 
\Bigg( 
2 c (c+3) \pa \hat{G}^i
+  6 i c    \epsilon^{ijk} \hat{J}^j \hat{G}^k \Bigg)
-  
\frac{1}{2}  C_{(\frac{3}{2}) (\frac{3}{2})}^{(2)} \hat{\phi}^{(\frac{5}{2}),i}
\Bigg](w) \nonu \\
& + & \cdots,
\nonu \\
\hat{\psi}^{(\frac{3}{2})}(z) \, \hat{\psi}^{(\frac{5}{2}),i}(w) & = &
\frac{1}{(z-w)^3} \,  6 \hat{J}^i(w)
\nonu \\
& + & \frac{1}{(z-w)} \Bigg[  
 \frac{1}{(c+1) (c+6) (2 c-3)} \Bigg(
-  18 i c (c+2)
\epsilon^{ijk} \hat{G}^j \hat{G}^k 
+  36 (c^2+3 c+6)
\hat{T} \hat{J}^i 
\nonu \\
& - & 6 (5 c^2+9 c+18) \pa^2 \hat{J}^i
-  54 (c+2) \hat{J}^i \hat{J}^j \hat{J}^j
+  18 i (c-6)
\epsilon^{ijk} \pa \hat{J}^j \hat{J}^k 
    \Bigg)
\nonu \\
&+ &  C_{(\frac{3}{2}) (\frac{3}{2})}^{(2)} \Bigg(
 \frac{9 (3 c-1)  }{5 (c-3) c} 
\hat{J}^i \hat{\psi}^{(2)}
+ \frac{3 (c-7) }{10 (c-3)}
  \hat{\psi}^{(3), i} \Bigg)
+ \frac{2}{5}   C_{(\frac{3}{2}) (3)}^{(\frac{5}{2})}  \hat{\phi}^{(3),i} 
\nonu \\
& + &  
 C_{(\frac{3}{2}) (\frac{5}{2})}^{(3)} \hat{\psi}^{(3),\alpha=i}
\Bigg](w) +\cdots,
\nonu \\
\hat{\psi}^{(\frac{3}{2})}(z) \, \hat{\phi}^{(3)}(w) & = &
 \frac{1}{(z-w)^2}  \Bigg[ \frac{36 c}{(c+1) (2 c-3)} \,
\hat{J}^i \hat{G}^i
+  C_{(\frac{3}{2}) (3)}^{(\frac{5}{2})} \hat{\psi}^{(\frac{5}{2})}
\Bigg](w) \nonu \\
& + & \frac{1}{(z-w)}
\Bigg[ 
\frac{1}{(c+1) (c+6) (2 c-3)} \Bigg(-36 c
\pa \hat{J}^i \hat{G}^i  + 12 c (c+3) \hat{J}^i \pa \hat{G}^i
 \Bigg)
\nonu \\
& + & 
C_{(\frac{3}{2}) (\frac{3}{2})}^{(2)} \Bigg(
-\frac{3 }{2 (c-3)}  \hat{J}^i \hat{\phi}^{(\frac{5}{2}),i}
- 
\frac{(c-12) }{5 (c-3)}   \hat{\phi}^{(\frac{7}{2})} \Bigg)
+  \frac{1}{5} C_{(\frac{3}{2}) (3)}^{(\frac{5}{2})} \pa \hat{\psi}^{(\frac{5}{2})}
\nonu \\
& - & 
\frac{1 }{4}  C_{(\frac{3}{2}) (\frac{5}{2})}^{(3)} \hat{\phi}^{(\frac{7}{2}),i,\alpha=i} 
\Bigg](w) 
+ \cdots.
\label{firstfour} 
\eea}
In particular,
the $\Psi(w)$ dependent-terms with $C_{(\frac{3}{2}) (\frac{3}{2})}^{(2)} $ 
structure constant appearing in Appendix $F$
do not appear in the third and fourth equations of (\ref{firstfour}). 

The next three types of OPEs can be described as
{\small
\bea
\hat{\phi}^{(2), i}(z) \, \hat{\phi}^{(2),j}(w)  & = &
\frac{1}{(z-w)^4} \, 2 c \, \delta^{ij}
+ \frac{1}{(z-w)^3} \, 6 i \epsilon^{ijk} \hat{J}^k(w)
\nonu \\
& + &  \frac{1}{(z-w)^2} \Bigg[ 
\frac{8 c (2 c+3) }{(c+1) (2 c-3)} \delta^{ij} \hat{T} 
-\frac{18}{ (2 c-3)} \hat{J}^i \hat{J}^j 
-\frac{6 c }{(c+1) (2 c-3)} \delta^{ij} \hat{J}^k \hat{J}^k
\nonu \\
& + & 
  \frac{6 i c}{(2c-3)} \epsilon^{ijk} \pa \hat{J}^k + \delta^{ij} 
 C_{(\frac{3}{2}) (\frac{3}{2})}^{(2)}
\hat{\psi}^{(2)}
\Bigg](w)
\nonu \\
&+& \frac{1}{(z-w)} \Bigg[ 
\frac{1 }{(c+1) (c+6) (2 c-3)} \Bigg(
6 c (5 c+6)   \hat{G}^i \hat{G}^j 
-54 i (c+2)  \epsilon^{ijk} \hat{J}^l \hat{J}^l \hat{J}^k
\nonu \\
&+& 2c(4c^2 +15c+18) \delta^{ij} \pa \hat{T} 
-  6c (c+6) \delta^{ij} \pa \hat{J}^k \hat{J}^k
\nonu \\
&+& 12 i (4 c^2+15 c+18)
\epsilon^{ijk} \hat{T} \hat{J}^k
-  
12 (c^2-9 c-18) \hat{J}^i \pa \hat{J}^j
\nonu \\
& - & 6 (c^2+24 c+36)  \pa \hat{J}^i \hat{J}^j
+ 2 i (c^3-9 c^2-45 c-54) 
\epsilon^{ijk} \pa^2 \hat{J}^k
\Bigg)
\nonu \\
&  + & C_{(\frac{3}{2}) (\frac{3}{2})}^{(2)} \Bigg(
\frac{9 i (3 c-1)  }{5 (c-3) c} 
 \epsilon^{ijk} \hat{J}^k \hat{\psi}^{(2)} +  
 \frac{1}{2} \delta^{ij} 
\pa \hat{\psi}^{(2)}
-  \frac{i (c+3)  }{5 (c-3)}  
\epsilon^{ijk} \hat{\psi}^{(3),k} \Bigg)
\nonu \\
& + & 
\frac{2}{5} i  C_{(\frac{3}{2}) (3)}^{(\frac{5}{2})} \epsilon^{ijk} \hat{\phi}^{(3),k}
+ i  C_{(\frac{3}{2}) (\frac{5}{2})}^{(3)} \epsilon^{ijk} \hat{\psi}^{(3),\alpha=k}
\Bigg](w) + \cdots,
\nonu \\
\hat{\phi}^{(2),i}(z) \, \hat{\psi}^{(\frac{5}{2}),j} & = &
\frac{1}{(z-w)^3} \, \frac{1}{(2c-3)} \Bigg[ 6 i (4 c-3)  \epsilon^{ijk}
\hat{G}^k  \Bigg](w)
\nonu \\
&+ & \frac{1}{(z-w)^2} \Bigg[ 
 \frac{1}{(c+1) (2 c-3)}
\Bigg(
\frac{72 c(c+1) }{(c+6)} \delta^{ij} \hat{J}^k \hat{G}^k
-18(c+1) \hat{J}^i \hat{G}^j
\nonu \\
& - &
 6 c  (\hat{J}^i \hat{G}^j-\hat{J}^j \hat{G}^i)
+  4 i c (2 c+3) \epsilon^{ijk} \pa \hat{G}^k \Bigg)
\nonu \\
& + & 2  C_{(\frac{3}{2}) (3)}^{(\frac{5}{2})} \delta^{ij} \hat{\psi}^{(\frac{5}{2})}
-\frac{i}{2}  C_{(\frac{3}{2}) (\frac{3}{2})}^{(2)} \epsilon^{ijk} 
\hat{\phi}^{(\frac{5}{2}),k}
\Bigg](w) \nonu \\
& + & \frac{1}{(z-w)} \Bigg[ 
\frac{1 }{(c+1) (c+6) (2 c-3)}
\Bigg(
-18 i(c+6) 
 \hat{J}^i (\hat{J}^{i+1} \hat{G}^{i+2}- \hat{J}^{i+2} \hat{G}^{i+1})
\nonu \\
&+& 24 c^2 \delta^{ij} \sum_k \hat{J}^k \pa \hat{G}^k 
+ 36 c (c+2) \delta^{ij} \sum_k \pa \hat{J}^k  \hat{G}^k 
+ 12 c (c+6) \hat{J}^i \pa \hat{G}^i
\nonu \\
& - & 36 (c+1)(c+6) \pa \hat{J}^i  \hat{G}^i
  \Bigg)_{j=i}
 \nonu \\
&+& \frac{1 }{(c+1) (c+6) (2 c-3)}
\Bigg(
- 162 i (c+2) \hat{J}^{i+1} \hat{J}^{i+1} \hat{G}^{i+2}
\nonu \\
& + & 108 i (c+2) \hat{J}^{i+1} \hat{J}^{i+2} \hat{G}^{i+1}
+36 i (3c^2 + 7c +6) \hat{T} \hat{G}^{i+2}
\nonu \\
& - & 6 (c^2 + 36c +36) \pa \hat{J}^{i} \hat{G}^{i+1}
- 6(c-18)(2c+3) \hat{J}^i \pa \hat{G}^{i+1}
\nonu \\
&-& 6c(5c+6) \pa \hat{J}^{i+1} \hat{G}^i
+ 6(4c^2 -21c-54) \hat{J}^{i+1} \pa \hat{G}^{i}
\nonu \\
&-& 72 i (2c+3) \hat{J}^i \hat{J}^i \hat{G}^{i+2}
+18i (5c+6) \hat{J}^i \hat{J}^{i+2} \hat{G}^i
-54 i (c+2) \hat{J}^{i+2} \hat{J}^{i+2} \hat{G}^{i+2}
\nonu \\
&+&  i (2c^3 -21c^2 -117c-162) \pa^2 \hat{G}^{i+2}
\Bigg)_{j=i+1}
\nonu \\
&+& \frac{1 }{(c+1) (c+6) (2 c-3)}
\Bigg(
162 i (c+2) \hat{J}^j \hat{J}^j \hat{G}^{j+2} 
-108 i (c+2) \hat{J}^j \hat{J}^{j+2} \hat{G}^j
\nonu \\
& + & 6 (4c^2 -21c-54) \hat{J}^j \pa \hat{G}^{j+1}
+ 72 i (2c+3) \hat{J}^{j+1} \hat{J}^{j+1} \hat{G}^{j+2}
\nonu \\
&- & 18 i (5c+6) \hat{J}^{j+1} \hat{J}^{j+2} \hat{G}^{j+1}
- 6 (c-18)(2c+3) \hat{J}^{j+1} \pa \hat{G}^j
\nonu \\
& + & 54 i (c+2) \hat{J}^{j+2} \hat{J}^{j+2} \hat{G}^{j+2}
- 36 i (3c^2 +7c +6) \hat{T} \hat{G}^{j+2}
\nonu \\
&-& 6 c (5c+6) \pa \hat{J}^j \hat{G}^{j+1}
- 6 (c^2 +36 c+36) \pa \hat{J}^{j+1} \hat{G}^j
\nonu \\
&-& i (2c^3 -21c^2 -117c-162) \pa^2 \hat{G}^{j+2}
\Bigg)_{j=i-1}
\nonu \\
& + & 
 C_{(\frac{3}{2}) (\frac{3}{2})}^{(2)} \Bigg(-
\frac{3 }{(c-3)} 
\delta^{ij} \hat{J}^k \hat{\phi}^{(\frac{5}{2}),k}
+ \frac{6 }{(c-3)}  
 \hat{J}^i \hat{\phi}^{(\frac{5}{2}),j} 
+   \frac{(c+3)  }{5 (c-3)}    
  \delta^{ij} \hat{\phi}^{(\frac{7}{2})}
\nonu \\
& - & \frac{9 (3c-1)  }{5 (c-3) c}   
(\hat{J}^i \hat{\phi}^{(\frac{5}{2}),j}-\hat{J}^j \hat{\phi}^{(\frac{5}{2}),i})
 +  \frac{9 i (3 c-1)  }{5 (c-3) c}  
\epsilon^{ijk} \hat{G}^k \hat{\psi}^{(2)} -
\frac{i(c+3) }{5(c-3)} 
\epsilon^{ijk} \pa \hat{\phi}^{(\frac{5}{2}),k}
\Bigg)
\nonu \\
& + &  C_{(\frac{3}{2}) (\frac{5}{2})}^{(3)}
\Bigg(-
\frac{1}{2}  \delta^{ij} \hat{\phi}^{(\frac{7}{2}),k,
\alpha=k}
+     \hat{\phi}^{(\frac{7}{2}),i,\alpha=j}
\Bigg) 
\nonu \\
&+ &  C_{(\frac{3}{2}) (3)}^{(\frac{5}{2})} \Bigg(
\frac{4}{5}   \delta^{ij} 
\pa \hat{\psi}^{(\frac{5}{2})} 
+ \frac{2}{5} i    \epsilon^{ijk}
\hat{\psi}^{(\frac{7}{2}),k} \Bigg)
\Bigg](w) +  \cdots,
\nonu \\
\hat{\phi}^{(2),i}(z) \, \hat{\phi}^{(3)}(w) &= & 
\frac{1}{(z-w)^4} \, 12 \hat{J}^i(w) 
+\frac{1}{(z-w)^2} \, \Bigg[ 
\nonu \\
&  
+ & \frac{1 }{(c+1) (c+6) (2 c-3)}
\Bigg(-
9 i c (5 c+6)  
\epsilon^{ijk} \hat{G}^j \hat{G}^k 
-72 (2 c+3) 
\hat{J}^i \hat{J}^j \hat{J}^j
\nonu \\
& + & 9 i (4 c^2+15 c+18)
\epsilon^{ijk}  \hat{J}^j \pa \hat{J}^k
+  
12 (8 c^2+15 c+18) 
\hat{T} \hat{J}^i
-  9 (6 c^2-c-6)
  \pa^2 \hat{J}^i \Bigg)
\nonu \\
&+&  C_{(\frac{3}{2}) (\frac{3}{2})}^{(2)} \Bigg(
\frac{3 (c-7)  }{20 (c-3)} 
\hat{\psi}^{(3),i}
+ \frac{3 (7 c-9)  }{5 (c-3) c}  
\hat{J}^i \hat{\psi}^{(2)} \Bigg)
\nonu \\
& + & 
\frac{6}{5}  C_{(\frac{3}{2}) (3)}^{(\frac{5}{2})} \hat{\phi}^{(3),i}
+\frac{1}{2}  C_{(\frac{3}{2}) (\frac{5}{2})}^{(3)} \hat{\psi}^{(3),\alpha=i}
\Bigg](w)
\nonu \\
&+& \frac{1}{(z-w)} \Bigg[ 
\frac{1}{(c+1) (c+6) (2 c-3)} 
\Bigg(
-3 i c (5 c+6)  \epsilon^{ijk} \pa (\hat{G}^j  \hat{G}^k)
\nonu \\
& + &  3 i (4 c^2+15 c+18)  
\epsilon^{ijk} \pa (\hat{J}^j \pa \hat{J}^k) 
 -  108 (c+2)  \hat{J}^i  \pa \hat{J}^j  \hat{J}^j
-36 c  \pa \hat{J}^i \hat{J}^j \hat{J}^j
\nonu \\
& - & 36 (5 c+6) 
\hat{T} \pa \hat{J}^i 
 +  12 (4 c^2+15 c+18)  \pa \hat{T} \hat{J}^i
-  (2 c-15) (5 c+6)
  \pa^3 \hat{J}^i
 \Bigg)
\nonu \\
&+&  C_{(\frac{3}{2}) (\frac{3}{2})}^{(2)}  \Bigg(
\frac{(c+3)  }{20 (c-3)} \pa \hat{\psi}^{(3),i}
+ \frac{3 i  }{2 (c-3)}  \epsilon^{ijk}
\hat{G}^j \hat{\phi}^{(\frac{5}{2}),k}
+\frac{3 (4 c-3)  }{5 (c-3) c}  
 \hat{J}^i \pa \hat{\psi}^{(2)}
\nonu \\
& - & \frac{3 i }{2 (c-3)} 
\epsilon^{ijk} \hat{J}^j \hat{\psi}^{(3),k}
-  \frac{3 (c+3)  }{5 (c-3) c} 
\pa \hat{J}^i \hat{\psi}^{(2)}
 \Bigg)
+\frac{2}{5}   C_{(\frac{3}{2}) (3)}^{(\frac{5}{2})} \pa \hat{\phi}^{(3),i}
\nonu \\
&+&  C_{(\frac{3}{2}) (\frac{5}{2})}^{(3)} \Bigg(
\frac{1}{4}  \pa \hat{\psi}^{(3),\alpha=i}
 +
\frac{i}{4}   \epsilon^{ijk} \hat{\psi}^{(4),j,\alpha=k}
\Bigg)
\Bigg](w) + \cdots.
\label{nextthree}
\eea}
In this case, the $\Psi(w)$ dependent terms appearing in 
Appendix $F$ are disappeared in the corresponding OPEs in 
(\ref{nextthree}).
The presentation in the first-order pole of 
the OPE between the higher spin-$2$ currents 
and the higher spin-$\frac{5}{2}$
currents is rather complicated.
For the indices $(i,j)=(1,1), (2,2)$ and $(3,3)$ of these OPEs,
we can cover them from the first piece with $j=i$ 
condition of the first-order pole.
In other words, the $(i,j)=(2,2)$ case can be obtained 
from the expression of $(i,j)=(1,1)$ by replacing 
$1\rightarrow 2, 2 \rightarrow 3$, and $3 \rightarrow 1$ simply.  
The $(i,j)=(3,3)$ case can be obtained similarly from the 
result of $(i,j)=(1,1)$ by 
changing  with $1\rightarrow 3, 2 \rightarrow 1$, and $3 \rightarrow 2$.

Furthermore, for the indices $(i,j)=(1,2), (2,3)$ and $(3,1)$
of these OPEs can be analyzed with the second piece with $j=i+1$ 
condition of the first-order pole.
The $(i,j)=(2,3)$ case can be obtained 
from the expression of $(i,j)=(1,2)$ by the above first replacement 
while 
the $(i,j)=(3,1)$ case can be obtained 
from the expression of $(i,j)=(1,2)$ by the above second replacement. 

For the remaining three cases where the indices are 
given by $(i,j)=(2,1), (3,2)$ and $(1,3)$, 
we have the first case $(i,j)=(2,1)$ from  the third piece 
with $j=i-1$ 
condition of the first-order pole. The case $(i,j)=(3,2)$
can be obtained from this by replacing the indices according to the 
above first replacement. The final case  $(i,j)=(1,3)$
can be obtained from the case $(i,j)=(2,1)$  
by the above second replacement.

Now the remaining three types of OPEs can be written as 
{\small
\bea
\hat{\psi}^{(\frac{5}{2}),i}(z) \, \hat{\psi}^{(\frac{5}{2}),j}(w) & = &
\frac{1}{(z-w)^5} \, 2(4c-3) \delta^{ij} 
+\frac{1}{(z-w)^4} \,  \frac{6 i (4 c-3) }{c} \epsilon^{ijk} \hat{J}^k(w)
\nonu \\
& + & \frac{1}{(z-w)^3} \Bigg[ \frac{1 }{(c+1) (2 c-3)} 
\Bigg(
4 (20 c^2+3 c-27)  \delta^{ij} \hat{T}
-6 (8 c+3)  \delta^{ij} \hat{J}^k \hat{J}^k
\nonu \\
& + & 
  \frac{18}{c} (c+1)(2 c+3) \hat{J}^i \hat{J}^j + 24 i ( c-3)(c+1)
\epsilon^{ijk} \pa \hat{J}^k \Bigg)
+  \frac{(2 c-3)}{c} C_{(\frac{3}{2}) (\frac{3}{2})}^{(2)} \delta^{ij} \hat{\psi}^{(2)} 
\Bigg](w)
\nonu \\
&+& \frac{1}{(z-w)^2} \Bigg[ 
\frac{1 }{(c+1) (c+6) (2 c-3)} \Bigg(
2 (20 c^2+3 c-27)(c+6)  \delta^{ij} \pa \hat{T}
\nonu \\
& - & 6 (8c+3)(c+6)  \delta^{ij} \pa \hat{J}^k \hat{J}^k
\nonu \\
& + & 
 \frac{18}{c}(c+1)(c+6)(2c+3)  \pa \hat{J}^i \hat{J}^j
+ 6 (13 c^2-3 c-18) 
(\hat{G}^i \hat{G}^j-\hat{G}^j \hat{G}^i)
\nonu \\
&-& \frac{3}{c} i (8 c^3+3 c^2-45 c-54)
\epsilon^{ijk} \hat{J}^l \hat{J}^l \hat{J}^k +
\frac{3}{c} i (8 c^3-93 c^2-135 c+54) \epsilon^{ijk} \hat{J}^k 
\hat{J}^l \hat{J}^l 
\nonu \\
&+& \frac{12}{c} i (20 c^3+45 c^2+9 c-54) 
\epsilon^{ijk} \hat{T} \hat{J}^k
\nonu \\
&+& \frac{i}{c} (8 c^4-132 c^3-297 c^2-189 c+162)
\epsilon^{ijk} \pa^2 \hat{J}^k \Bigg)
\nonu \\
&+&    C_{(\frac{3}{2}) (\frac{3}{2})}^{(2)} 
\Bigg( 
 \frac{(2 c-3)}{2c}
\delta^{ij} \pa \hat{\psi}^{(2)}
-\frac{i (c+3) }{5 c}  \epsilon^{ijk} 
 \hat{\psi}^{(3),k}
+   \frac{3 i (14 c-3)  }{5 c^2}
  \epsilon^{ijk} \hat{J}^k \hat{\psi}^{(2)} \Bigg)
\nonu \\
& + &  \frac{6 i (2 c-1) }{5 c}
  C_{(\frac{3}{2}) (3)}^{(\frac{5}{2})}
\epsilon^{ijk}  \hat{\phi}^{(3),k}
+ \frac{i (c-3) }{c}  
\epsilon^{ijk} C_{(\frac{3}{2}) (\frac{5}{2})}^{(3)} \hat{\psi}^{(3),\alpha=k}
\Bigg](w)
\nonu \\
& + & \frac{1}{(z-w)} \Bigg[ \frac{1 }{(c+1) (c+6) (2 c-3)}
\Bigg(
- 324 i (c+2) 
  \hat{J}^i \hat{G}^{i+1} \hat{G}^{i+2}
\nonu \\
& - &   36 i (c+6)
  \hat{J}^i \hat{J}^{i+1}  \pa \hat{J}^{i+2}
+\frac{36}{c} i (c^2-21 c-54)  \hat{J}^i \pa \hat{J}^{i+1}   \hat{J}^{i+2}
\nonu \\
& + &  144 i (2 c+3)
 \hat{J}^{i+1} \hat{G}^i \hat{G}^{i+2}
- 144 i (2 c+3)
 \hat{J}^{i+2} \hat{G}^i \hat{G}^{i+1}
\nonu \\
&+& 72 (3 c^2+7 c+6) 
\delta^{ij} \hat{T} \hat{T}
-36(c+6)  \hat{T} \hat{J}^i \hat{J}^i
-  144 (2 c+3)  \delta^{ij} \sum_{k} \hat{T} \hat{J}^k \hat{J}^k
\nonu \\
& - & 6 (8 c+9)(c+6)
\pa \hat{G}^i \hat{G}^i
+ 18 (3 c^2-10 c-24) 
\delta^{ij} \sum_{k} \pa \hat{G}^k \hat{G}^k
\nonu \\
& - & 
18 ( c^2-7 c-6)
\delta^{ij} \sum_{k} \pa \hat{J}^k \pa \hat{J}^k
-\frac{18}{c} (c^3+6 c^2+48 c+72)
 \pa \hat{J}^i \pa \hat{J}^i
\nonu \\
&+ &  \frac{3}{c} (10 c^3+48 c^2-243 c-378)  \pa^2 \hat{J}^i \hat{J}^i
- 3 (c-6) (4 c+15)  \delta^{ij} \sum_{k} \pa^2 \hat{J}^k \hat{J}^k
\nonu \\
&+& 6(2c+3)(c^2-3c+18) \pa^2 \hat{T}
\Bigg)_{j=i}
\nonu \\
& + &
 \frac{1 }{(c+1) (c+6) (2 c-3)}
\Bigg( 
 6 (17 c^2+33 c+18) \hat{G}^i \pa \hat{G}^{i+1}
+ 18 i (c+6) \hat{J}^i \hat{G}^i \hat{G}^{i+2}
\nonu \\
& - &  
18 i (7 c+6) \hat{J}^i \hat{J}^i \pa \hat{J}^{i+2}
-
9 (2 c^2-47 c-66) \hat{J}^i \pa^2 \hat{J}^{i+1}
- 18 i (c+6) \hat{J}^{i+1} \hat{G}^{i+1} \hat{G}^{i+2}
\nonu \\
&- &  162 i (c+2) \hat{J}^{i+1} \hat{J}^{i+1} \pa \hat{J}^{i+2} 
- 36 (c+6) \hat{T} \hat{J}^i \hat{J}^{i+1}
+ 24 i (5 c^2+15 c+18) \hat{T} \pa \hat{J}^{i+2}
\nonu \\
&+& 18 (3 c^2-8 c-12) \pa \hat{G}^i \hat{G}^{i+1}
-\frac{18}{c} i (17 c^2+27 c-18) \pa \hat{J}^i \hat{J}^i \hat{J}^{i+2}
\nonu \\
& - & \frac{18}{c} (c^3-11 c^2-33 c-18) \pa \hat{J}^i \pa \hat{J}^{i+1} 
-\frac{54}{c} i (5 c^2+7 c+6) \pa \hat{J}^{i+1} \hat{J}^{i+1} \hat{J}^{i+2}
\nonu \\
&-& 216 i (2c+3) \pa \hat{J}^{i+2} \hat{J}^{i+2}  \hat{J}^{i+2} 
+ 6 i (20 c^2+57 c+54) \pa \hat{T} \hat{J}^{i+2}
\nonu \\
& + & \frac{3}{c} (16 c^3+9 c^2-99 c-54) \pa^2 \hat{J}^i \hat{J}^{i+1}
+ i (2 c^3-77 c^2-249 c-234) \pa^3 \hat{J}^{i+2}
\Bigg)_{j= i+1} \nonu \\
& - & \frac{3}{c} \hat{\phi}^{(2),i} \hat{\phi}^{(2),j}
\nonu \\
&+&  C_{(\frac{3}{2}) (\frac{3}{2})}^{(2)}
\Bigg(
-  \frac{9 (3 c-1) }{5 (c-3) c}  
\delta^{ij} \hat{G}^k \hat{\phi}^{(\frac{5}{2}),k}
-  \frac{3 (c+3)  }{5 (c-3) c}  
\delta^{ij} \hat{J}^k \hat{\psi}^{(3),k}
+ \frac{18 (3 c-1)  }{5 (c-3) c} 
 \delta^{ij} \hat{T} \psi^{(2)} 
\nonu \\
& + & \frac{3 (c-7)  }{10 (c-3)} 
\delta^{ij} \pa^2 \hat{\psi}^{(2)}
+ \frac{3  }{(c-3)} 
\hat{G}^i \hat{\phi}^{(\frac{5}{2}),j}
+\frac{3  }{(c-3)} 
\hat{G}^j \hat{\phi}^{(\frac{5}{2}),i}
\nonu \\
& + & \frac{3  }{(c-3)} 
\hat{J}^i \hat{\psi}^{(3),j} +
\frac{3  }{(c-3)} 
\hat{J}^j \hat{\psi}^{(3),i}
+  \frac{3 i (7 c-9)  }{5 (c-3) c}  
\epsilon^{ijk} \pa (\hat{J}^k  \hat{\psi}^{(2)})
\nonu \\
& - & 
  \frac{i (c+3) }{10 (c-3)}  
\epsilon^{ijk} \pa \hat{\psi}^{(3),k} \Bigg)
+  C_{(\frac{3}{2}) (3)}^{(\frac{5}{2})} \Bigg(
\frac{6}{5} i   \epsilon^{ijk} \pa \hat{\phi}^{(3),k}
 + \frac{4}{5}   \delta^{ij}   \hat{\phi}^{(4)} \Bigg) 
\nonu \\
&+ & 
C_{(\frac{3}{2}) (\frac{5}{2})}^{(3)} \Bigg(
\frac{1}{2} i   
\epsilon^{ijk} \pa \hat{\psi}^{(3),\alpha=k} 
+\frac{1}{2}   \hat{\psi}^{(4),i,\alpha=j}
+\frac{1}{2}  \hat{\psi}^{(4),j,\alpha=i}
\Bigg)
\Bigg](w)
+ \cdots,
\nonu \\
\hat{\psi}^{(\frac{5}{2}),i}(z) \, \hat{\phi}^{(3)}(w) & = &
\frac{1}{(z-w)^4}  \frac{12 (5 c-6)}{(2 c-3)}  \hat{G}^i(w) \nonu \\
& + & \frac{1}{(z-w)^3} \Bigg[ \frac{1 }{(c+1) (2 c-3)}
\Bigg(
12 i (2 c+3)
\epsilon^{ijk} \hat{J}^j \hat{G}^k
+  4 c (5 c+3)
\pa \hat{G}^i \Bigg)
\nonu \\
& - & 
\frac{(c+3)  }{2 c}
  C_{(\frac{3}{2}) (\frac{3}{2})}^{(2)} \hat{\phi}^{(\frac{5}{2}),i}
\Bigg](w)
\nonu \\
&+& \frac{1}{(z-w)^2}  \Bigg[ \frac{1 }{(c+1) (c+6) (2 c-3)}
\Bigg(
9 (37 c+54)  \hat{J}^i \hat{J}^j \hat{G}^j
-252 (2 c+3)  \hat{J}^j \hat{J}^j \hat{G}^i
\nonu \\
&+ & 6 i (11 c^2-81 c-126)
\epsilon^{ijk} \hat{J}^j \pa \hat{G}^k
+ 
6 (55 c^2+87 c+54) 
\hat{T} \hat{G}^i
\nonu \\
& - & 3 i (25 c^2-144 c-252)
  \epsilon^{ijk}
\pa \hat{J}^j \hat{G}^k
\nonu \\
& + & 
 (5 c^3-72 c^2-279 c-270)
\pa^2 \hat{G}^i \Bigg)
-\frac{9}{2 c} \hat{\psi}^{(\frac{3}{2})} \hat{\phi^{(2),i}}
\nonu \\
& + & 
 C_{(\frac{3}{2}) (\frac{3}{2})}^{(2)} \Bigg(
\frac{3 (23 c-21)  } 
{10 (c-3) c}  \hat{G}^i \hat{\psi}^{(2)}
+  \frac{3 i (13 c-21)  }{10 (c-3) c}
 \epsilon^{ijk} \hat{J}^j \hat{\phi}^{(\frac{5}{2}),k}
 -\frac{(2 c^2+21 c-45)}{10 (c-3) c}
\pa \hat{\phi}^{(\frac{5}{2}),i} \Bigg)
\nonu \\
& + &  
\frac{7  }{5}  C_{(\frac{3}{2}) (3)}^{(\frac{5}{2})} \hat{\psi}^{(\frac{7}{2}),i}
-\frac{1}{2} i 
 C_{(\frac{3}{2}) (\frac{5}{2})}^{(3)} \epsilon^{ijk} \hat{\phi}^{(\frac{7}{2}),j,\alpha=k}
\Bigg](w)
\nonu \\
& + & \frac{1}{(z-w)} \Bigg[ 
\frac{1}{(c+1) (c+6) (2 c-3)}
\Bigg(
9 (17 c+30)
\pa \hat{J}^i \hat{J}^j  \hat{G}^j
+ 9 (17 c+30)
\hat{J}^i \hat{J}^j \pa \hat{G}^j
\nonu \\
&+ & 9 (13 c+6) 
\hat{J}^i \pa \hat{J}^j \hat{G}^j
- 216 (2c+3) \pa \hat{J}^j  \hat{J}^j \hat{G}^i
- 108 (2c+3)  \hat{J}^j  \hat{J}^j \pa \hat{G}^i
\nonu \\
& + &  6 i (5 c^2-33 c-54) 
\epsilon^{ijk} \hat{J}^j \pa^2 \hat{G}^k
+
 54 (3 c^2+7 c+6)
\pa \hat{T}  \hat{G}^i
\nonu \\
&+& 
6 (19 c^2+3 c-18)  \hat{T}  \pa \hat{G}^i
- 21 i c(c+6)
\epsilon^{ijk} \pa \hat{J}^j \pa \hat{G}^k
-  27 i (c^2-8 c-12)  \epsilon^{ijk} 
\pa^2 \hat{J}^j \hat{G}^k
\nonu \\
& + & (c^3-32 c^2-75 c-54)
\pa^3 \hat{G}^i \Bigg)
+ \frac{3}{2 c} \pa \hat{\psi}^{(\frac{3}{2})} \hat{\phi}^{(2),i}
- \frac{9}{2 c}  \hat{\psi}^{(\frac{3}{2})} \pa \hat{\phi}^{(2),i}
\nonu \\
&+&   C_{(\frac{3}{2}) (\frac{3}{2})}^{(2)}
\Bigg(
\frac{9 (4 c-3)  }{10 (c-3) c} 
\hat{G}^i \pa \hat{\psi}^{(2)}
-\frac{3 i  }{2 (c-3)}   \epsilon^{ijk}
\hat{G}^j \hat{\psi}^{(3),k}
-  \frac{3  }{(c-3)}  \hat{J}^i \hat{\phi}^{(\frac{7}{2})}
\nonu \\
& - & \frac{3  }{(c-3)}  \hat{T} \hat{\phi}^{(\frac{5}{2}),i}
+ \frac{3 i ( c-12)  }{5 (c-3) c}  
\epsilon^{ijk} \pa \hat{J}^j \hat{\phi}^{(\frac{5}{2}),k}
+\frac{3 (7 c-9)  }{10 (c-3) c}  
\pa \hat{G}^i \hat{\psi}^{(2)}
\nonu \\
& - &  
\frac{(c^2+18 c-45)  }{20c (c-3)} 
\pa^2 \hat{\phi}^{(\frac{5}{2}),i} +  \frac{3 i (7 c-9)  }{10 (c-3) c}  
\epsilon^{ijk} \hat{J}^j \pa \hat{\phi}^{(\frac{5}{2}),k}\Bigg)
\nonu \\
&+& \frac{3}{5}  C_{(\frac{3}{2}) (3)}^{(\frac{5}{2})} \pa \hat{\psi}^{(\frac{7}{2}),i}
+ C_{(\frac{3}{2}) (\frac{5}{2})}^{(3)} \Bigg(
-\frac{1}{4} i   \epsilon^{ijk} 
\pa \hat{\phi}^{(\frac{7}{2}),j,\alpha=k} -\frac{1}{2}  
\hat{\phi}^{(\frac{9}{2}),\alpha=i} \Bigg)
\Bigg](w)
+\cdots,
\nonu \\
\hat{\phi}^{(3)}(z) \, \hat{\phi}^{(3)}(w) & = & 
\frac{1}{(z-w)^6} \, 2(5c-3) +\frac{1}{(z-w)^4} \Bigg[ 
\frac{1 }{(c+1) (2 c-3)} \Bigg(
24 (5 c^2-9)  \hat{T}
\nonu \\
& - &  36 (c-1) \hat{J}^i \hat{J}^i \Bigg)
+ \frac{3 (c-7) }{2 c}
   C_{(\frac{3}{2}) (\frac{3}{2})}^{(2)} \hat{\psi}^{(2)}
\Bigg](w)
\nonu \\
&+& \frac{1}{(z-w)^3} 
 \Bigg[  \frac{1}{2} \pa (\mbox{pole four})
\Bigg](w)
\nonu \\
& + & \frac{1}{(z-w)^2} \Bigg[ 
\frac{1 }{(c+1) (c+6) (2 c-3)} \Bigg(
-144 i (2 c+3)  \epsilon^{ijk} \hat{J}^i \hat{G}^j \hat{G}^k
\nonu \\
&+ & 12 (7 c^2-42 c-72)  \pa \hat{G}^i \hat{G}^i
-\frac{18}{c}  (3 c^3-8 c^2+36 c+72)   \pa \hat{J}^i \pa \hat{J}^i 
\nonu \\
&+& 
18 (c^2+15 c+18) 
\pa^2 \hat{J}^i \hat{J}^i
- 288 (2 c+3) \hat{T} \hat{J}^i \hat{J}^i
\nonu \\
& + & 18 (c^3-4 c^2+18 c+36)
\pa^2 \hat{T} 
+24 (16 c^2+27 c+18) \hat{T} \hat{T}
\Bigg)
-\frac{18}{c} \pa \hat{\psi}^{(\frac{3}{2})} \hat{\psi}^{(\frac{3}{2})} 
\nonu \\
&+&  C_{(\frac{3}{2}) (\frac{3}{2})}^{(2)} \Bigg(
-\frac{3 (11 c-12)  }{10 (c-3) c} 
\hat{G}^i \hat{\phi}^{(\frac{5}{2}),i}
+  \frac{9 (c-2)  }{5 (c-3) c}   \hat{J}^i 
\hat{\psi}^{(3),i}
+ \frac{9 (c-5) (c+3) }{40 (c-3) c}
\pa^2 \hat{\psi}^{(2)} \nonu \\
& + & \frac{12 (4 c-3) }{5 (c-3) c}
\hat{T} \hat{\psi}^{(2)} \Bigg)
+  \frac{8}{5} C_{(\frac{3}{2}) (3)}^{(\frac{5}{2})} \hat{\phi}^{(4)}
+ \frac{1}{4}    C_{(\frac{3}{2}) (\frac{5}{2})}^{(3)}  \hat{\psi}^{(4),i,\alpha=i}
\Bigg](w) 
\nonu \\
&+& \frac{1}{(z-w)} \Bigg[  -\frac{1}{24} \pa^3 (\mbox{pole four}) + \frac{1}{2} \pa  (\mbox{pole two} )
\Bigg](w)
+\cdots.
\label{remainingthree}
\eea}
There are also 
the nonlinear terms between the higher spin currents
in (\ref{remainingthree}).
Compared to the previous OPEs in this Appendix,
the coefficients appearing in the higher spin currents
in (\ref{remainingthree}) 
are not simply equal to those in Appendix $F$. 
This implies that 
the extra terms appearing on the right hand sides of (\ref{relation})
can contribute to the higher spin current dependent terms in 
(\ref{remainingthree}).  
The presentation in the first-order pole of 
the OPE between the higher spin-$\frac{5}{2}$ currents 
is rather complicated and we can analyze the notations here 
by doing similar procedures in (\ref{nextthree}).
The cases $(i,j)=(2,2)$ and $(3,3)$
can be read off from the case $(i,j)=(1,1)$
while the cases  $(i,j)=(2,3)$ and $(3,1)$
can be obtained from the case $(i,j)=(1,2)$.


\section{Further ${\cal N}=3$ description for low $(N,M)$ cases }

In this Appendix, we describe the ${\cal N}=3$ OPEs for different
$(N,M)$ cases which are not explained in the main text.
There is no higher spin-$1$ current 
for all of the 
following cases as before and 
it is identically and trivially zero.
Then we construct the nontrivial lowest higher spin-$\frac{3}{2}$ current 
for each case.

\subsection{ The  $(N,M)=(2,1)$ case}
 
Since $M=1$, there are only $SU(N)$ adjoint fermions.
One can check that
there exists the higher spin-$\frac{3}{2}$ current 
and it is similar to the one for $(N,M)=(2,2)$ case
by removing the $SU(M)$ adjoint fermions.
From the general expression 
(\ref{spin3halfexpression}), one has
\bea
\psi^{(\frac{3}{2})}(z) \sim \Bigg(  a J_1^{\alpha} 
\Psi^{\alpha}+ J_2^{\alpha} \Psi^{\alpha}+ 
j^{\alpha} \Psi^{\alpha} \Bigg)(z),
\label{3halfexpress}
\eea
where 
$a$ 
is some constant $a(2,1)$ defined in Appendix (\ref{nmdependent})
and the index $\alpha$ is $SU(2)$ adjoint index.
The spin-$\frac{1}{2}$ and spin-$1$ currents are given in Appendix $C$.
Then we can calculate the various OPEs based on 
the higher spin-$\frac{3}{2}$ current (\ref{3halfexpress}) 
and it turns out that 
there are no other higher spin currents in the 
OPE ${\bf \Phi}^{(\frac{3}{2})}(Z_1) \, {\bf \Phi}^{(\frac{3}{2})}(Z_2)$
and can be summarized as 
follows in the simplified notation
\bea
\left[ {\bf \Phi}^{(\frac{3}{2})} \right] \cdot \left[  {\bf \Phi}^{(\frac{3}{2})}
\right] =  \left[ {\bf I}\right]. 
\label{opefirst}
\eea
Here $ \left[ {\bf I}\right]$ stands for 
the ${\cal N}=3$ superconformal family of the identity operator.  
The explicit result can be seen from  (\ref{finalPhiPhi}) 
by  neglecting all the higher spin currents appearing in the 
right hand side.
Furthermore, we can try to find whether the higher spin-$2$ current
exists or not.
If we require that the general higher spin-$2$ ansatz  should satisfy 
the primary conditions given in Appendix $B$, 
this higher spin-$2$ current vanishes identically.
This argument also holds for the other higher spin-$\frac{5}{2}$ current.

\subsection{ The $(N,M)=(3,1), (4,1), (5,1)$ cases}

What happens for other $N$ values for fixed $M=1$?
Let us increase the $N$ value for fixed $M=1$.
As in  the above case, the higher spin-$\frac{3}{2}$ current  
is given by 
\bea
\psi^{(\frac{3}{2})}(z) \sim \Bigg(  a J_1^{\alpha} \Psi^{\alpha}+ J_2^{\alpha} \Psi^{\alpha}+ 
j^{\alpha} \Psi^{\alpha} \Bigg)(z),
\label{spin3halfother}
\eea
which comes from (\ref{spin3halfexpression}) or (\ref{3halfexpress}).
Again, 
$a$ is some constant $a(N,1)$ in Appendix (\ref{nmdependent}) 
and the index $\alpha$ is $SU(N=3,4,5)$ 
adjoint index.
For $M=1$, there are no $SU(M)$ adjoint fermions in (\ref{spin3halfother}).
Again, there are no higher spin-$2$ currents for each case in general.
However, there exists the higher spin-$\frac{5}{2}$ current
for each case. 
Moreover, the ${\cal N}=3$ OPE between the lowest higher spin-$\frac{3}{2}$
current 
can be summarized by
\bea
\left[ {\bf \Phi}^{(\frac{3}{2})} \right] \cdot \left[  {\bf \Phi}^{(\frac{3}{2})}
\right] =  \left[ {\bf I}\right] +\left[ {\bf \Phi}^{(\frac{5}{2})} 
\right],
\label{opesecond}
\eea
from (\ref{spin3halfother}).
The explicit result is the same as (\ref{finalPhiPhi}) once 
we ignore the higher spin currents in the right hand side of 
(\ref{finalPhiPhi}) 
except ${\bf \Phi}^{(\frac{5}{2})}(Z_2)$.

When $M=1$, our coset is (that is, 
we can ignore the $\hat{SU}(M)_{2N+M}$ factor 
in the denominator of (\ref{KScoset}))
given by 
\bea
\frac{\hat{SU}(N+M)_{N+M} \oplus \hat{SO}(2 N M)_{1}}{\hat{SU}(N)_{N+2M} 
\oplus \hat{U}(1)_{2NM(N+M)^2}}.
\label{noSUMcoset}
\eea
When the free fermions, $\Psi^{\rho}(z)$ and $\Psi^{u(1)}(z)$, 
are decoupled, 
the above coset (\ref{noSUMcoset}) is dual to the following coset 
(see $(4.13)$ in \cite{CHR1406})
\bea
\frac{\hat{SU}(N+2M)_{N} }{\hat{SU}(N)_{N} 
\oplus \hat{U}(1)_{2N^2 M(N+2M)}}.
\label{Dualcoset}
\eea
In the large $N$ limit, 
the vacuum character of (\ref{Dualcoset}) is (see $(3.13)$ 
with replaced $M$ by $2M$ 
in \cite{CHR1406})
\bea
\chi_0^{2M}= \Bigg( \prod_{s=2}^{\infty} z_B^{(s)}(q) z_F^{(s)}(q) \Bigg)^{4M^2}
\Bigg( z_B^{(1)}(q) \Bigg)^{4M^2-1},
\label{vacuumcha}
\eea
where  $z_B^{(s)}(q) \equiv \prod_{n=s}^{\infty} \frac{1}{(1-q^n)}$
and   $z_F^{(s)}(q) \equiv  \prod_{n=s}^{\infty} (1+q^{n-\frac{1}{2}})$ 
are defined in $(3.7)$ and $(3.13)$ of \cite{CHR1406}.
When $M=1$, the vacuum character (\ref{vacuumcha}) 
implies that 
there are three spin-$1$ currents and  four spin-$s$ currents
for every half-integer spin $s$ greater than $1$. 
This interpretation  partly matches with our spin content
in the lower spin cases when we ignore 
spin-$\frac{1}{2}$ current $\Psi (z)$($=\sqrt{\frac{N M }{2}} \Psi^{u(1)}(z)$) 
in the ${\cal N}=3$ superconformal algebra (\ref{nonKB}). 
Note that there are no $\hat{SU}(M)$ currents
$\Psi^{\rho}(z)$ because of $M=1$.
The previous $(N,M)=(2,1)$ 
case matches with (\ref{vacuumcha}) for spin $s=1, \frac{3}{2}$, and $2$.
Recall that the ${\cal N}=3$ superconformal algebra 
contains three spin-$1$, three spin-$\frac{3}{2}$ and one spin-$2$
currents while the higher spin-$\frac{3}{2}$ current 
${\bf \Phi}^{(\frac{3}{2})}(Z)$ 
contains
the one higher spin-$\frac{3}{2}$ and three higher spin-$2$ currents. 

The present case  matches with (\ref{vacuumcha}) for spin $s=1, 
\frac{3}{2}, 2, \frac{5}{2}$, and $3$.
In this case, there are three higher spin-$\frac{5}{2}$ currents 
from the lowest higher spin-$\frac{3}{2}$ current ${\bf \Phi}^{(\frac{3}{2})}(Z)$
and one  higher spin-$\frac{5}{2}$ current from 
${\bf \Phi}^{(\frac{5}{2})}(Z)$. Moreover,
one sees that one higher spin-$3$ current from 
${\bf \Phi}^{(\frac{3}{2})}(Z)$
and  three higher spin-$3$ currents 
from 
${\bf \Phi}^{(\frac{5}{2})}(Z)$.
We expect that if we increase the $N$ values and 
find more higher spin currents, 
then  the spin content will match further beyond the higher spin-$3$ 
current.

\subsection{ The $(N,M)=(2,2)$ case}

What happens for different $M$ value? 
Let us increase the $M$ value.
For $M=2$, the nontrivial case arises for $N=2$.
From the general expression (\ref{spin3halfexpression}), one has 
\bea
\psi^{(\frac{3}{2})}(z)
&\sim&   d \Bigg(  a J_1^{\alpha} \Psi^{\alpha}+ J_2^{\alpha} \Psi^{\alpha}+ 
j^{\alpha} \Psi^{\alpha} \Bigg)(z)
+  \Bigg( 
b J_1^{\rho} \Psi^{\rho} + J_2^{\rho} \Psi^{\rho}+  j^{\rho} \Psi^{\rho} \Bigg)(z).
\label{3halftwoterms}
\eea
Note that the index $\rho$ runs over $1, 2, \cdots, M^2-1$
for general $M$.
For $M \neq 1$, we have the terms having $SU(M)$ adjoint $\rho$ indices in 
(\ref{3halftwoterms}).
There are new higher spin currents 
${\bf \Phi}^{(2)}(Z)$, $ {\bf \Phi}^{(\frac{5}{2})}(Z)$,
and ${\bf \Phi}^{(3), i}(Z)$
 which are not present for $M=1$ cases described before
 \footnote{We can also construct the new higher spin currents 
${\bf \Phi}^{(\frac{3}{2}), i}(Z)$ for $(N,M)=(2,2)$ case.
However,  they do not appear in the OPE 
${\bf \Phi}^{(\frac{3}{2})}(Z_1) \, {\bf \Phi}^{(\frac{3}{2})}(Z_2)$.}.
Compared to the previous cases (\ref{opefirst}) and (\ref{opesecond}),
we have the following OPE
\bea
\left[ {\bf \Phi}^{(\frac{3}{2})} \right] \cdot \left[  {\bf \Phi}^{(\frac{3}{2})}
\right] =  \left[ {\bf I}\right] +
\left[ {\bf \Phi}^{(2)} \right] +  \left[ {\bf \Phi}^{(\frac{5}{2})} 
\right]
+  \theta^{3-i} \left[ {\bf \Phi}^{(3), i} \right],
\label{opethird}
\eea
which appeared in section $7$.
So far, the vacuum character in this case corresponding to 
(\ref{vacuumcha}) is not known.
It would be interesting to obtain this vacuum character 
by dividing the $\hat{SU}(M)$ factor from the result in \cite{CV1312}. 

\subsection{ The $(N,M)=(3,2)$ case}

Let us increase the $N$ value for fixed $M=2$.
Once again, the higher spin-$\frac{3}{2}$ current is given by
(\ref{spin3halfexpression}) for $(N,M)=(3,2)$
\bea
\psi^{(\frac{3}{2})}(z)
&\sim&   d \Bigg(  a J_1^{\alpha} \Psi^{\alpha}+ J_2^{\alpha} \Psi^{\alpha}+ 
j^{\alpha} \Psi^{\alpha} \Bigg)(z)
+  \Bigg( 
b J_1^{\rho} \Psi^{\rho} + J_2^{\rho} \Psi^{\rho}+  j^{\rho} \Psi^{\rho} \Bigg)(z).
\label{finalspin3half}
\eea
In general, the new higher spin fields 
can arise in the coset when $(N,M)$ increase. 
We can investigate whether new higher spin fields arise in the OPE 
${\bf \Phi}^{(\frac{3}{2})}(Z_1) {\bf \Phi}^{(\frac{3}{2})}(Z_2)$ or not.
It is sufficient to see 
the singular terms of the OPEs between the lowest component field
$\psi^{(\frac{3}{2})}(z)$  and all the other component fields of 
${\bf \Phi}^{(\frac{3}{2})}(Z)$ with (\ref{finalspin3half}) using the 
${\cal N}=3$ supersymmetry. 
It turns out that there are no new higher spin currents. Again, 
we summarize the OPE as follows as in (\ref{opethird}):
\bea
\left[ {\bf \Phi}^{(\frac{3}{2})} \right] \cdot \left[  {\bf \Phi}^{(\frac{3}{2})}
\right] =  \left[ {\bf I}\right] +
\left[ {\bf \Phi}^{(2)} \right] +  \left[ {\bf \Phi}^{(\frac{5}{2})} 
\right]
+  \theta^{3-i} \left[ {\bf \Phi}^{(3), i} \right].
\label{finalfusion}
\eea
The explicit check for the other values of $(N,M)$ is computationally 
involved. If there are extra higher spin currents in the above 
(\ref{finalfusion})
for large $(N,M)$ values, 
we expect that 
they (and their descendant fields) will appear linearly \cite{BS}. 


\end{document}